\documentclass{ar-1col}

\usepackage{natbib}
\usepackage{amssymb}

\usepackage[total={30.5pc,47pc},centering]{geometry}

\usepackage[colorlinks,citecolor=blue]{hyperref} 
\usepackage[all]{hypcap} 

\PassOptionsToPackage{unicode}{hyperref}
\PassOptionsToPackage{naturalnames}{hyperref}

\newcommand{\pasp}{Pub. Astron. Soc. Pac.}
\newcommand{\apj}{Ap. J.}
\newcommand{\apjl}{Ap. J. Let.}
\newcommand{\aj}{Astron. J.}

\newcommand{\apjs}{Ap. J. Suppl.}
\newcommand{\mnras}{MNRAS}
\newcommand{\aap}{Astron. Astrophys.}
\newcommand{\aaps}{Astron. Astrophys. Suppl.}
\newcommand{\araa}{Annu. Rev. Astron. Astrophys.}
\newcommand{\nat}{Nature}

\newcommand{\zap}{Zeitschrift f\"ur Astroph.}
\newcommand{\aapr}{Astron. Astrophys. Rev.}
\newcommand{\ssr}{Space Sci. Rev.}

\newcommand{\sauron}{\texttt{SAURON}}
\newcommand{\atl}{ATLAS$^{\rm 3D}$}
\newcommand{\kms}{\mbox{km s$^{-1}$}}
\newcommand{\msun}{\mbox{$\rm M_\odot$}}

\newcommand{\re}{\mbox{$R_{\rm e}$}}
\newcommand{\rmaj}{\mbox{$R_{\rm e}^{\rm maj}$}}
\newcommand{\se}{\mbox{$\sigma_{\rm e}$}}
\newcommand{\mjam}{\mbox{$M_{\rm JAM}$}}

\newcommand{\vse}{\mbox{$(V/\sigma,\varepsilon)$}}
\newcommand{\lre}{\mbox{$(\lambda_{R_{\rm e}},\varepsilon)$}}
\newcommand{\lr}{\mbox{$\lambda_{R_{\rm e}}$}}
\newcommand{\vs}{\mbox{$V/\sigma$}}
\newcommand{\mlpop}{\mbox{$(M_{\textstyle \ast}/L)_{\rm pop}$}}
\newcommand{\mldyn}{\mbox{$(M_{\textstyle \ast}/L)_{\rm dyn}$}}

\newcommand{\mdyn}{\mbox{$M_{\textstyle \ast}^{\rm dyn}$}}
\newcommand{\mstar}{\mbox{$M_{\textstyle \ast}$}}
\newcommand{\bigast}{\textstyle \ast}

\newcommand{\dd}{\,\mathrm{d}}

\newcommand\farcs{\mbox{$.\!\!^{\prime\prime}$}}
\let\la=\lesssim
\let\ga=\gtrsim 

\jname{Annu. Rev. Astron. Astrophys.}
\jyear{2016}
\jvol{54}
\doi{\href{http://www.annualreviews.org/doi/abs/10.1146/annurev-astro-082214-122432}{10.1146/annurev-astro-082214-122432}}

\begin{document}

\title{Structure \& Kinematics of Early-Type Galaxies from Integral-Field Spectroscopy}

\markboth{M.~Cappellari}{Structure of early-type galaxies}

\author{Michele Cappellari
\affil{Sub-department of Astrophysics, Department of Physics, University of Oxford, \\
Denys Wilkinson Building, Keble Road, Oxford OX1 3RH}}

\begin{abstract}
Observations of galaxy isophotes, longs-slit kinematics and high-resolution photometry suggested a possible dichotomy between two distinct classes of E galaxies. But these methods are expensive for large galaxy samples.
Instead, integral-field spectroscopic can efficiently recognize the shape, dynamics and stellar population of complete samples of early-type galaxies (ETGs). These studies showed that the two main classes, the fast and slow rotators, can be separated using stellar kinematics. We showed there is a dichotomy in the dynamics of the two classes. The slow rotators are weakly triaxial and dominate above $M_{\rm crit}\approx2\times10^{11} \msun$. Below $M_{\rm crit}$, the structure of fast rotators parallels that of spiral galaxies. There is a smooth sequence along which, the metals content, the enhancement in $\alpha$-elements, and the ``weight'' of the stellar initial mass function, all increase with the {\em central} mass density slope, or bulge mass fraction, while the molecular gas fraction correspondingly decreases.
The properties of ETGs on galaxy scaling relations, and in particular the $(\mstar,\re)$ diagram, and their dependence on environment, indicate two main independent channels for galaxy evolution. Fast rotators ETGs start as star forming disks and evolve trough a channel dominated by gas accretion, bulge growth and quenching. While slow rotators assemble near the center of massive halos via intense star formation at high redshift, and remain as such for the rest of their evolution via a channel dominated by gas poor mergers. This is consistent with independent studies of the galaxies redshift evolution.
\end{abstract}

\begin{keywords}
galaxies: elliptical and lenticular, cD --
galaxies: evolution --
galaxies: formation --
galaxies: structure --
galaxies: kinematics and dynamics
\end{keywords}
\maketitle

\section{INTRODUCTION}

Our knowledge of the structure of galaxies in general, and of early-type galaxies (ETGs: ellipticals Es and S0s) in particular, has evolved in parallel with the technological advances in the instrumentation used to study them. 
Galaxy photographic plates formed the basis of the classic and still widely popular galaxy morphological classification scheme by \citet{Hubble1926}.
From the late 50s rotation curves were obtained from ionized gas emission in external spiral galaxies \citep[e.g.][]{Burbidge1959}. This revealed flat rotation curves out to large radii \citep{Rubin1970}, subsequently strengthened by radio determinations \citep{Bosma1978}, indicating dark matter surrounding the galaxies \citep[see][for a review]{Courteau2014}. Dark matter is one of the pillars of the current paradigm of how galaxies form \citep{White1978,Blumenthal1984}.

\begin{marginnote}[120pt]
\entry{E}{Elliptical galaxy. Displays elliptical isophotes on the sky}
\entry{S0}{Has an outer disk like spiral galaxies but smooth appearance, without spiral arms}
\end{marginnote}
 
In the 70s long-slit spectrographs were used measure stellar rotation in elliptical galaxies. This revealed that the massive Es tend to rotate slowly, while less massive ones and galaxy bulges appeared to rotate faster (\autoref{sec:vs_eps_before_ifs}). In the 80s photometry of Es using CCD detectors showed that faint ellipticals can have disky isophotes, suggesting the presence of stellar disks embedded in dominant stellar spheroids (\autoref{sec:isophotal_shapes}).
The {\em Hubble Space Telescope} (HST) in the 90s showed that the steepness of the inner galaxy profiles also relates to the other galaxy properties (\autoref{sec:nuclear_profiles}). HST also revolutionized our understanding of the link between galaxies and BHs evolution \cite[see][for a review]{Kormendy2013review}.

In parallel to detailed studies of nearby galaxies, deep multi-band photometric galaxy surveys (e.g.\ GOODS \citealt{Giavalisco2004} and COMSOS \citealt{Scoville2007}), and multiplexed spectroscopic surveys (e.g. DEEP2 \citealt{Newman2013deep2}) were delivered. These provided the ability to trace galaxy evolution back in time \citep[see][for a review]{Conselice2014}.

The present review focuses on the progress in our knowledge of the structure of ETGs, and of galaxies in general, brought by the next major technological advance: integral-field spectroscopy (IFS). This brings the ability to obtain a spectrum at every position on a grid of sky coordinates covering the galaxy image. Effective IFS prototypes appeared on the scene in the 90s \citep[e.g.][]{Bacon1995}, but it took a decade for them to reach maturity. IFS has recently become a standard asset on all major telescopes and an essential tool of astrophysics research.

\begin{marginnote}[120pt]
\entry{IFS}{Integral-field spectroscopy}
\end{marginnote}

This review summarizes the status of our knowledge before another instrumental revolution. This is the time when the first generation of IFS surveys, which targeted one galaxy at a time (\sauron\ \citealt{deZeeuw2002}, \atl\ \citealt{Cappellari2011a}, CALIFA \citealt{Sanchez2012}, DiskMass \citealt{Bershady2010}) have been completed and a new generation of multiplexed IFS surveys, which can observe multiple galaxies simultaneously, is starting (SAMI \citealt{Bryant2015} and MaNGA \citealt{Bundy2015} galaxy surveys).

\section{STRUCTURE FROM PHOTOMETRY}
\label{sec:photometry}

This section gives a concise summary of the main findings on the structure of ETGs before the publication of the first large set of IFS kinematics for ETGs in \citet{Emsellem2004}. Most early developments were based on photometry alone. Extensive reviews were written on the classification \citep{Sandage1975,Sandage2005} and photometry \citep{Kormendy2009,Kormendy2012,Graham2013} of ETGs. These include a summary of the historic developments. We will not duplicate that material here but rather refer the reader to those publications. We only mention the results and definitions which are essential to understand the findings from two-dimensional spectroscopy, which is the focus of this review.

\subsection{Definition of early-type galaxy}

\begin{marginnote}[120pt]
\entry{ETG}{Early-Type Galaxy: E or S0. Defined by the lack of spiral arms in optical images}
\end{marginnote}

The focus of this review are ETGs, although it will be clear in what follows that one cannot understand ETGs without linking their structure to the general galaxy population. ETGs can be broadly characterized by their old population, red colors, small amount of gas and dust, and lack of spiral arms. These characteristics however are not equivalent. In fact applying different selection criteria for ETGs leads to quite different sets of galaxies \citep{Strateva2001,Conselice2006,vandenBergh2007,Bernardi2010}. 

Here we adopt the standard definition of ETGs, as consisting the galaxies in the handle of \citet{Hubble1936} empirical tuning-fork diagram. In the {\em revised} Hubble classification system \citep{Sandage1961}, the separation between ETGs and spiral galaxies is entirely based on the presence of spiral arms (or extended dust lanes for edge-on galaxies).
This separation was adopted unchanged by \citet{deVaucouleurs1959,deVaucouleurs1963} and in the very popular Third Reference Catalogue of Bright Galaxies \citep[hereafter RC3]{deVaucouleurs1991}. The same criterion was used in the RC3 follow-up HyperLeda \citep{Paturel2003}. This nearly universal definition of ETGs is also  adopted in this review.

\subsection{Global galaxy profiles}
\label{sec:sersic_profiles}

\begin{marginnote}[120pt]
\entry{Sersic index $n$}{Increases with the galaxy concentration}
\end{marginnote}

For a number of years elliptical galaxies were thought to be well described by the \citet{deVaucouleurs1948} $R^{1/4}$ profile.
The work by \citet{Caon1993} discovered that the profiles of many Es require the more general parametrization proposed by \citet{Sersic1968}:
\begin{equation}
I(R)=I_{\rm e} \exp\left\{-b(n) \left[ \left(\frac{R}{R_{\rm e}}\right)^{1/n} - 1\right]\right\},
\label{eq:sersic}
\end{equation}
where $b(n)\approx2n-0.327$ \citep{Capaccioli1989}. Most importantly, it was found that the Sersic index $n$, related to the galaxy concentration, increases with increasing total galaxy luminosity. This result was confirmed by a number of authors \citep[e.g.][]{DOnofrio1994,Bertin2002fp,Graham2003,Ferrarese2006acs,Kormendy2009}.

The reported correlations were interpreted as due to a systematic change in the intrinsic properties of elliptical galaxies with luminosity. However it can also be partially explained by a systematic variation in the fraction of hidden disks in ellipticals. Most likely {\em both} effects play a role. In fact the Sersic index of bulges of spiral galaxies and S0s tends to be different from that of massive E \citep[e.g.][]{Kormendy2009,Krajnovic2013p17} and systematic variations in the Sersic index of spiral bulges also exist \citep{Andredakis1995}.

\subsection{Isophotal shape}
\label{sec:isophotal_shapes}

\begin{figure}
\centering
\begin{minipage}{.49\textwidth}
\includegraphics[width=\columnwidth]{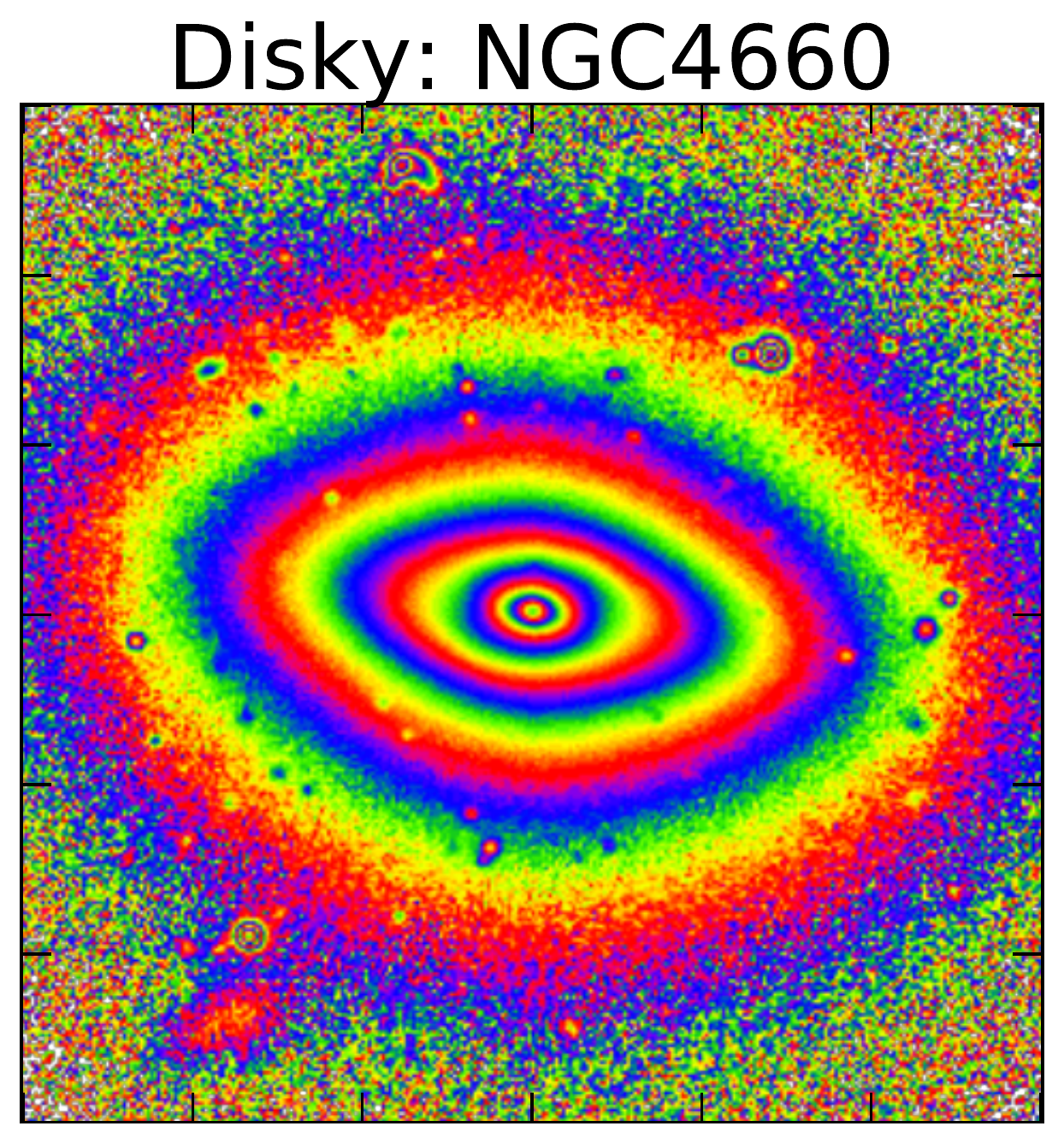}
\end{minipage}
\begin{minipage}{.49\textwidth}
\includegraphics[width=\columnwidth]{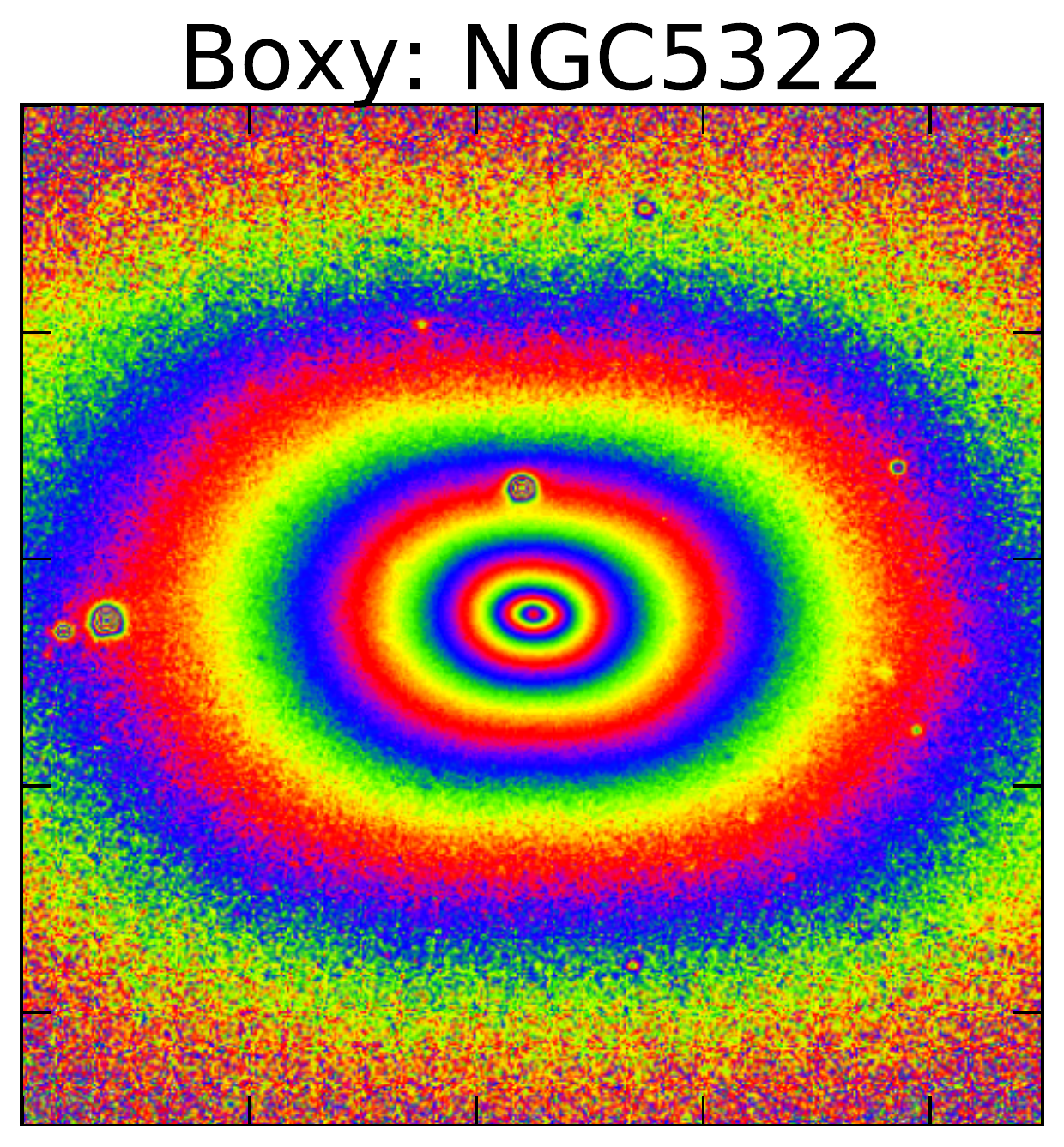}
\end{minipage}
\caption{{\bf Disky versus boxy isophotes.} {\em Left:} NGC~4660 has disky isophotes with $a_4\approx3\%$. {\em Right:} NGC~5322 has boxy isophotes with $a_4\approx-1\%$. The photometry is taken from the SDSS. These galaxies were chosen as representative of the two classes in \citet{Bender1988phot}.}
\label{fig:isophotal_shapes}
\end{figure}

\begin{marginnote}[120pt]
\entry{$a_4$}{Isophotal shape parameter. $a_4>0$ is ``boxy'' and $a_4<0$ is ``disky''}
\end{marginnote}

The shape of Es isophotes is quantified by finding the best-fitting ellipse, which provides a good first-order approximation, and then measuring the deviations from that ellipse. In the standard algorithm \citep{Carter1978,Jedrzejewski1987,Bender1988phot,Franx1989,Peletier1990}, the galaxy surface brightness, sampled along the ellipse, is fitted by the truncated Fourier expansion via linear least squares minimization
\begin{equation}
\Sigma(\psi) = \Sigma_0 + A_1\sin(\psi) + B_1\cos(\psi) + A_2\sin(2 \psi) + B_2\cos(2 \psi),
\label{eq:ellipse_fitting_phot}
\end{equation}
where $\psi$ is the eccentric anomaly, so that the angles are equally spaced when the ellipse is projected onto a circle. The best-fitting ellipse is {\em defined} as the one for which $A_1, B_1, A_2, B_2$ are zero within numerical accuracy. Once the best-fitting ellipse has been determined, the surface brightness along that ellipse is parametrized by the next higher term in the Fourier expansion: 
\begin{equation}
\Sigma(\psi) = \Sigma_0 + A_3\sin(3 \psi) + B_3\cos(3 \psi) + A_4\sin(4 \psi) + B_4\cos(4 \psi),
\label{eq:ellipse_residuals_phot}
\end{equation}
When the profile is properly sampled, the higher terms are mathematically orthogonal to the lower order ones.
The amplitude of the axially-symmetric fourth Fourier coefficient $a_4\equiv\sqrt{A_4^2+B_4^2}$ measures whether an isophote is ``boxy'' ($a_4>0$) or ``disky'' ($a_4<0$; \autoref{fig:isophotal_shapes}). For physical interpretation, the quantity $a_4$ is typically divided by $(a\,\dd\Sigma/\dd a)$, where $a$ is the ellipse semi-major axis, in such a way that $a_4$ represents fractional deviations of the isophote from the best fitting ellipse.

These studies led to the discovery that disky Es appeared to rotate faster than non-disky or boxy Es \citep{Bender1988}. Moreover radio-loud E were found to only be present in E without disky isophotes, showing that isophotal shape was an intrinsic parameter in galaxy structure \citep{Bender1989}. Ultimately, the realization of the important connection of isophotal shape and other global galaxy properties led to the proposal of a new classification scheme for Es \citep{Kormendy1996}.

\begin{marginnote}[120pt]
\entry{Disky elliptical E(d)}{Intermediate between E and S0. The disk dominates the central regions, while the spheroid dominates at large radii}
\end{marginnote}

\subsection{Nuclear galaxy profiles}
\label{sec:nuclear_profiles}

\begin{figure}
\centering
\begin{minipage}{.44\textwidth}
\includegraphics[width=\columnwidth]{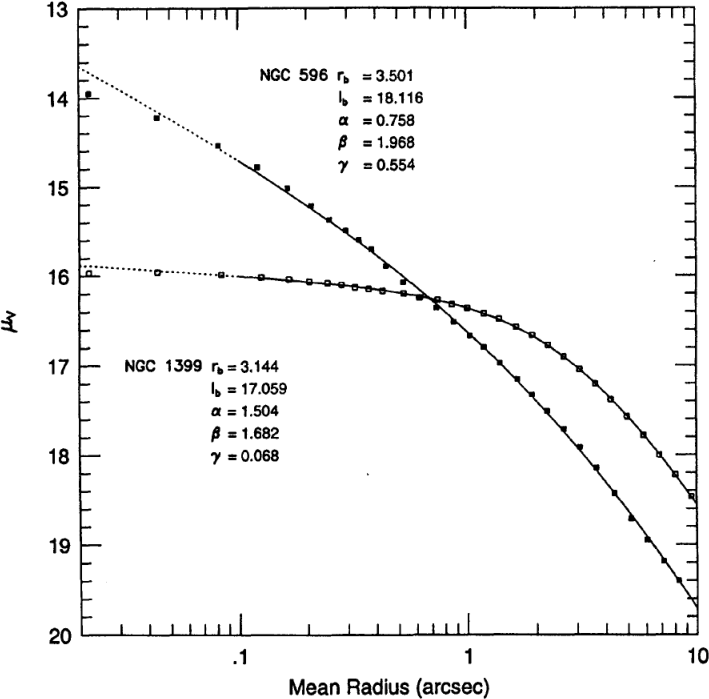}
\end{minipage}
\begin{minipage}{.55\textwidth}
\includegraphics[width=\columnwidth]{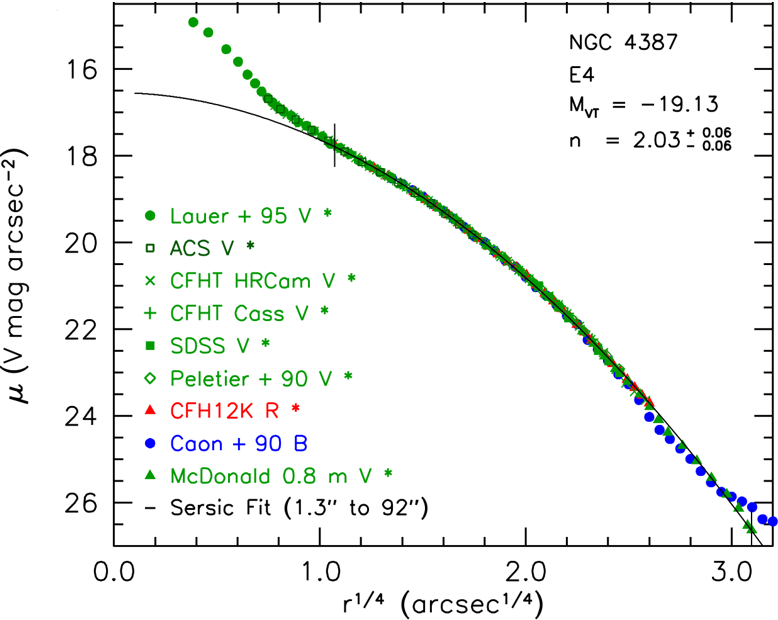}
\end{minipage}
\caption{{\bf Nuclear profiles of ETGs.} {\em Left:} The break in the surface brightness profile of a core galaxy is contrasted to the smooth steep rise of a power-law one  \citep[from][]{Lauer1995}. Both are fitted by \autoref{eq:nuker}; {\em Right:} ``extra light'' in the inner surface brightness profile, with respect to a Sersic profile fit. This figure is taken from  \citet{Kormendy2009}, where  these are interpreted as the extra light components that are produced by gas accretions or mergers, when cold gas dissipates and falls to the center producing a starburst.}
\label{fig:nuclear_profiles}
\end{figure}

The launch of the {\em Hubble Space Telescope} (HST) in 1990 revolutionized the study of inner profiles in Es by firmly establishing that the centers of Es do not posses flat ``cores'' \citep{Crane1993}.
Additionally it was found that the profiles of Es could be broadly separated into two classes. The ``core galaxies'' showed clear breaks in the surface brightness, with a steeper outer profile followed by a more shallow inner one. While the ``power-law galaxies'' showed  modest changes in profile slope (\autoref{fig:nuclear_profiles}; \citealt{Ferrarese1994,Lauer1995}).

A popular parametrization which was used to quantify the surface brightness profile shapes is the following double power-law (informally dubbed ``Nuker law'' by its proposers \citealt{Lauer1995})
\begin{equation}\label{eq:nuker}
\Sigma(R)=\Sigma_{\rm b} \left(\frac{R_{\rm b}}{R}\right)^{\gamma}
\left[
\frac{1}{2} + \frac{1}{2} \left(\frac{R}{R_{\rm b}}\right)^\alpha
\right]^{(\gamma-\beta)/\alpha},
\end{equation}
where $\gamma$ and $\beta$ represent the (positive) asymptotic slope for radii respectively smaller and larger than the break radius $R_{\rm b}$, where the surface brightness is $\Sigma_b$, while $\alpha$ determines the sharpness of the transition. From this relation the slope $\gamma'$ is typically determined as the analytic value inferred from \autoref{eq:nuker} at HST resolution limit $R_0\approx0\farcs1$.
According to this definition, core galaxies are those with $\gamma'\le0.3$ \citep{Lauer1995,lauer07prof}.

Alternative ways of quantifying inner profiles is by measuring either``extra light'' or a ``deficit'' with respect to a global Sersic fit to the outer profile \citep[e.g.][]{Ferrarese2006acs,Kormendy2009}, or using the ``Core-Sersic'' parametrization \citep{Graham2003core,Trujillo2004}. Reassuringly, the different profile classifications agree with each other with good accuracy \citep{Krajnovic2013p23}. 

Some studies have emphasized a bi-modality, or dichotomy, in the distribution of the profile slopes within the general galaxy population \citep{Lauer1995,Byun1996,Lauer2005,lauer07prof}, while others find a smooth continuity \citep{Carollo1997,Ferrarese2006acs,Cote2007}. However, the key and remarkable result on which all studies agree is that the inner slope is closely related to the global galaxy properties measured spatial scales more than one order of magnitude larger \citep{Faber1997,Lauer2005,Lauer2007,Ferrarese2006acs,Kormendy2009}. This shows that nuclear slopes provide key physical insight on galaxy structure.

\begin{textbox}[h]
\section{ELLIPTICALS DICHOTOMY BEFORE INTEGRAL-FIELD SPECTROSCOPY}
To define our knowledge of the structure of ETGs before the advent of IFS, we consider papers published before the two companion \sauron\ papers introducing a {\em kinematic} classification \citep{Emsellem2007,Cappellari2007}. Key was the recognition of two classes, and possibly a dichotomy, among Es:

{\bf Giant Ellipticals ($M_V\la-21.5$):} (i) Have Sersic function outer profiles with $n\ga4$ \citep{Caon1993}; (ii) have cores in their nuclear profiles \citep{Ferrarese1994,Lauer1995}; (iii) rotate slowly \citep{Illingworth1977}; (iv) are anisotropic and triaxial \citep{Binney1978}; (v) tend to be rounder \citep{Tremblay1996}; (vi) can have boxy isophotes \citep{Bender1988}; (vii) contain X-ray emitting gas \citep{Bender1989}; (viii) have old and $\alpha$-elements enhanced stellar population \citep{Thomas2005daniel}.

{\bf Normal-luminosity Ellipticals ($M_V\ga-21.5$):} (i) Have Sersic profiles with $n\la3$ \citep{Graham2003}; (ii) are core-less \citep{Faber1997}; (iii) rotate rapidly \citep{Davies1983}; (iv) are nearly isotropic and oblate \citep{Kormendy1996}; (v) can be quite flat; (vi) can have disky isophotes; (vii) rarely contain X-ray gas; (viii) can have young population and are not $\alpha$ enhanced.

The physical relevance of this possible dichotomy was summarized in \citet{Kormendy1996}, who proposed a revision to Hubble's classification of Es based on isophotal shapes, and by \citet{Faber1997}; but see  \citet{Ferrarese2006acs} for a different view. More recent, post-IFS, reviews of the dichotomy are given in \citet{Kormendy2012,Kormendy2016}.
\end{textbox}


\section{STRUCTURE FROM KINEMATICS}
\label{sec:kinematics}

\subsection{Visual classification of kinematic maps}
\label{sec:visual_kin_class}

\begin{figure}
\centering
\begin{minipage}{0.49\textwidth}
\includegraphics[width=\columnwidth]{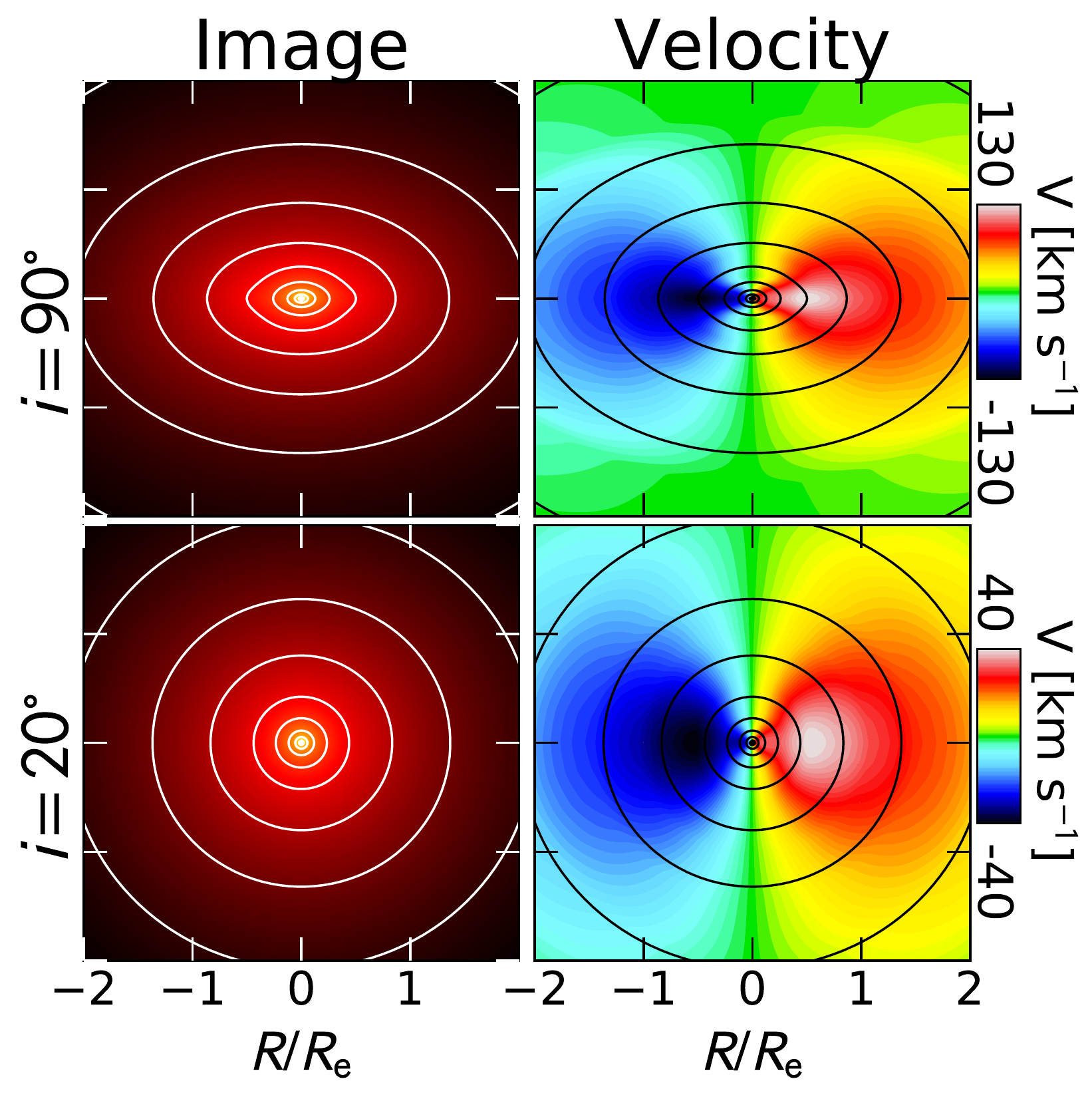}
\end{minipage}
\begin{minipage}{0.49\textwidth}
\includegraphics[width=\columnwidth]{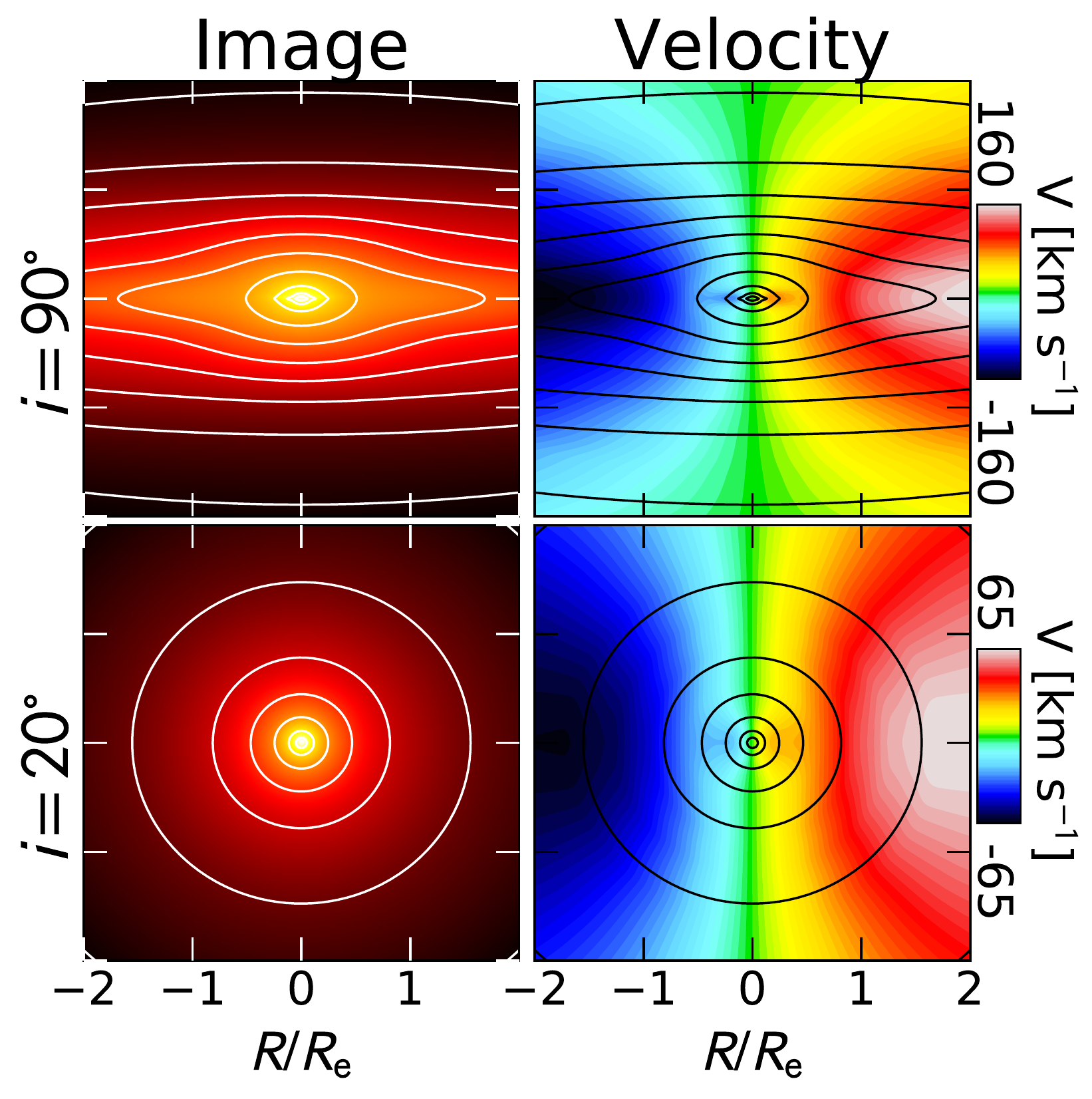}
\end{minipage}
\caption{{\bf Recognizing face-on stellar disks.} {\em Left:} photometric models for the disky E galaxy NGC~821, projected at two inclinations. The rainbow-colored panels show the corresponding stellar velocity predicted by the models. {\em Right:} as in the left plot, for the S0 galaxy NGC~5308. The presence of a disk can be visually recognized from the kinematics even near face-on inclination, while the photometric evidence disappears below $i\la60^\circ$. The input MGE surface brightness was taken from \citet{Cappellari2006} and \citet{Scott2009} respectively, while the adopted model parameters from \citet{Cappellari2008}.
Note the different behavior of the velocity in the two types of galaxies. In the disky E the stellar disk dominates the surface brightness only out to $R\la\re$ and the velocity sharply drops beyond that radius. In the S0 the disk extends to the edge of the galaxy and the velocity is still rising at $R\ga2\re$. This difference is encoded in the photometry and is not due to differences in the galaxies anisotropy, which is here assumed constant throughout the galaxy. Importantly, in both cases, the rotation is {\em not} limited to the disks, however the spheroids rotate more slowly due to their rounder shapes.
}
\label{fig:recognizing_disks}
\end{figure}

The photometric approaches to recognize different types of Es,  described in \autoref{sec:photometry}, suffer from two limitations: (i) measuring nuclear profiles require sub-arcsec spatial resolution; (ii) deviations from elliptical isophotes are only visible near edge-on orientations. This prevents the applicability of either technique to the large galaxy samples at significant distances.

A solution to both limitations is provided by IFS. In fact IFS observations of the stellar kinematics provide the long-sought ability to recognize the presence of stellar disks at virtually any inclination. This is illustrated in \autoref{fig:recognizing_disks}, which predicts, using dynamical models \citep{Cappellari2008}, how two nearly edge-on galaxies, the disky E (as classified by \citealt{Bender1994}) NGC~821 and the S0 galaxy NGC~5308, would appear when seen close to face on. The justification for the adopted models will be given in \autoref{sec:jam_results}. The plot shows that both the centrally concentrated disk of a disky E, and the extended disk of an S0 galaxy, produce clear observable signatures in the kinematics at nearly all inclinations. In both cases the velocity fields display extended ordered rotation, with the kinematic position angle PA$_{\rm kin}$ aligned with the photometric major axis PA$_{\rm phot}$. In contrast, the measurable effects on the isophote shape are hidden in the noise for inclinations $i\la60^\circ$. This is quantified in figure~8 of \citet{Krajnovic2013p17}, which shows that even the strong disk of an edge-on S0 galaxy produces nearly elliptical isophotes ($a_4/a_0\la2\%$).

The earliest IFS observations of the stellar kinematics used a simple but very time-consuming $y$-scanning approach, where a long slit is moved across the field to map a two dimensional field of the nearest E, Centaurus A \citep{Wilkinson1986}. This was followed by proof-of-concept large-scale observations of the stellar kinematics of individual ETGs with real IFS units like the TIGER integral-field unit \citep{Bacon1995} on the CFHT \citep{Emsellem1996,Emsellem1999} or the MPFS integral-field unit \citep{Silchenko1997} on the 6m telescope of the Special Astrophysical Observatory \citep{Silchenko1999}.

A breakthrough came with the introduction of the \sauron\ \citep{Bacon2001} IFS, due to the dramatic improvement of the instrument data quality with respect to previous prototypes. 
The \sauron\ survey \citep{deZeeuw2002} was the first project to map the two-dimensional stellar kinematics, ionized gas and stellar population of a significant sample of 48 nearby ETGs with total absolute magnitudes $M_B<-18$.

\begin{figure}
\centering
\includegraphics[width=0.9\columnwidth, trim={0 3.3cm 0 2.6cm}, clip]{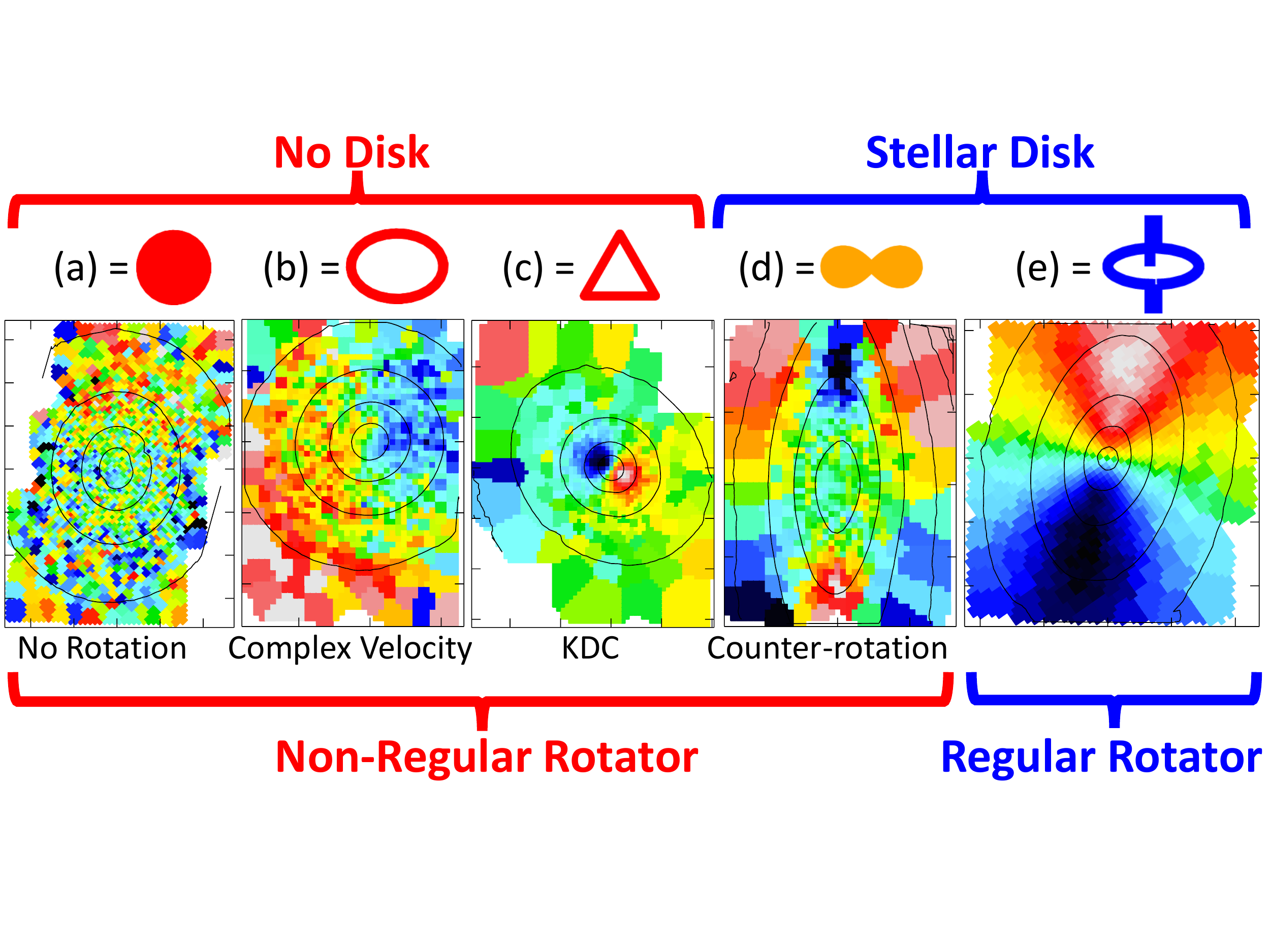}
\caption{{\bf Morphological classification of stellar kinematic.} The features in large samples of ETGs can all be qualitatively described by five classes. (a) No clearly detectable rotation (NGC~4374); (b) Clear but not regular rotation (NGC~4552); (c) kinematically distinct core (KDC; NGC~5813); (d) Counter rotating disks (NGC~4550); (e) Regular extended disk-like rotation (NGC~2974). The five classes were introduced by \citet{Krajnovic2011}. The Voronoi binned \citep{Cappellari2003} kinematics were taken from \citet{Emsellem2004}. The symbols above the maps are used consistently throughout this review.}
\label{fig:kinematic_classes}
\end{figure}

A striking feature which became immediately apparent at a simple visual inspection of the kinematic maps of the sample galaxies \citep{Emsellem2004} was the qualitative separation between two classes of ETGs: on one side were galaxies consistent with the models of \autoref{fig:recognizing_disks}, namely with what one would expect for disks seen at various inclinations (\autoref{fig:kinematic_classes}e), while on the other side were galaxies clearly inconsistent with simple disks (\autoref{fig:kinematic_classes}a--d). In two companion papers from the survey, this initial insight led to the proposal of a quantitative kinematic classification of ETGs, which is virtually independent on inclination effects \citep{Emsellem2007,Cappellari2007}.

The \sauron\ survey was followed by the \atl\ project \citep{Cappellari2011a}, which, rather than being focused on IFS alone, was a multiwavelength survey combined with a theoretical modeling effort. It targeted a complete sample of 260 ETGs, extracted from a volume-limited sample of 871 galaxies brighter than $M_{K_s}<-21.5$, within $D\la 42$ Mpc. The observations of the sample spanned from the radio to the millimeter and optical. The survey includes galaxies with a minimum stellar mass of $\mstar\ga6\times10^9$. 

\atl\ confirmed the striking visual distinction between the kinematics of ``regular rotator'' or ``non-regular rotator'' \citep{Krajnovic2011}. It additionally defined four sub-classes of the non-regular class as illustrated in \autoref{fig:kinematic_classes}. Non-regular rotators were found to (a) either not rotate at all, (b) to show clear but not bi-symmetric or irregular rotation, (c) to present a kinematically decoupled cores (KDCs, these features were discovered by \citealt{Bender1988,Jedrzejewski1988,Franx1988,Franx1989}) or (d) to indicate the presence of two counter-rotating disks (like the prototypical S0 NGC4550 discovered by \citealt{Rubin1992,Rix1992}).

\subsection{Generalizing photometry to kinematics maps}
\label{sec:kinemetry}

\begin{marginnote}[120pt]
\entry{LOSVD}{Line-of-sight velocity distribution}
\end{marginnote}

A galaxy image is only the 0th moment of the line-of-sight stellar velocity distribution (LOSVD). This suggests one may use an approach similar to that described in \autoref{sec:isophotal_shapes} to measure the the shape of the higher moments of the LOSVD, and in particular the mean velocity field, which is the best measured quantity. Like in the photometric case, the approach will work as long as one can find a good zero-order description of the velocity along the best-fitting ellipse.

In the case of photometry, the zero-order approximation along a best-fitting ellipse is a constant. \citet{Krajnovic2006,Krajnovic2008} discovered that the velocity field of ETGs along the best-fitting ellipse, is approximated by a cosine law $V(\psi)=V_0+B_1\cos(\psi)$ with better than 2\% accuracy. This is the same form one would expect if the kinematics was the one of an infinitesimally thin disk. The fact that this law is able to describe the stellar kinematics of ETGs was thus unexpected. This finding motivated the extension of photometry to velocity fields (and other velocity moments) called \textsc{kinemetry} \citep{Krajnovic2006}.

A popular, so called tilted-ring method, to perform ellipse fitting to gas velocity fields existed well before \textsc{kinemetry} and was implemented in the \textsc{rotcur} program \citep{Begeman1989}. \citet{Schoenmakers1997} used  \textsc{rotcur} to fit ellipses to the gas velocity field of spiral galaxies. The observed velocity along those ellipses was subsequently measured and interpreted using a Fourier expansion like in \textsc{kinemetry}. However, the key difference is that \textsc{rotcur} defines the best-fitting ellipse as the one which minimizes the squared deviations between the cosine law and the observed velocity samples along the ellipse, while \textsc{kinemetry} makes the low-order Fourier coefficient equal to zero. This makes \textsc{kinemetry} more robust and ensures that the low-order Fourier terms cannot affect the higher ones.

\begin{marginnote}[120pt]
\entry{Kinemetry}{Generalizes photometry to IFS kinematics}
\end{marginnote}

\begin{figure}
\centering
\includegraphics[width=0.9\columnwidth]{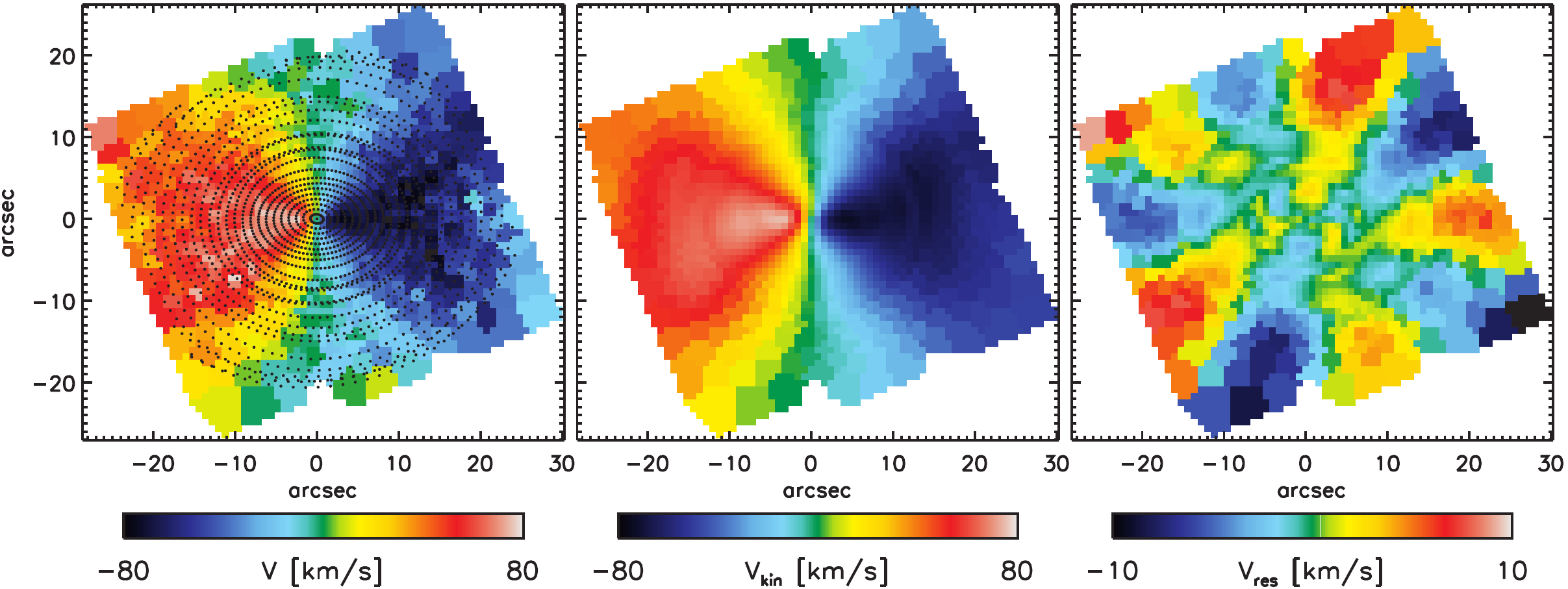}
\caption{{\bf Applying \textsc{kinemetry} to stellar velocity maps.} (a) Data with best-fitting ellipses overlaid. (b) Reconstructed velocity field adopting a pure cosine law along every ellipse. (c) Residuals between the observed and reconstructed velocity field. Note the 5-fold symmetry, which implies a significant $k_5$ term, indicative of a secondary kinematic component \citep[from][]{Krajnovic2006}.}
\label{fig:kinemetry}
\end{figure}

In practice, \textsc{kinemetry} was designed to be a direct generalization of the Fourier approach used for photometry. In close analogy to \autoref{eq:ellipse_fitting_phot} and \autoref{eq:ellipse_residuals_phot}, \textsc{kinemetry} samples the galaxy velocity (or other odd moments of the velocity) along an ellipse using the truncated Fourier expansion
\begin{equation}
V(\psi) = V_0 + A_1\sin(\psi) + B_1\cos(\psi) + A_2\sin(2 \psi) + B_2\cos(2 \psi) + A_3\sin(3 \psi) + B_3\cos(3 \psi),
\label{eq:ellipse_fitting_kin}
\end{equation}
where $\psi$ is the eccentric anomaly. Now the best-fitting ellipse is {\em defined} as the one for which $A_1, A_2, B_2, A_3, B_3$ are zero within numerical accuracy, while $B_1$ is allowed to be non-zero. Once the best-fitting ellipse has been determined, the velocity along that ellipse is again parametrized by the next higher term in the Fourier expansion 
\begin{equation}
V(\psi) = V_0 + A_4\sin(4 \psi) + B_4\cos(4 \psi) + A_5\sin(5 \psi) + B_5\cos(5 \psi).
\label{eq:ellipse_residuals_kin}
\end{equation}
In this case the deviations of the ellipse from the pure $V(\psi)=V_0+B_1\cos(\psi)$ curve are quantified by the $k_5\equiv\sqrt{A_5^2+B_5^2}$ term. Similarly to the photometric $a_4$ parameter, which is sensitive to hidden disks, also $k_5$ is useful to determine the presence of multiple kinematic components. An application of the method to the E galaxy NGC~4473, which is known from dynamical modelling to contain two counter-rotating disks \citep{Cappellari2007}, is illustrated in \autoref{fig:kinemetry}.

\textsc{kinemetry} was used to describe all velocity maps from the \atl\ survey. \citet{Krajnovic2011} found that the regular rotators are characterized by having residuals from the cosine law, quantified by the radially-averaged ratio $k_5/B_1\la0.04$. This ratio measures the fractional amplitude of the 5th Fourier term, with respect to the peak velocity amplitude along the same ellipse.

\subsection{Intrinsic shapes of early-type galaxies}
\label{sec:intrinsic_shapes}

\begin{figure}
\centering
\includegraphics[width=0.75\columnwidth]{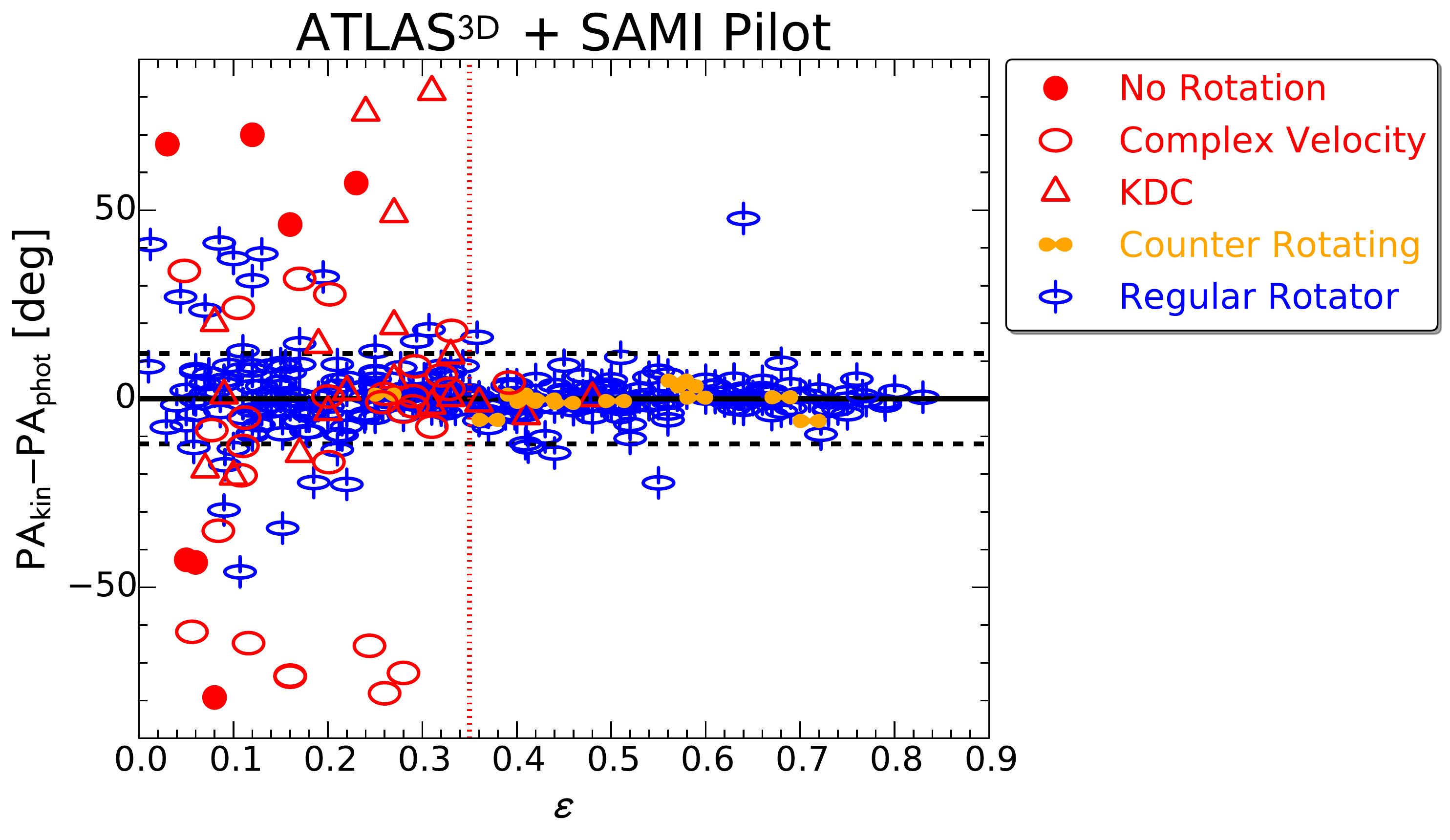}
\caption{{\bf Kinematic misalignment.} Difference between the photometric major axis PA$_{\rm phot}$, measured around $R\approx3\re$, and the kinematic major axis PA$_{\rm kin}$, measured around $R\approx\re$. The plot includes data for 340 ETGs of which 260 were taken from \citet{Krajnovic2011} and 80 from \citet{Fogarty2015}. The meaning of the symbols is illustrated in \autoref{fig:kinematic_classes}.}
\label{fig:kin_mis_a3d}
\end{figure}

\begin{figure}
\centering
\includegraphics[width=\textwidth]{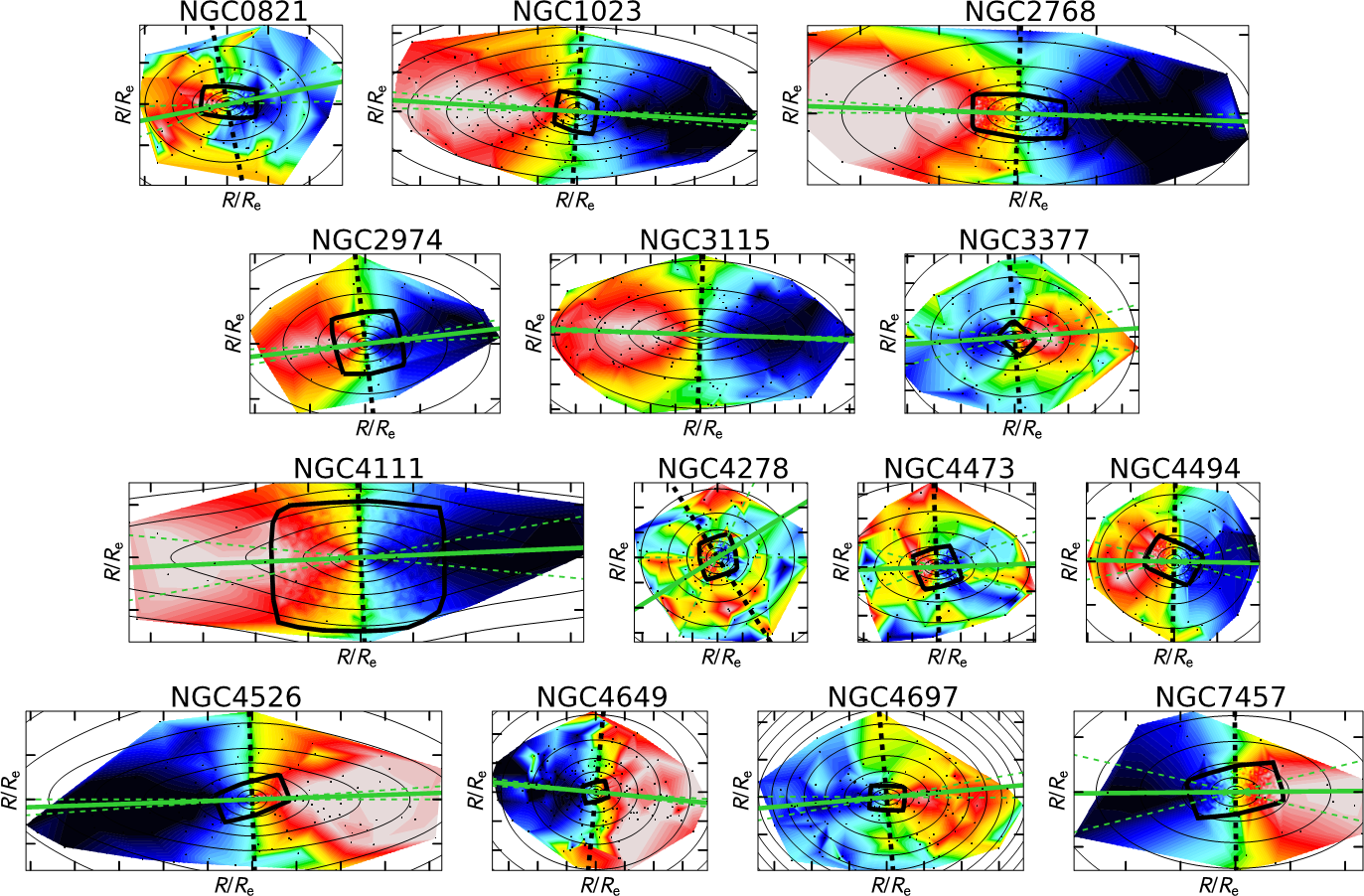}
\caption{{\bf Kinematic misalignment at large radii.} The maps show mean stellar velocity fields of regular rotators and one $2\sigma$ galaxy (NGC~4473) measured by the SLUGGS survey and taken from \citet{Arnold2014}. The data are linearly interpolated from the measured positions, which are indicated by the small black dots. The \sauron\ IFS velocity at smaller radii are enclosed by the solid thick black line (from \citealt{Emsellem2004} and \citealt{Cappellari2011a}). The fields are oriented in such a way that PA$_{\rm phot}$, measured around 3\re\ in \citet{Krajnovic2011}, is horizontal. The solid green line is the best fitting global kinematic PA$_{\rm kin}$, fitted {\em only} to the SLUGGS data, and the dashed line indicate the measurement uncertainties. Isophotes of the galaxy surface brightness are overlaid, in 1 mag intervals. Tick marks are separated by 1\re, from \citet{Cappellari2011a}.}
\label{fig:kin_mis_sluggs}
\end{figure}

Following the first study of the intrinsic shape of E galaxies by \citet{Hubble1926}, a large number of papers have investigated the intrinsic shape of E and S0 galaxies by statistical inversion of their apparent shape distribution \citep[e.g.][]{Sandage1970,Binney1981,Fasano1991,Lambas1992,Ryden1992}.
An intrinsic limitation of these studies, is that the recovery of a generally triaxial shape distribution (a 2-dimensional function of the two axial ratios) from the distribution of the observed apparent ellipticity (a 1-dimensional function) is non unique. This remains true even when stellar kinematics is available \citep{Franx1991}.

\begin{marginnote}[120pt]
\entry{PA$_{\rm kin}$}{Position angle of the axis where the projected velocities reach the maximum absolute values}
\entry{PA$_{\rm phot}$}{Position angle of the axis along which the surface brightness reaches the maximum values}
\end{marginnote}

Luckily, the vast majority of ETGs turns out to be much simpler than the purely dynamical consideration would permit them to be. This allows one to measure their shape, even in the presence of degeneracies. In fact IFS data showed that {\em all} regular rotators have kinematic axes PA$_{\rm kin}$ (measured for $R\approx\re$) essentially aligned with the photometric one PA$_{\rm phot}$ (at much larger radii $R\approx3\re$) \citep{Cappellari2007,Krajnovic2011,Fogarty2015}. This is illustrated in \autoref{fig:kin_mis_a3d} for a combined sample of 340 ETGs. The 1$\sigma$ rms (biweight) scatter of $4^\circ$ is almost at the level of the combined measurement errors of PA$_{\rm kin}$ and PA$_{\rm phot}$. The few deviant objects appear to be either interacting systems or strongly barred.

The {\em only} way to observe such a tight alignment, for such a large sample of ETGs, is if regular rotators are axisymmetric out to their stellar halos, up to about 3\re. This is because in triaxial systems: (i) the intrinsic symmetry axes needs not be aligned with the projected photometric major axis \citep{Contopoulos1956,Stark1977} and (ii) the intrinsic angular momentum needs not be aligned with the intrinsic symmetry axes \citep[e.g.][]{Statler1987}. This implies that, in triaxial galaxies, kinematic misalignments are unavoidable, except in very special configurations. The lack of misalignment for a large sample then unambiguously implies axisymmetry for the whole class.

The alignment between the kinematics and photometry of regular rotators, out to a median radius of 4\re, can be seen directly, albeit for a much smaller sample, from the stellar kinematics obtained by the SLUGGS survey \citep{Brodie2014}. The data were presented in \citet{Arnold2014} and are reproduced in \autoref{fig:kin_mis_sluggs} together with the \sauron\ kinematics in the central parts. The kinematics were oriented in such a way that the PA$_{\rm phot}$ from \citet{Krajnovic2011} are horizontal. Overlaid are the kinematic axes, and errors, measured with the procedure \textsc{fit\_kinematic\_pa}\footnote{Available from \url{http://purl.org/cappellari/software}} described in \citet{Krajnovic2006}. Although the SLUGGS data are in some cases rather noisy, the plot shows that in all cases where it can be measured, the kinematic major axis agrees within the errors with the photometric one, consistently with the global axisymmetry of the stellar halos of regular rotators.

The ability of the stellar kinematics to separate the class of axisymmetric regular rotators from the rest, allows one to revisit the statistical inversion of the observed shape distribution. This is motivated by the fact that, once a family of objects has been proven to be axisymmetric, then the photometric statistical inversion procedure {\em does} provide a unique and well defined solution. This study of the intrinsic shape of regular rotators was performed using the volume-limited \atl\ sample by \citet{Weijmans2014}. They found that the intrinsic axial ratio of regular rotators, in their outer disks, can be described by a nearly Gaussian distribution with mean axial ratio $\langle q\rangle=0.25$ and dispersion $\sigma_q=0.14$. This distribution is consistent with that of spiral galaxies \citep[e.g.][]{Lambas1992,Padilla2008}. One should note that this flattening mainly refers to the outer parts where disks dominate, but ETGs have larger bulges than spirals as will be discussed later.

The shape inversion remains non unique for the non-regular rotators, which show evidence of kinematic misalignment, implying triaxiality. However the kinematic data show that the intrinsic ratio between the smallest and largest axis of the triaxial ellipsoid, measured around 3\re, must be $c/a\ga0.65$. This is demonstrated by the fact that no kinematic misalignment are observed for the 38 non-regular rotators with $\varepsilon\la0.35$ (\autoref{fig:kin_mis_a3d}). Even ignoring the extra information provided by the misalignment, which is not always well defined for non-regular rotators, a stringent limit is placed by the fact that all non-regular rotators have $\varepsilon\la0.4$ (also see \autoref{sec:lam_eps_morph}). The small ellipticity of slow rotating galaxies is also confirmed using VIMOS kinematic observations of 7 additional slow rotator BCGs \citep{Jimmy2013}, using the Mitchell Spectrograph (formerly VIRUS-P) for 11 more massive slow rotators \citep{Raskutti2014}. Overall, out of the 73 slow rotators observed so far, {\em all} are rounder than $\varepsilon\la0.4$. This shows that non-regular rotators as a class are quite close to spherical and only weakly triaxial. 

\subsection{Dynamical modeling of stellar kinematics}
\label{sec:dynamical_models}

\begin{figure}
\centering
\includegraphics[width=0.9\columnwidth, trim={0 3cm 0 3cm}, clip]{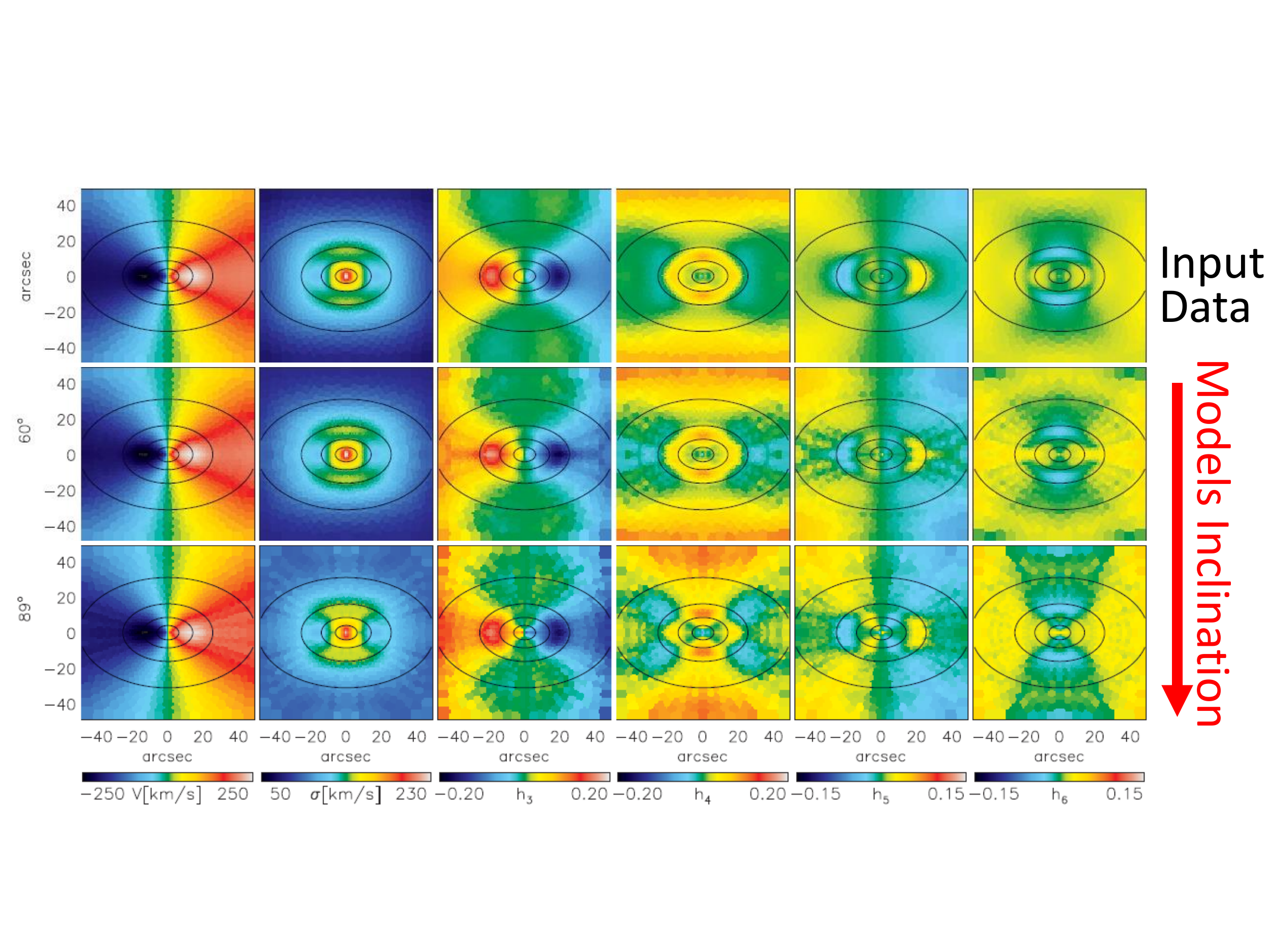}
\caption{{\bf Inclination degeneracy.} The top row shows the mean stellar velocity, the velocity dispersion $\sigma$ and the higher Gauss-Hermite moments of the velocity, for the input simulated data, for an axisymmetric galaxy seen at an inclination $i=60^\circ$. These data were fitted by Schwarzschild models at different inclinations. At every inclination the models surface brightness is constrained to agree with the input one. To eliminate the effect of the intrinsic degeneracy in the deprojection, the surface brightness is deprojected in such a way that at the correct inclination the intrinsic density is the same as the input one. Even in these idealized conditions the differences in the model fits at different inclinations are at the level of the systematic errors affecting real kinematic observations. (taken from \citealt{Krajnovic2005})}
\label{fig:inclination_degeneracy}
\end{figure}

\subsubsection{Techniques and degeneracies}

A complementary way of quantifying the information content of kinematic maps is via dynamical modeling of the stellar kinematics. A key assumption of the models is that the galaxies are in a steady state. The models also generally assume simple spherical, axisymmetric or triaxial shapes. Under the steady state assumption, the galaxy dynamics is fully specified by (i) the six-dimensional stellar distribution function (DF), which describes the distribution of the positions and velocities of stars in the galaxy, and (ii) by the gravitational potential, or equivalently the total mass distribution, which may include the stellar contribution, as well as a dark matter halo and supermassive black hole. A recent review of this topic was given by \citet{Courteau2014}. Here we focus on results specific to IFS observations.

Three major methods have been used in the past three decades: (i) The equations of stellar hydrodynamics, first applied to galaxies by \citet{Jeans1922}; (ii) The numerical orbit-superposition method by \citet{Schwarzschild1979}; and (iii) the made-to-measure N-body models by \citet{Syer1996}. The first has the advantage that it has predictive power and that one can compute reproducible results to numerical accuracy. While the latter two methods have the advantage of generality, which is required for unbiased results. However these two methods cannot make predictions, and depend on implementation details.

\begin{marginnote}[120pt]
\entry{DF}{Stellar distribution function. Describes the orbital distribution in a galaxy}
\end{marginnote}

The recovery of the DF and the mass distribution using only line-of-sight quantities, is an intrinsically degenerate and non-unique problem. This  is because the DF is a function of the three isolating integrals of motion \citep{Jeans1915} and one cannot expect to uniquely constrain {\em both} the 3-dim DF and the 3-dim mass distribution using only a 3-dim observable, as provided by the LOSVD at every spatial location \citep[e.g.][section~3]{Valluri2004}.

Moreover, already the deprojection of the stellar surface brightness into an intrinsic stellar luminosity density is known to be mathematically non unique, even when assuming axisymmetry, unless the galaxy is known to be edge-on \citep{Rybicki1987}. This degeneracy increases rapidly at lower inclinations \citep{Gerhard1996,vandenBosch1997,Romanowsky1997} and represents a fundamental barrier to detailed models of external galaxies. Relaxing the axisymmetry assumption further changes the dimension of the problem, and vastly increase the room for degeneracies \citep{Gerhard1996triax}, making the recovery of general triaxial stellar densities impossible without strong assumptions.

Empirical explorations of the model degeneracies using IFS data have revealed that in practice, even in an ideal case in which (i) one uses simulated noiseless 2-dim data, (ii) one artificially removes the deprojection degeneracy and (iii) one assumes there is no dark matter and the gravitational potential is produced by the stars alone, still, also a basic parameter like the galaxy inclination, or equivalently its shape, is virtually unconstrained by the IFS data (\citealt{Krajnovic2005,vandenBosch2009}; \autoref{fig:inclination_degeneracy}). Tests on the ability of the models to constrain the mass profiles have also revealed ample room for degeneracies in the model parameters even with excellent data \citep{Gerhard1998,deLorenzi2009}.

\subsubsection{Results from integral-field data using general models}
\label{sec:schwarzschild_models}

With the previous caveats in mind, key results were obtained using dynamical models fitted to stellar kinematics. \citet{Schwarzschild1979} models, generalized to fit kinematic data \citep{Richstone1988,Rix1997,vanDerMarel98} have provided most of the mass determinations of supermassive black holes in galaxies \citep[e.g., some examples from different groups are][]{vanDerMarel98,Gebhardt2000,Cappellari2002bh,Valluri2005,McConnell2011,vandenBosch2012,Rusli2013}. These measurements have driven much of our understanding of the connection between supermassive black holes and galaxy evolution. The important topic was reviewed by \citet{Kormendy2013review} and will not be addressed here.

Other significant results are the determination of stellar mass-to-light ratios and  of total mass profiles. According to these studies, dark matter appears to play a minor role within 1\re\ (\autoref{sec:imf}), which implies that, at those radii, the shape of the total mass density is close to that of the luminous density (\autoref{sec:total_density_profiles}).
Under this assumption, the availability of the first set of IFS data from the \sauron\ survey \citep{deZeeuw2002}, opened up the possibility for a unique inversion of the datacube into the DF. This was attempted by \citet{Cappellari2007}, using the axisymmetric \citet{Schwarzschild1979} implementation, optimized for IFS data, described in \citet{Cappellari2006}.
The dynamical models provide the stellar mass orbiting along the half a million orbits which approximate the whole galaxy. 
To quantify this large amount of information, three anisotropy parameters were defined \citep[equation~4.265]{Cappellari2007,Binney2008}
\begin{equation}
    \beta_z\equiv 1 - \frac{\Pi_{zz}}{\Pi_{RR}}, \quad
    \gamma\equiv 1 - \frac{\Pi_{\phi\phi}}{\Pi_{RR}}, \quad 
    \delta\equiv 1 - \frac{\Pi_{zz}}{\Pi_{xx}},
    \label{eq:delta}
\end{equation}
where $(R, z, \phi)$ are the standard cylindrical coordinates, $z$ coincides with the symmetry axis of an axisymmetric galaxy, and $x$ is any direction orthogonal to it. Here
\begin{equation}
    \Pi_{k k}=\int\nu\,\sigma_k^2\, d^3\mathbf{x},
    \label{eq:pi}
\end{equation}
with $\sigma_k$ the velocity dispersion along the direction $k$ at a given location inside the galaxy and $\nu$ the stellar density. The numerical integral extends to the region covered by the IFS observations.
$\beta_z$ describes the global shape of the velocity dispersion tensor in the $(v_R,v_z)$ plane. If the anisotropy is spatially constant then $\beta_z=1-(\sigma_z/\sigma_R)^2$. 
The anisotropy $\gamma$ describes the global shape of the velocity dispersion tensor in a plane orthogonal to $v_z$. If the anisotropy is spatially constant then $\gamma=1-(\sigma_\phi/\sigma_R)^2$. For an isotropic system (spherical velocity ellipsoid) one has $\beta_z=\gamma=\delta=0$.
Integrating over the azimuthal angle one finds that the three anisotropy parameters are related by $\delta=(2\beta_z-\gamma)/(2-\gamma)$.
In the case $\gamma=0$ the simple relation $\beta_z=\delta$ applies.

The anisotropy parameters $\beta_z$, $\gamma$ and $\delta$ converge to zero in the spherical non-rotating limit, for symmetry. For this reason, to quantify the anisotropy of nearly spherical galaxies a complementary anisotropy parameter was defined, in spherical coordinates:
\begin{equation}
    \beta_r\equiv 1 - \frac{\Pi_{tt}}{\Pi_{rr}} =
    1 - \frac{\Pi_{\theta\theta} + \Pi_{\phi\phi}}{2\,\Pi_{rr}},
    \label{eq:beta_r}
\end{equation}
where $(r,\theta,\phi)$ are the standard spherical coordinates. In the spherical limit, assuming the galaxy is non rotating, $\Pi_{\theta\theta} = \Pi_{\phi\phi}$ by symmetry.
The parameter is $\beta_r=0$ for an isotropic galaxy and is positive (negative) when the luminosity-weighted average dispersion along the radial direction is larger (smaller) than the average dispersion along any direction orthogonal to it.

The result of the calculation of the four anisotropy parameters for 25 galaxies from the \sauron\ sample consistent with axisymmetry, showed that {\em on average}: (i) regular rotators have significant anisotropy $\delta$; (ii) the velocity ellipsoid is {\em oblate} with $\delta\sim\beta_z$ and $\gamma\sim0$; (iii) $\beta_z\ga0$ even tough no limits are enforced by the models on this parameter \citep{Cappellari2007}. This picture was independently confirmed using long-slit kinematics of a different galaxy sample and a different modeling code \citep{Thomas2009}.

The two galaxies NGC~4550 and NGC~4473 stood out for a significant tangential anisotropy ($\gamma<0$). In both cases the IFS data shows a characteristic and peculiar enhancement of the stellar velocity dispersion $\sigma$  along the galaxy major axis, with two symmetric peaks in $\sigma$ along the major axis, qualitatively suggesting the presence of counter-rotating disks (for NGC~4550 it confirms the result by \citealt{Rubin1992} and \citealt{Rix1992}). This interpretation was quantitatively confirmed by the dynamical models which recovered two clearly distinct population of stars rotating in opposite direction \citep{Cappellari2007}. The strong tangential anisotropy is precisely what one would have expected given the fact that the two counter-rotating stellar disks strongly increase the random motions in the tangential direction. These two well-studied prototypes of counter-rotating disks motivated the definition of a more general class of qualitatively similar galaxies, which \citet{Krajnovic2011} named $2\sigma$ galaxies, because of their distinctive double peaks in $\sigma$ along the major axis. They were found to constitute 4\% of the \atl\ sample \citep{Krajnovic2011}.

\begin{marginnote}[120pt]
\entry{KDC}{Kinematically decoupled core}
\entry{$2\sigma$ galaxies:}{Contain two counter-rotating stellar disks}
\end{marginnote}

The situation is different for the rounder non-regular rotators. They were found to span a smaller range of anisotropies than the regular rotators and their anisotropy scatters around zero, implying that on average they are close to isotropic. They cannot be precisely so, given their triaxial shapes but, adopting either cylindrical or spherical coordinates, the global anisotropy parameters indicate deviations $\la10\%$ of their velocity ellipsoid from a sphere. This agrees with results obtained using long-slit kinematics of nearly round Es \citep{Gerhard2001,Gebhardt2003}.

An interesting class of non-regular rotators are those with central KDC. Detailed Schwarzschild  dynamical models using \sauron\ data for NGC~4365 \citep{vanDenBosch2008} revealed that the KDC is not spatially distinct in terms of orbital distribution. The ``apparent'' KDC arises because of the superposition of two populations of counter-rotating large-scale tube orbits, with mean velocities canceling out over most of the field, except near the center. The large extent of the stellar orbits producing the KDC, is consistent with the observed homogeneity in the stellar population of the KDC in this galaxy \citep{Davies2001}, and of the large KDCs of non-regular rotators in general \citep{McDermid2006}. A similar example of Schwarzschild modeling of a KDC in the non-regular rotator NGC~5813 was presented by \citet{Krajnovic2015} using very high-quality MUSE \citep{Bacon2010} observations. The study confirmed the ``apparent'' nature of the KDC in this other non-regular rotator. The similarity in the density distribution of the two counter-rotating orbits, which is needed to produce the observed negligible mean stellar velocity outside the KDC, is reminiscent of the ``Separatrix crossing'' mechanism proposed by \citet{Evans1994ngc4550}. \citet{Krajnovic2015} summarizes possible KDCs formation mechanisms.

\begin{figure}
\centering
\begin{minipage}{0.5\textwidth}
\includegraphics[width=\columnwidth]{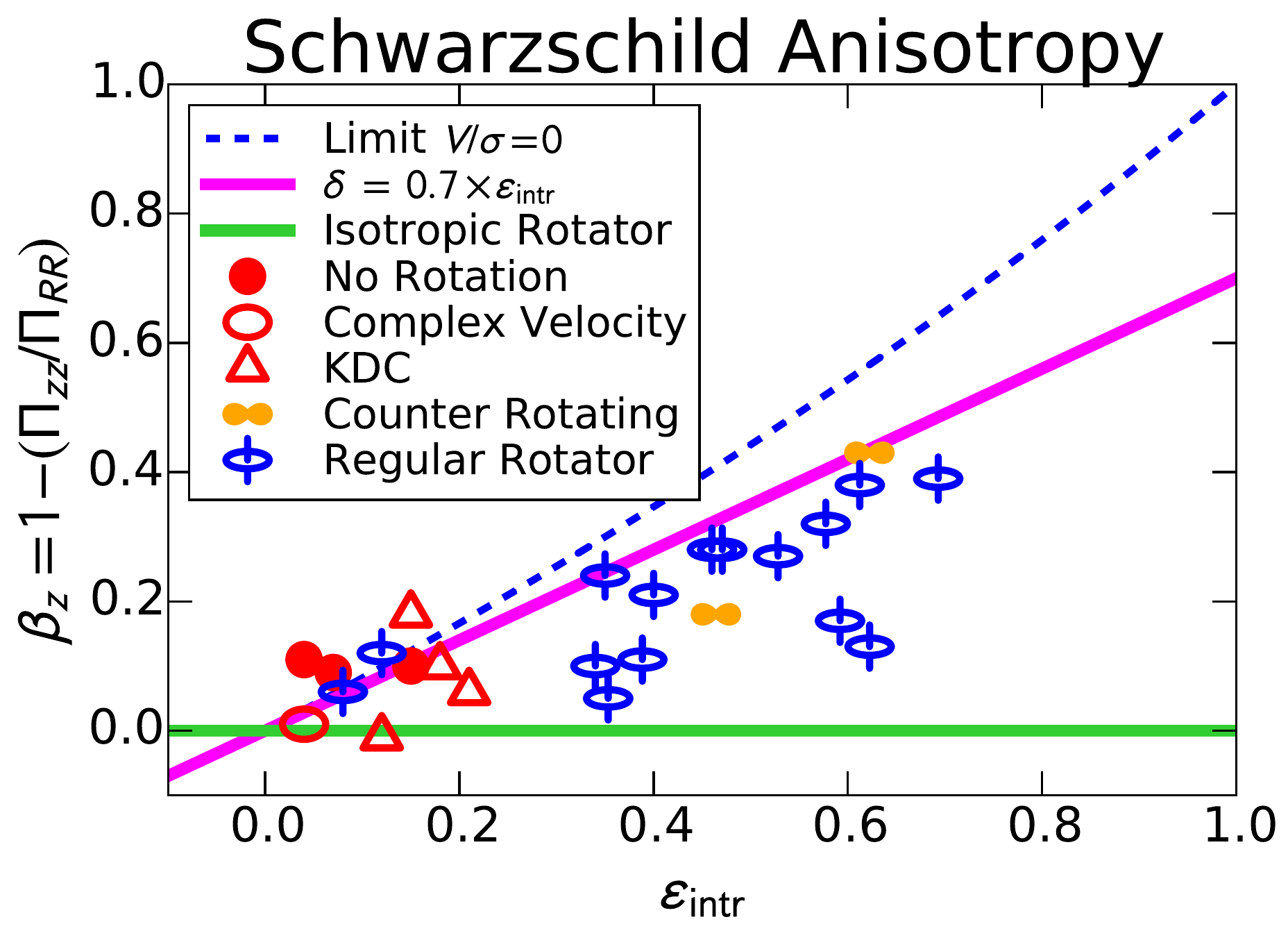}
\end{minipage}%
\begin{minipage}{0.5\textwidth}
\includegraphics[width=\columnwidth]{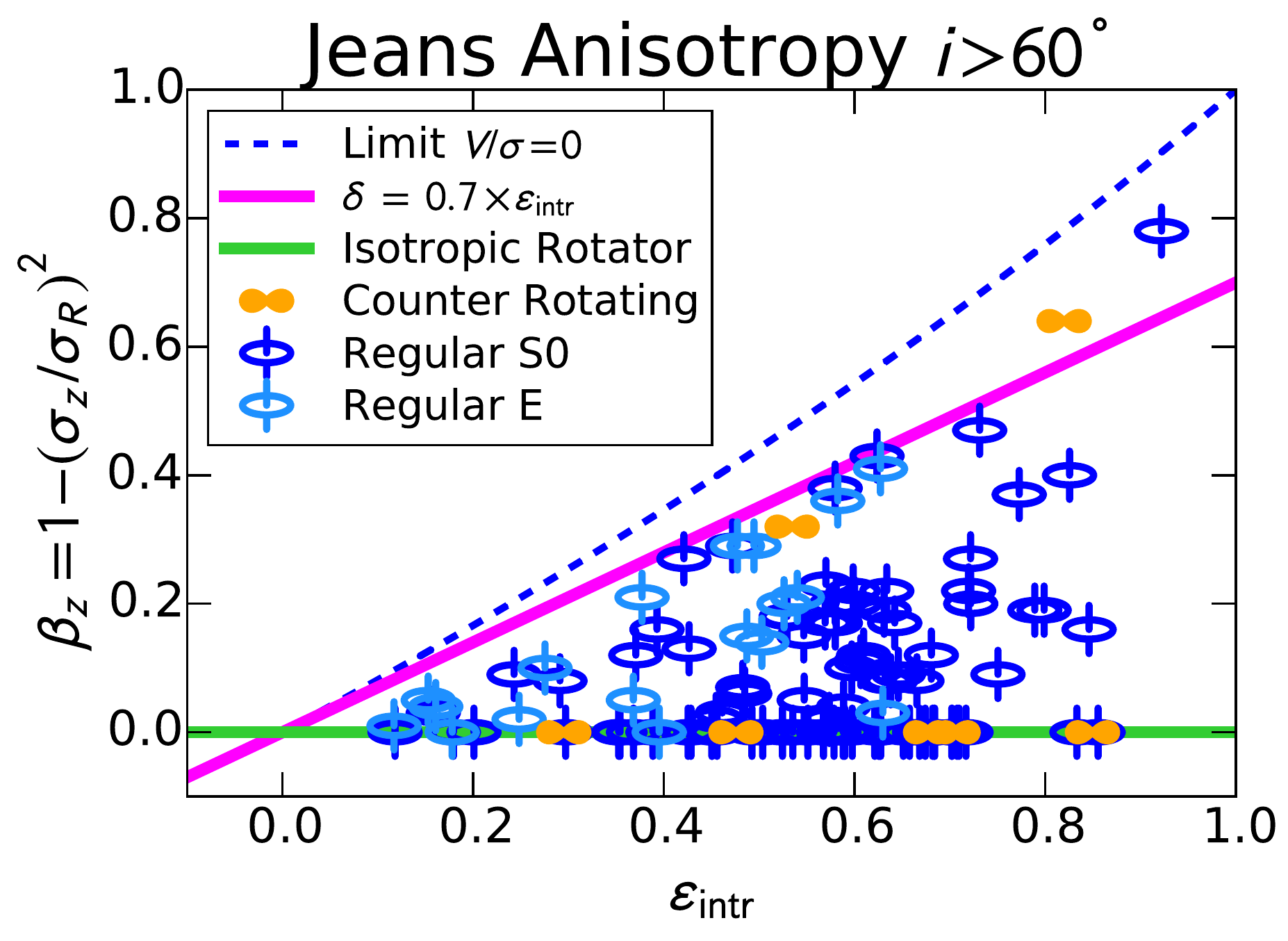}
\end{minipage}
    \caption{{\bf Anisotropy versus intrinsic flattening.} {\em Left:} The anisotropy derived via Schwarzschild models is plotted against the deprojected intrinsic ellipticity of the galaxies around 1\re. Non-regular rotators tend to be close to isotropic in their central regions. While regular rotators span a range of anisotropies, but appear bounded by a relation of the form $\beta_z\approx0.7\times\varepsilon_{\rm intr}$ (magenta line). The green line indicates $\sigma_z=\sigma_R$, while the dashed line is the limit $\beta_z=1-1/\Omega(e)$ (see \autoref{eq:e}) set by the tensor virial equations for objects with oblate velocity ellipsoid. This plot was adapted from \citet{Cappellari2007}. {\em Right:} As in the left panel, for the anisotropy determined via JAM models. Only galaxies with $i>60^\circ$ were included to reduce the effect of the inclination-anisotropy degeneracy. The symbols are defined in \autoref{fig:kinematic_classes}. }
\label{fig:anisotropy_vs_flattening}
\end{figure}

The anisotropy $\beta_z$, namely the $z$-flattening of the velocity ellipsoid, of regular rotators was found to be related to their intrinsic axial ratios $\varepsilon_{\rm intr}$. This is illustrated in the left panel of \autoref{fig:anisotropy_vs_flattening}. Here the flattening is the intrinsic one, deprojected from the observed one $\varepsilon$ within an isophote enclosing half of the total light using ($i=90^\circ$ being edge-on)
\begin{equation}
\varepsilon_{\rm intr}=1-\sqrt{1+\varepsilon(\varepsilon-2)/\sin^2 i}.
\label{eq:eps_intr}
\end{equation}
Regular rotators are found to span a range of $\beta_z$ at given $\varepsilon_{\rm intr}$, with the range increasing with $\varepsilon_{\rm intr}$. For an oblate velocity ellipsoid ($\sigma_\phi=\sigma_R$) the tensor virial theorem \citep{Binney2008} already sets an upper limit $\beta_z<1-1/\Omega(e)$, corresponding to $V/\sigma=0$ in \autoref{eq:vs_tvt}. However the observed values do not span the full space allowed by virial equilibrium, and in particular are found to lie approximately below the linear relation \citep{Cappellari2007}
\begin{equation}
\delta\approx\beta_z=0.7\times\varepsilon_{\rm intr}.
\label{eq:beta_vs_eps}
\end{equation}
In the next section we show that these results, obtained from detailed dynamical models of a small set of galaxies, are fully consistent with what one can infer, on a much larger sample of galaxies, using the very different approach provided by the \citet{Jeans1922} equations.

\begin{marginnote}[120pt]
\entry{$\varepsilon$}{Galaxy observed ellipticity}
\entry{$\varepsilon_{\rm intr}$}{Intrinsic ellipticity}
\end{marginnote}

\subsubsection{Results from integral-field data using Jeans models}
\label{sec:jam_results}

Motivated by the finding that the anisotropy of fast rotator ETGs is {\em on average} best approximated as a flattening of the velocity ellipsoid $\sigma_z<\sigma_R$, \citet{Cappellari2008} developed an accurate and efficient method to solve the axisymmetric Jeans equations allowing for a cylindrically-aligned velocity ellipsoid with general axial ratios $\sigma_R\ne\sigma_z\ne\sigma_\phi$. As discussed in detail in the paper, the cylindrical alignment is only an approximation, as it cannot be accurately satisfied in real galaxies.
This Jeans Anisotropic Modeling (JAM) formalism is an anisotropic (three-integral) generalization of the semi-isotropic (two-integral) formalism ($\sigma_R=\sigma_z$) presented in \citet{Emsellem1994}. It uses the Multi-Gaussian Expansion \citep[MGE;][]{Emsellem1994,Cappellari2002mge} to parametrize the observed galaxy surface brightness. The MGE allows for an analytic deprojection of the observed surface brightness \citep{Bendinelli1991,Monnet1992}. One can employ an arbitrary number of Gaussians to reproduce all features of a galaxy image. In particular one can describe in detail multiple photometric components, including bulges, disks, inner disks and general ellipticity variations.

\begin{marginnote}[120pt]
\entry{MGE}{Multi-Gaussian expansion}
\entry{JAM}{Jeans anisotropic modeling}
\entry{$\kappa$}{Ratio $V_{\rm obs}/V(\sigma_\phi=\sigma_R)$ of the observed velocities and a model with oblate velocity ellipsoid}
\end{marginnote}

\begin{figure}
\centering
\begin{minipage}[b]{.154\textwidth}
\includegraphics[width=\textwidth]{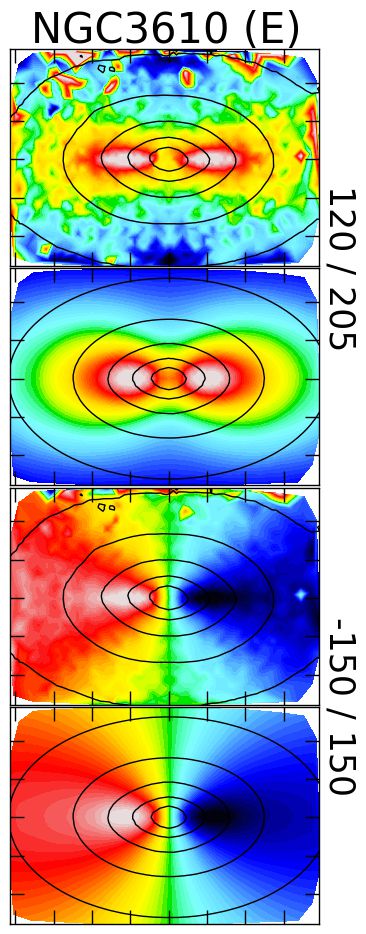}
\end{minipage}
\begin{minipage}[b]{.13\textwidth}
\includegraphics[width=\textwidth]{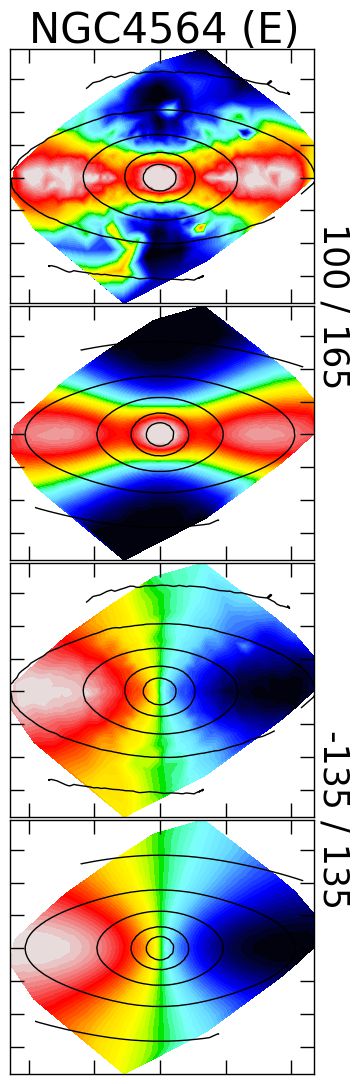}
\end{minipage}
\begin{minipage}[b]{.128\textwidth}
\includegraphics[width=\textwidth]{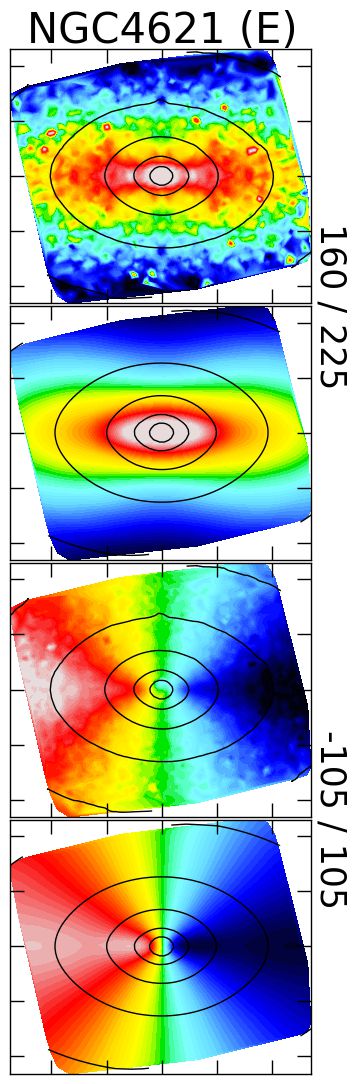}
\end{minipage}
\begin{minipage}[b]{.136\textwidth}
\includegraphics[width=\textwidth]{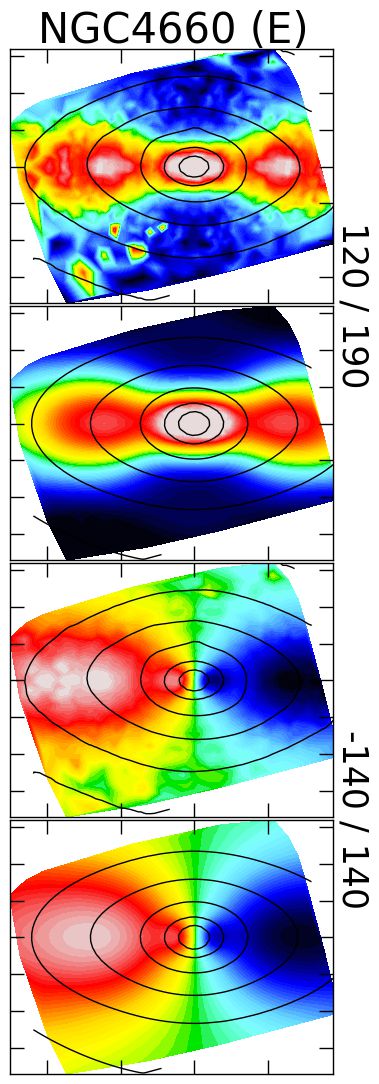}
\end{minipage}
\begin{minipage}[b]{.153\textwidth}
\includegraphics[width=\textwidth]{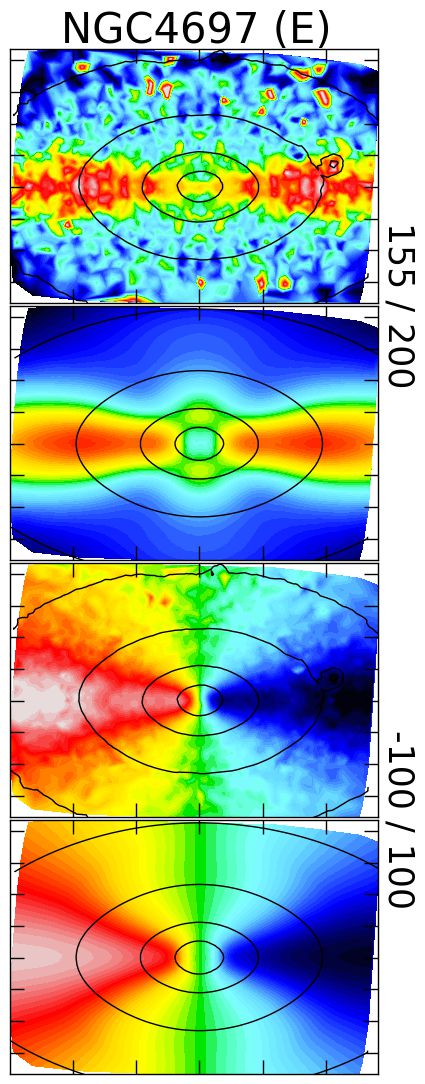}
\end{minipage}
\begin{minipage}[b]{.155\textwidth}
\includegraphics[width=\textwidth]{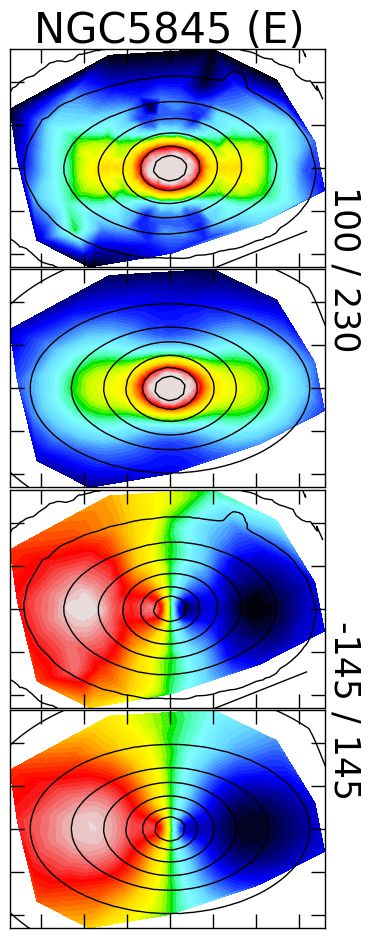}
\end{minipage}

\begin{minipage}[b]{.137\textwidth}
\includegraphics[width=\textwidth]{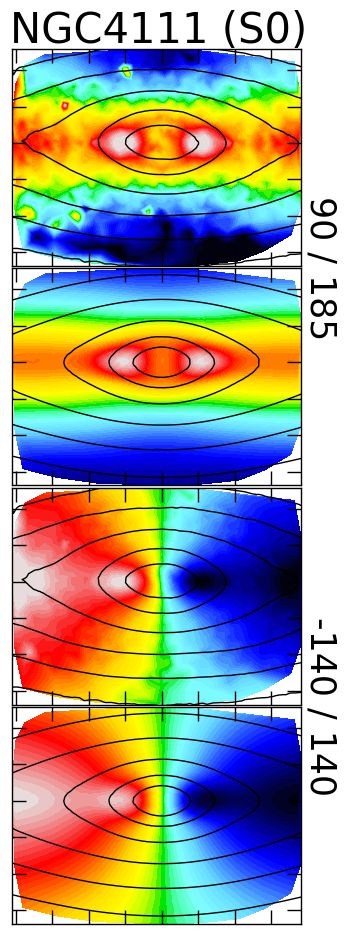}
\end{minipage}
\begin{minipage}[b]{.143\textwidth}
\includegraphics[width=\textwidth]{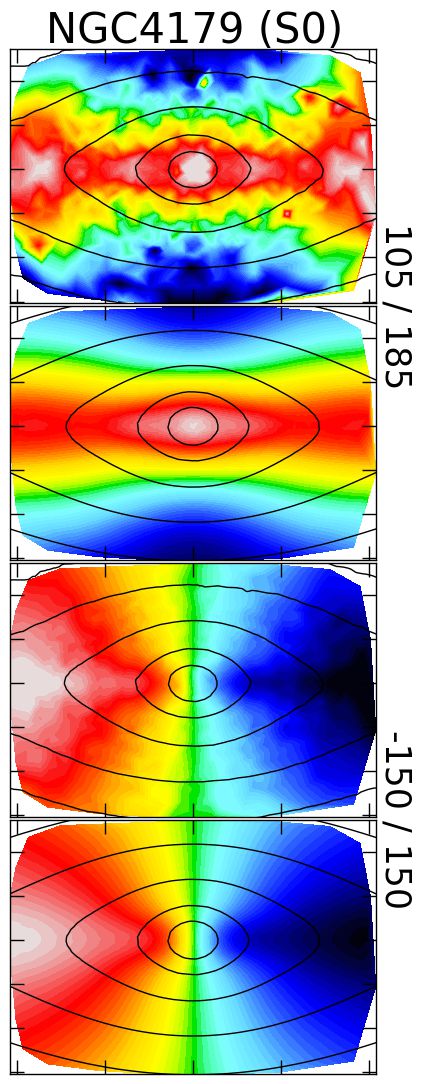}
\end{minipage}
\begin{minipage}[b]{.14\textwidth}
\includegraphics[width=\textwidth]{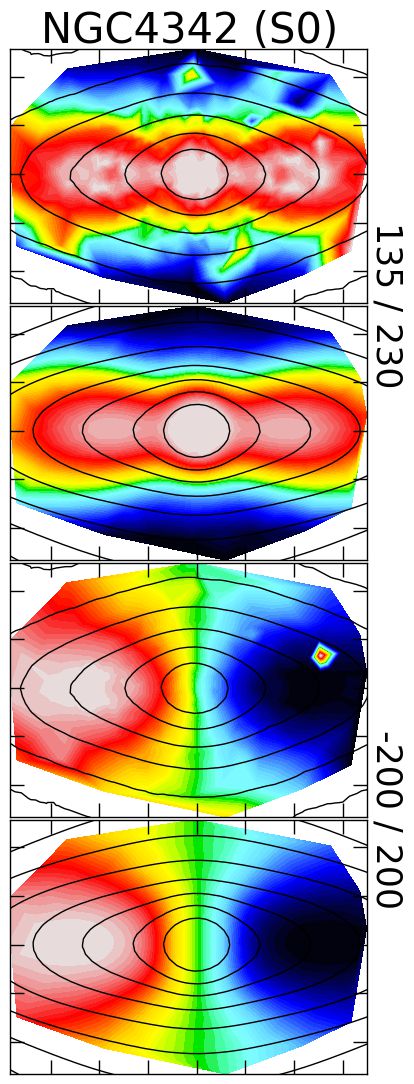}
\end{minipage}
\begin{minipage}[b]{.157\textwidth}
\includegraphics[width=\textwidth]{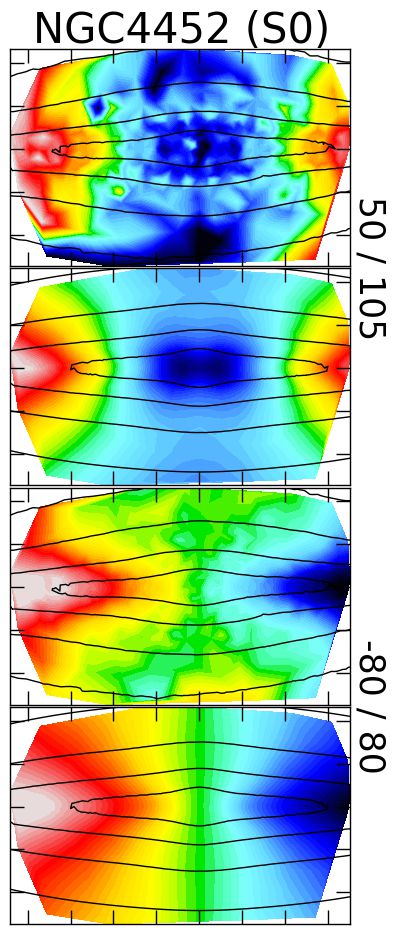}
\end{minipage}
\begin{minipage}[b]{.128\textwidth}
\includegraphics[width=\textwidth]{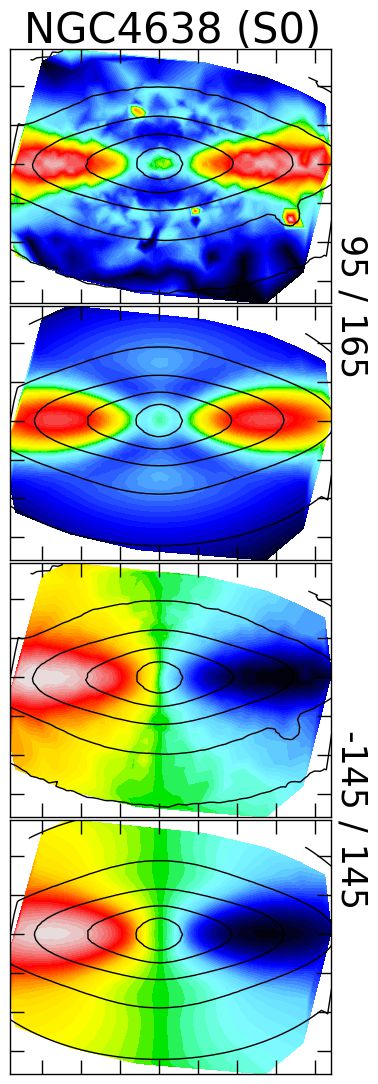}
\end{minipage}
\begin{minipage}[b]{.138\textwidth}
\includegraphics[width=\textwidth]{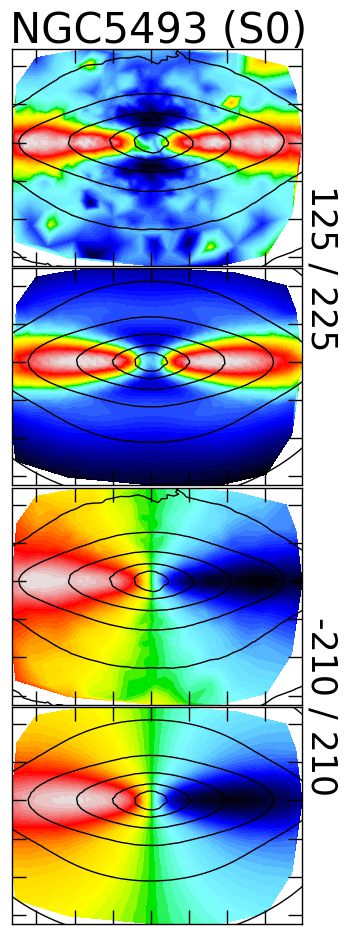}
\end{minipage}
\caption{{\bf JAM models of regular rotators.} The top rows show six Es, while the bottom one six S0s (from RC3). As expected for regular rotators, most Es have disky isophotes and are better morphologically classified as E(d). In each plot, the top panel shows the symmetrized \sauron\ stellar $V_{\rm rms}\equiv\sqrt{V^2+\sigma^2}$, the second panel is the best fitting JAM model to the $V_{\rm rms}$, the third panel is the \sauron\ mean stellar velocity $V$ and the bottom panel is the best fit to $V$ obtained by keeping the best fitting anisotropy $\beta_z$, inclination $i$ and $M/L$ from the previous fit to the $V_{\rm rms}$. Only an overall scaling of $V$ is done here.
The \sauron\ data come from \citet{Emsellem2004} or \citet{Cappellari2011a}. The MGE surface brightness is from \citet{Scott2013p21}. Once an accurate description of the surface brightness is given, the observed shape of the kinematics of each galaxy can be 
predicted with remarkable accuracy by varying the single physical parameter $\beta_z$, and by choosing an inclination. This illustrates the homogeneity in the dynamics of the regular rotators family. Given that these models assume a mass distribution following the light, the good predictive power suggests this assumption should also be nearly correct.
}
\label{fig:jam_models}
\end{figure}

The comparison between the JAM models and real state-of-the-art IFS observations of ETGs confirms the expectations from the Schwarzschild models. It shows that, assuming a constant-anisotropy, cylindrically-aligned velocity ellipsoid, one can ``predict'' the kinematics with remarkable accuracy. \autoref{fig:jam_models} shows that the large variety in the observed shapes of the velocity second moment $V_{\rm rms}\equiv\sqrt{V^2+\sigma^2}$ maps of fast rotator ETGs are properly captured by the simple JAM models. Once the photometry of the galaxies is given as input, the models are fully specified by the single physical parameter $\beta_z\equiv1-(\sigma_z/\sigma_R)^2$ and by the inclination, as well as by the overall mass scaling, which is parametrized via the total mass-to-light ratio ($M/L$). 
Moreover, the distribution of $\beta_z$ for the regular rotators, as a function of the intrinsic ellipticity $\varepsilon_{\rm intr}$, is consistent with the one observed from the Schwarzschild models. It shows that real galaxies have $\beta_z\la0.7\times\varepsilon_{\rm intr}$. The JAM models show that this trend is not due to anisotropic disks embedded in isotropic bulges, in fact the best fits to the kinematics are obtained with a constant anisotropy for both components.

Once the JAM models have been fitted to the $V_{\rm rms}$, the intrinsic velocity second moment $\overline{v_\phi^2}$ in the tangential direction are uniquely defined by the given assumptions. However, to estimate the first velocity moment, the mean stellar velocity $V$, one needs to make an extra assumption about how the $\overline{v_\phi^2}$ splits into ordered and random motion, as defined by
\begin{equation}\label{eq:v_phi2}
\overline{v_\phi^2}=\overline{v_\phi}^2 + \sigma_{\phi}^2.
\end{equation}

The JAM method allows for a general tangential anisotropy $\gamma=1-(\sigma_{\phi}/\sigma_R)^2$, however this generality is not actually needed to describe the kinematics of real galaxies, at least within about 1\re, where good data have been obtained for many galaxies. In fact, if one makes the simplest assumption of an oblate velocity ellipsoid $\gamma=0$ (or $\sigma_{\phi}=\sigma_R$), the shape of the mean stellar velocity field can also be quite accurately predicted, without the need to invoke extra parameters (\autoref{fig:jam_models}).

\begin{figure}
\centering
\includegraphics[width=0.7\textwidth]{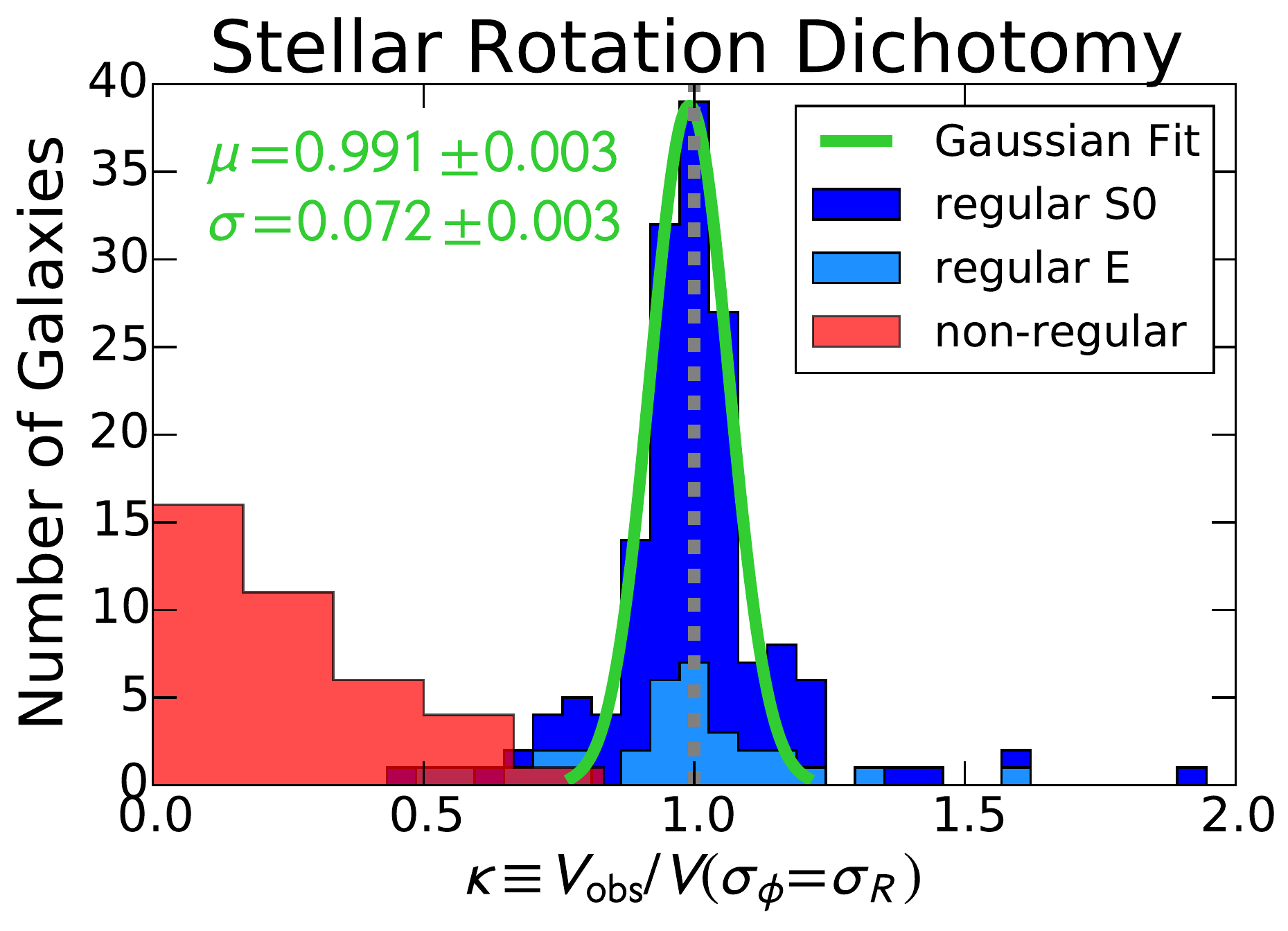}
\caption{{\bf Stellar rotation dichotomy.} Histogram of the ratio $\kappa$ between the observed velocity $V_{\rm obs}$ and the velocity $V(\sigma_\phi=\sigma_R)$ predicted by a JAM model with an oblate velocity ellipsoid. The values are extracted from models (A) of \citet{Cappellari2013p15}. More specifically this ratio is determined by linearly fitting the full JAM velocity field to the observed one. The green line is a Gaussian fit to the histogram, for the regular rotators only. The distribution peaks with high accuracy at the value $\kappa\approx1$, which corresponds to an oblate velocity ellipsoid on average. No statistically significant difference is observed (via K-S test), in the distribution of the E and S0 regular rotators. However the non-regular rotators have a completely different distribution, which peaks at $k\approx0$, with a broad tail. The regular and non-regular rotators are welll separated in this diagram around $k\approx0.65$, the transition region being due to counter-rotation in disk galaxies. Only non-regular rotators have $\kappa\la0.5$ and only regular ones have $\kappa\ga0.75$. This diagram demonstrates a clean dichotomy, rather than a continuity between the two classes of ETGs.}
\label{fig:kappa_histogram}
\end{figure}

A remarkable finding is that, if one takes as reference the mean projected velocity field $V(\sigma_\phi=\sigma_R)$ predicted by a JAM model with a perfectly oblate velocity ellipsoid, the overall scaling $\kappa$ required to best fit the observed velocity $V_{\rm obs}$ is close to $\kappa\approx1$ with an observed rms scatter of only 7\% (\autoref{fig:kappa_histogram}). This already small observed scatter also includes the effect of measurement errors in $\beta_z$, the inclination and $M/L$, which are independently fitted to the $V_{\rm rms}$ and used as input to predict the $V(\sigma_\phi=\sigma_R)$. This result confirms that an oblate velocity ellipsoid provides a good approximation to the observed galaxy dynamics within 1\re. 

\begin{figure}
\centering
\begin{minipage}[b]{.16\textwidth}
\includegraphics[width=\textwidth]{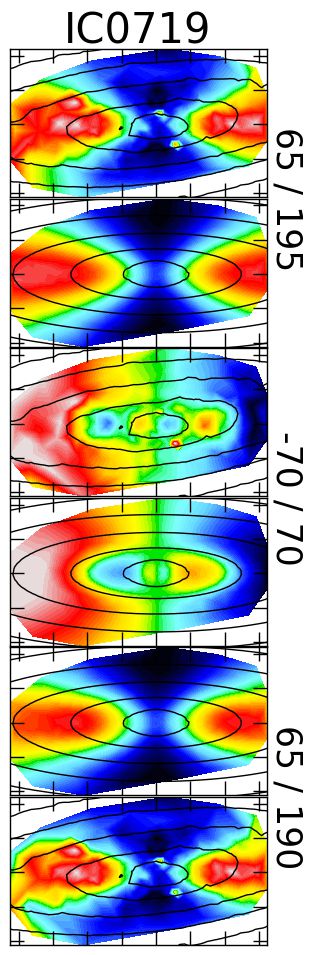}
\end{minipage}
\begin{minipage}[b]{.14\textwidth}
\includegraphics[width=\textwidth]{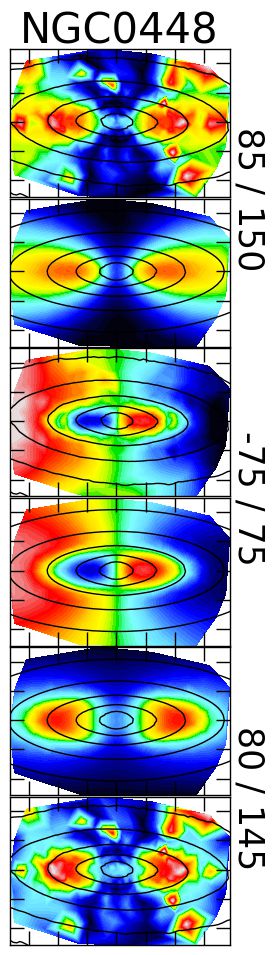}
\end{minipage}
\begin{minipage}[b]{.128\textwidth}
\includegraphics[width=\textwidth]{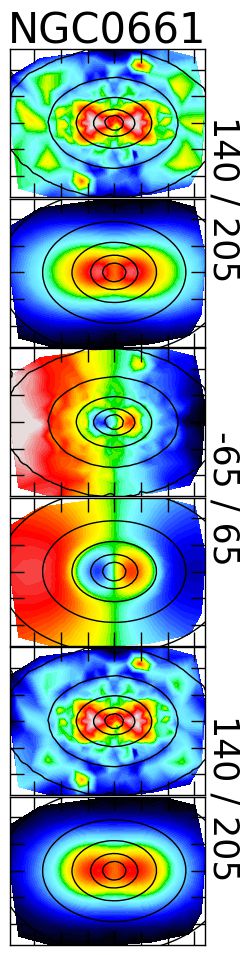}
\end{minipage}
\begin{minipage}[b]{.138\textwidth}
\includegraphics[width=\textwidth]{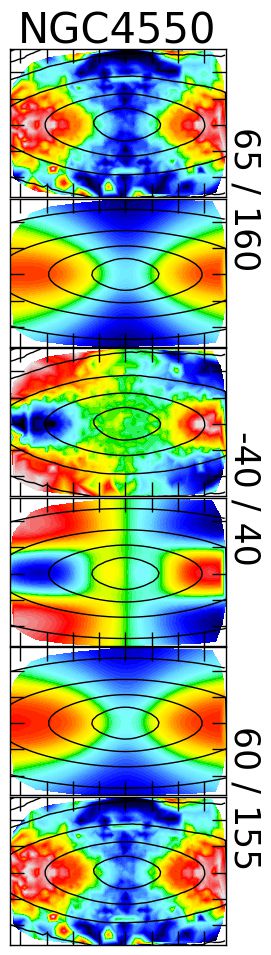}
\end{minipage}
\begin{minipage}[b]{.125\textwidth}
\includegraphics[width=\textwidth]{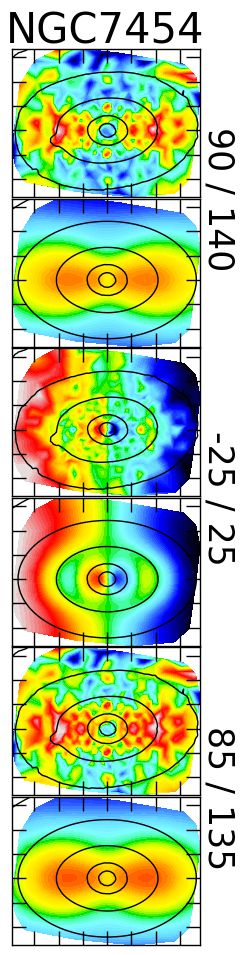}
\end{minipage}
\begin{minipage}[b]{.2\textwidth}
\includegraphics[width=\textwidth]{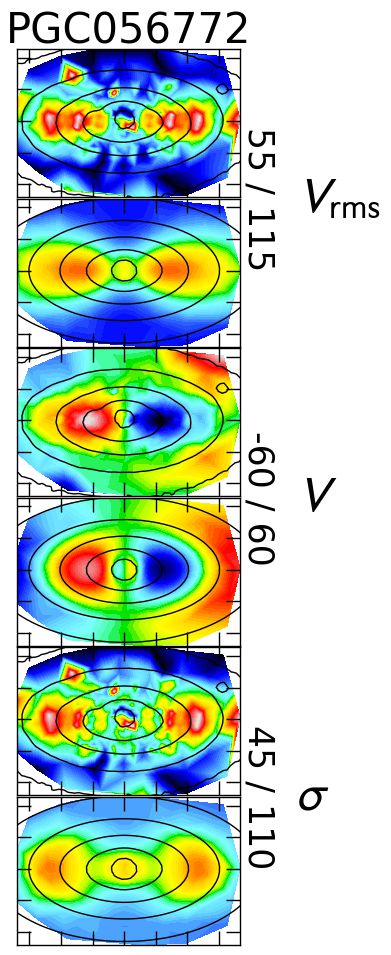}
\end{minipage}
\caption{{\bf JAM models of counter-rotating disks.} These are classified as $2\sigma$ by \citet{Krajnovic2011}. In each plot, the top panel shows the symmetrized \sauron\ stellar $V_{\rm rms}\equiv\sqrt{V^2+\sigma^2}$, the second panel is the best fitting JAM model to the $V_{\rm rms}$, the third panel is the \sauron\ mean stellar velocity $V$, the fouth panel is the best fit to $V$ obtained by keeping the best fitting anisotropy $\beta_z$, inclination $i$ and $M/L$ from the previous fit to the $V_{\rm rms}$. The penultimate panel is the velocity dispersion $\sigma$, which is characterized by two maxima at the opposite sides of the nucleus, along the major axis. The bottom panel is the JAM model $\sigma=\sqrt{V_{\rm rms}^2-V^2}$. The \sauron\ data come from \citet{Cappellari2011a}, except for NGC~4550, which comes from \citet{Emsellem2004}. The MGE surface brightness comes from \citet{Scott2013p21}. The only difference from what was done in \autoref{fig:jam_models}, is that the fit to $V$ allows for opposite sense of rotation for different Gaussians in the MGE. This reversal leaves the model $V_{\rm rms}$ rigorously unchanged. According to the models, the amount of mass that is affected by the counter-rotation ranges from about 10\% for NGC~661 to about 50\% for NGC~4550. This plot illustrates the physical continuity between the regular rotators and the counter-rotating disks.}
\label{fig:jam_kdc}
\end{figure}

The class of counter-rotating disks (\autoref{fig:kinematic_classes}d) necessarily cannot be described by models with oblate velocity dispersion tensor. This is because counter-rotating stars produce strong tangential anisotropy. However, it turns out that these ETGs have similar dynamics to the other regular rotators, once the counter-rotation is taken into account. This is illustrated in \autoref{fig:jam_kdc}, where the JAM models are used to describe their kinematics. One can see that their $V_{\rm rms}$ is still reasonably well predicted by the models, once the photometry is given, with a similar range of $\beta_z$ anisotropy. 

In this case however, to fit the velocities, the sense of rotation of the stars enclosed within some of the MGE Gaussians was reversed, by allowing $\kappa$ to have different signs for different Gaussians, while still being relatively close to unity. The stellar velocity dispersion was then given by $\sigma=\sqrt{V_{\rm rms}^2 - V^2}$.
Using this simple approach one can naturally describe the observations of the counter-rotating disks. This shows that these $2\sigma$ galaxies form a physically homogeneous family with the rest regular rotators. 

However, not all galaxies are well described by JAM models with oblate velocity ellipsoid. This is only true for the ETGs with evidence for stellar disks (\autoref{fig:kinematic_classes}d--e). The situation is dramatically different for the non-regular rotators (\autoref{fig:kinematic_classes}a--c). The $V_{\rm rms}$ of these object can generally still be well approximated by JAM, however the shape of the predicted $V(\sigma_\phi=\sigma_R)$ is, even qualitatively, {\em very} different from the observed velocity field. A simple way to quantify the difference between the dynamics of the regular versus the non-regular rotators is to look at the distribution of their measured $\kappa$ parameters (\autoref{fig:kappa_histogram}). This shows that, while the regular rotators show a nearly Gaussian distribution with average $\kappa\approx1$, the non-regular rotators are clearly distinct. Importantly, the kinematics shows a real dichotomy, not a smooth transition, between these two classes of galaxies, suggesting that they must follow different formation channels in their evolution.

\subsection{Understanding the $(V/\sigma,\varepsilon)$ diagram}
\label{sec:vs_eps}

\subsubsection{Before integral-field kinematics}
\label{sec:vs_eps_before_ifs}

Before observations of the stellar kinematics of ETGs became possible, these objects were thought to constitute a class of homogeneous systems, with an isotropic velocity dispersion tensor. A revolution was started by the first observations of the stellar kinematics in ETGs  \citep{Bertola1975}, which found much lower velocities than predicted by isotropic models \citep{Illingworth1977,Schechter1979}. To quantify this discrepancy, the now-classic $(V/\sigma, \varepsilon)$ diagram was proposed \citep{Binney1978}. It quantifies the ratio between the ordered rotation and the random motion in a stellar system, as a function of the observed (i.e.\ apparent) ellipticity $\varepsilon$ of a galaxy.

When the observations could be extended to galaxy bulges \citep{Kormendy1982,Kormendy1982a,Kormendy1982review} and fainter E ($M_B\ga-21$, \citealt{Davies1983}), it was found these were instead more consistent isotropic rotators. A separation around $M_B\approx-20.5$ was suggested between (i) brighter ellipticals, with slow rotation and triaxial shapes, and (ii) fainter ellipticals with faster rotation. The former were thought to be significantly anisotropic and likely triaxial, while the latter were interpreted as nearly isotropic stellar systems ``flattened by rotation''. The subsequent discoveries of a connection between E stellar rotation, isophotal shape (\autoref{sec:isophotal_shapes}) and nuclear profile slopes (\autoref{sec:nuclear_profiles}) significantly strengthened the case for two different types of E \citep{Kormendy1996,Faber1997}.

The theoretical interpretation of the $(V/\sigma, \varepsilon)$ diagram is based on the tensor virial theorem, which relates the kinetic and potential-energy tensors in a stellar system \citep[section~4.8.3]{Binney2008}. The tensors are formally integrated over the full extent of the galaxies. However, for decades the $V/\sigma$ could only to be measured in galaxies using long slit kinematics, which provide only a crude approximation of the global galaxy kinematics. 

\subsubsection{Formalism for integral-field kinematics}

The advent of IFS motivated a more rigorous and robust formulation of the theoretical diagram in which the $V$ and $\sigma$ are luminosity-weighted quantities integrated over the full extend of the system \citep{Binney2005}. The first application of Binney's revised $(V/\sigma,\varepsilon)$ formalism, performed the luminosity-weighting within an ellipse which encloses half of the projected total galaxy light \citep{Cappellari2007}. In this case the observed quantity becomes
\begin{equation}
\frac{\langle V^2 \rangle}{\langle \sigma^2 \rangle}\approx
\left(\frac{V}{\sigma}\right)_{\!e}^{\!2}\equiv
\frac{\sum_{n=1}^N F_n V_n^2}{\sum_{n=1}^N F_n \sigma_n^2},
\label{eq:vs}
\end{equation}
where $V_n$ and $\sigma_n$ are the mean stellar velocity and dispersion within a given spatial bin, $F_n$ is the flux enclosed within that bin, and the sum is performed over all bins falling within the half-light ellipse. One can verify using theoretical models that, by limiting the sum to 1\re, the measured values are no more than $\Delta(V/\sigma)\la0.1$ lower than the theoretical ones, extended to infinite radii \citep{Cappellari2007,Emsellem2011}.

\begin{figure}
\flushright 
\begin{minipage}{.49\textwidth}
\includegraphics[width=\columnwidth]{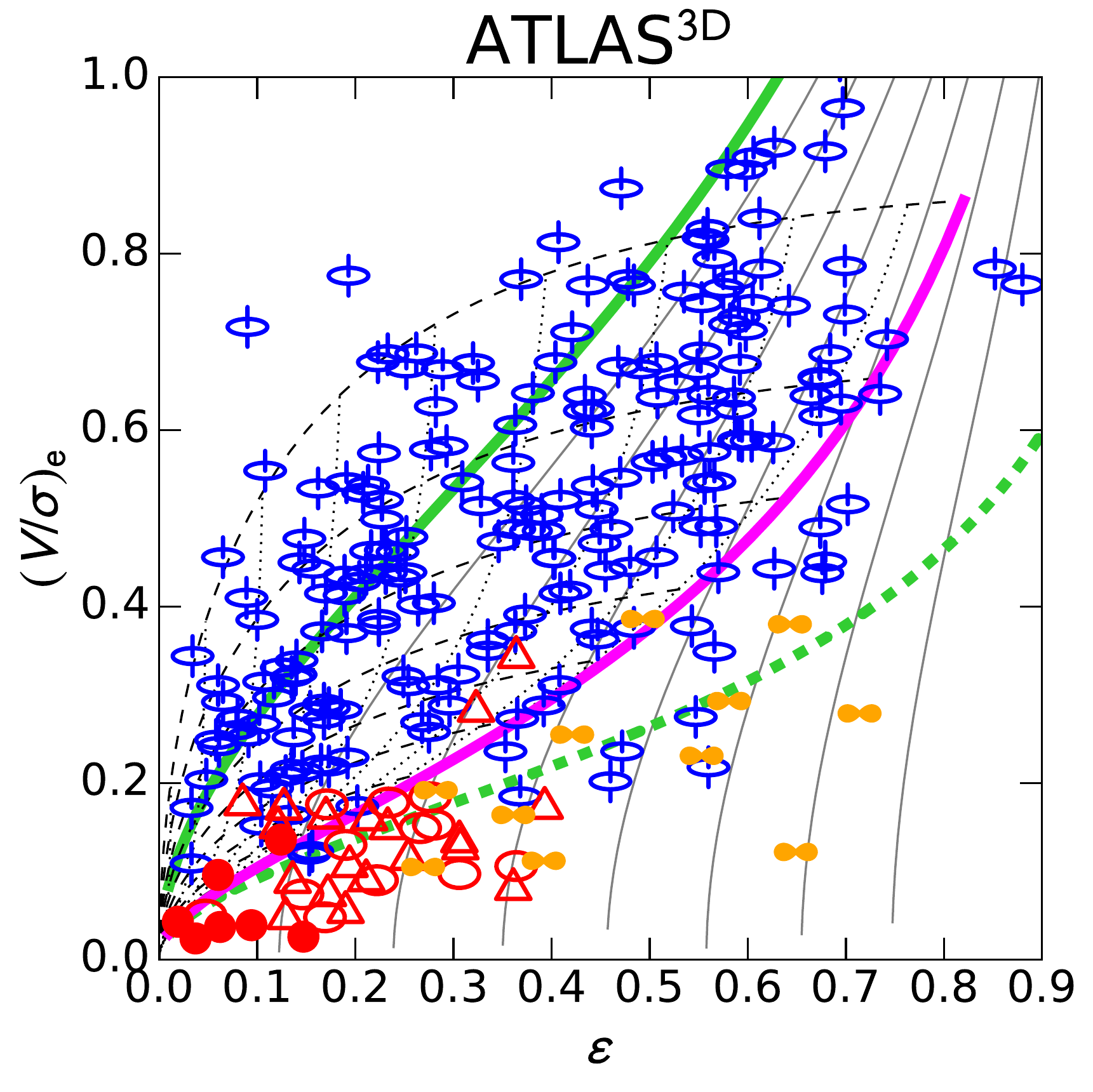}
\end{minipage}
\begin{minipage}{.49\textwidth}
\includegraphics[width=\columnwidth]{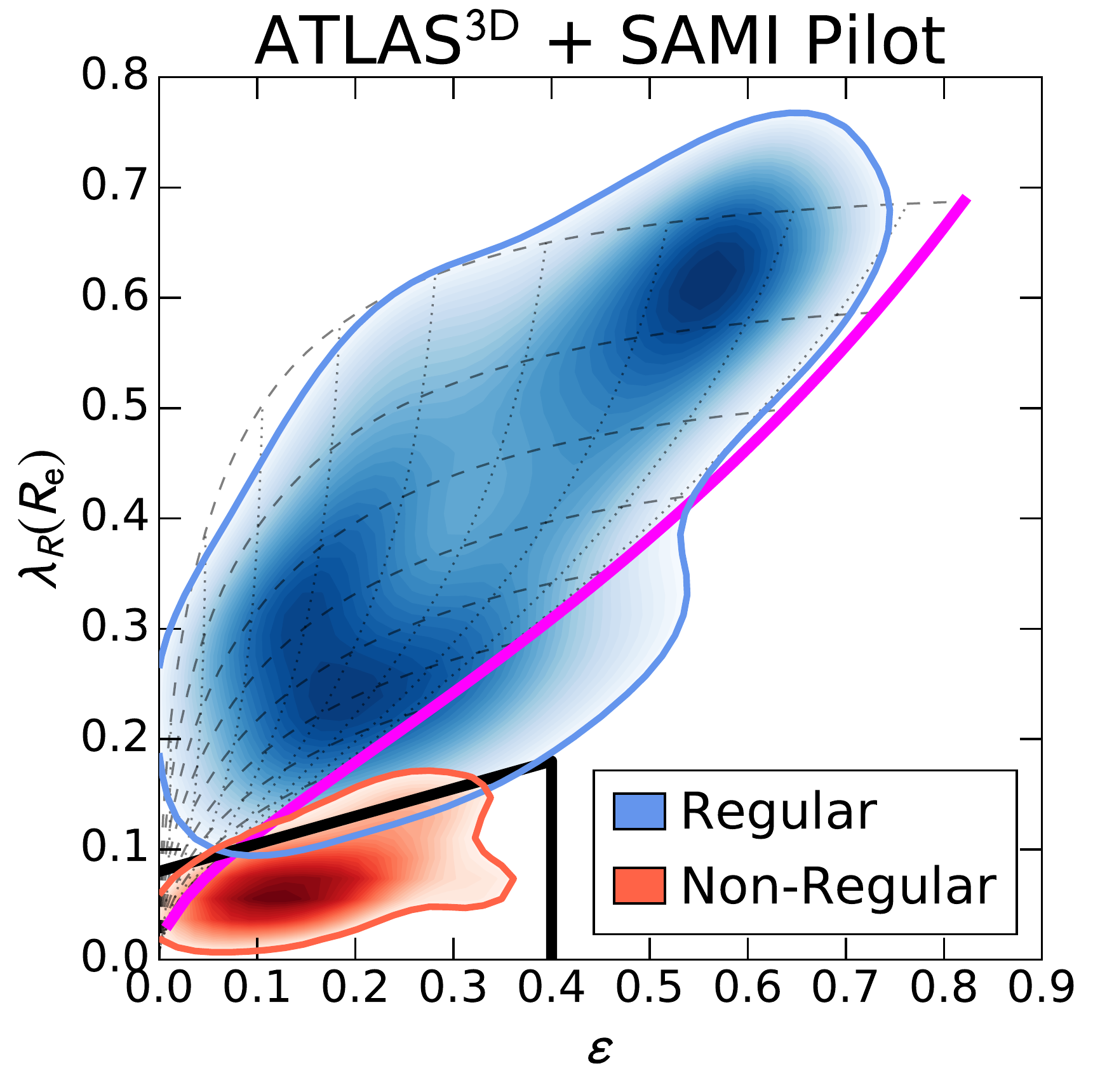}
\end{minipage}

\begin{minipage}{.74\textwidth}
\includegraphics[width=\columnwidth]{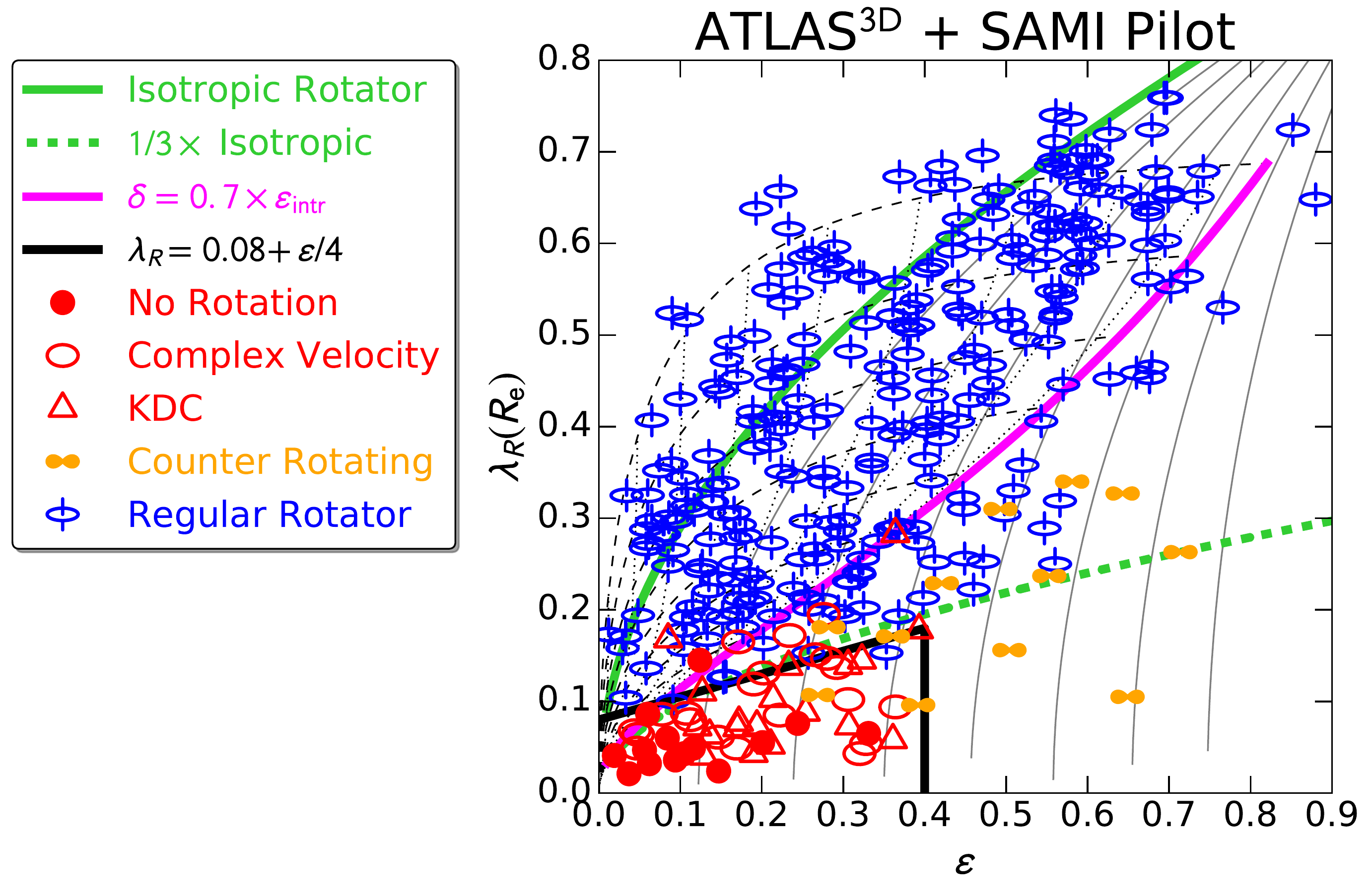}
\end{minipage}
\caption{{\bf The \vse\ and \lre\ diagrams.} {\em Top Left:} The symbols are the \vse\ values for 260 ETGs from \citet{Emsellem2011}, classified as in \autoref{fig:kinematic_classes}. The green line is the prediction for an edge-on isotropic rotator, while the thin lines are separated by $\Delta\delta=0.1$ in anisotropy, using \autoref{eq:vs_tvt} \citep{Binney2005}. The magenta line is the edge-on relation of \autoref{eq:beta_vs_eps} \citep{Cappellari2007}, while the dotted lines indicate how this relation transforms at different inclinations using \autoref{eq:vs_proj}. The distribution of regular rotators is broadly consistent with a family of axisymmetric systems following the anisotropy trends of \autoref{fig:anisotropy_vs_flattening}, seen at random orientations. Instead, the non-regular rotators tend to lie well below the magenta line. The two classes can be approximately separated by the line $(\vs)^{\bigast}=1/3$ (dashed green line). {\em Bottom Right:} The symbols are the \lre\ values for 340 ETGs from \citet{Emsellem2011} and \citet{Fogarty2015}. The lines are the same as in the left panel, with the projected \vs\ values transformed into \lr\ using \autoref{eq:from_vse_to_lre} \citep{Emsellem2007}. The black line is the best empirical separation between regular and non-regular rotators of \autoref{eq:fast_slow_divide}. The black line nearly overlaps with the $(\vs)^{\bigast}=1/3$ line on this diagram. {\em Top Right:} same as in the bottom panel, but using kernel density estimates for the regular/non-regular distributions. For each class, the thick solid lines enclose 80\% of the total probability.
}
\label{fig:vs_eps}
\end{figure}

\begin{marginnote}[120pt]
\entry{$i$}{Galaxy inclination, with $i=90^\circ$ being edge-ony}
\end{marginnote}

As shown by \citet{Binney2005}, for oblate galaxies with different anisotropies $\delta$ (\autoref{eq:delta}) and intrinsic ellipticity $\varepsilon_{\rm intr}$, the theoretical prediction for an edge-on view is 
\begin{equation}
\frac{\langle V^2 \rangle}{\langle\sigma^2\rangle} = 
\frac{(1 - \delta)\Omega(e) - 1}{\alpha(1 - \delta)\Omega(e) + 1}
\label{eq:vs_tvt}
\end{equation}
with 
\begin{equation}
\Omega(e)=\frac{0.5\left[(\arcsin e)/\sqrt{1 - e^2} - e\right]}{e - (\arcsin e)\, \sqrt{1 - e^2}},
\qquad e = \sqrt{1 - (1 - \varepsilon_{\rm intr})^2}
\label{eq:e}
\end{equation}
and $\alpha$ is a parameter which depends on the shape, but not the amplitude, of the galaxy's intrinsic rotation curve, and its radial luminosity profile. A fixed value $\alpha\approx0.15$ was found to provide a good representation of real galaxies \citep{Cappellari2007} and will be adopted in what follows. The edge-on isotropic line and the corresponding edge-on relations for different anisotropies, are shown in the \vse\ diagram of \autoref{fig:vs_eps}.
Given the edge-on ($i=90^\circ$) values of $(V/\sigma,\varepsilon_{\rm intr})$, their projection at different inclinations are
\begin{equation}
\left(\frac{V}{\sigma}\right)_{\rm\! e}^{\rm\! obs} = 
\left(\frac{V}{\sigma}\right)_{\rm\! e}  \frac{\sin i}{\sqrt{1 - \delta\cos^2 i}},
\qquad \varepsilon = 1 - \sqrt{1 + \varepsilon_{\rm intr}(\varepsilon_{\rm intr}-2)\sin^2 i}.
\label{eq:vs_proj}
\end{equation}

\subsubsection{Results from integral-field data}
\label{sec:vs_from_ifu}

The first application of this formalism to 66 galaxies with \sauron\ IFS revealed that, galaxies with the kinematic morphology of non-regular rotators (\autoref{fig:kinematic_classes}a--d), tend to lie well below the isotropic line in the $(V/\sigma,\varepsilon)$ diagram \citep{Cappellari2007}. With the exception of the special class of counter-rotating disks (d), these slow rotating ETGs were generally bright \citep{Emsellem2007} and consistent with the triaxial class of E galaxies found by the previous long-slit studies. However they were found to be only weakly triaxial and actually close to spherical and isotropic within 1\re\ \citep{Cappellari2007}. Their intrinsic ellipticity was found to be smaller than $\varepsilon\la0.4$, as evidenced by the fact that all slow rotating galaxies had observed $\varepsilon$ smaller than that value. The only significantly flattened slow rotating galaxies were found to be the counter-rotating disks (d). The regular rotators (e) showed a completely different distribution on the $(V/\sigma,\varepsilon)$. They were found to be broadly consistent with a population of randomly oriented axisymmetric galaxies, with oblate velocity ellipsoids ($\gamma\approx0$), satisfying the anisotropy condition $\beta_z\la0.7\times\varepsilon_{\rm intr}$ of \autoref{eq:beta_vs_eps}, suggested by the detailed dynamical models (\autoref{sec:schwarzschild_models}, \autoref{sec:jam_results}). In fact the regular rotators on the \vse\ diagram were found to be broadly distributed within the envelope defined by the edge-on relation of \autoref{eq:beta_vs_eps} and its projection at different inclinations  \citep{Cappellari2007}. These relations are shown in \autoref{fig:vs_eps}. The anisotropy of regular rotators, within 1\re, was found to span a larger range than that of the non-regular rotators. They are generally not close to isotropic, but often fall close to the edge-on isotropic line due to projection effects. This explain previous observations suggesting the fast rotating galaxies are more isotropic than the slow rotating ones. In practice, starting from a range of $\varepsilon$ values, the edge-on magenta relation can be plotted on the \vse\ diagram using \autoref{eq:vs_tvt} and \autoref{eq:e}, and then the different projections are obtained from \autoref{eq:vs_proj}. 

Also shown in \autoref{fig:vs_eps} is the relation $(V/\sigma)^{\bigast}=1/3$, where $(V/\sigma)^{\bigast}\equiv (V/\sigma)/(V/\sigma)_{\rm iso}$ is the ratio between $V/\sigma$ and the corresponding theoretical value for an isotropic galaxy with the same $\varepsilon$. This parameter was called ``anisotropy parameter'' \citep[e.g.][]{Kormendy1982,Davies1983,Bender1988,Bender1992,Bender1994,Naab2003} although is clear from \autoref{fig:vs_eps}, that galaxies with the same $(V/\sigma)^{\bigast}$ can span the full range of anisotropies. Nonetheless, $(V/\sigma)^{\bigast}$ is still a very useful to quantify the global dynamics of ETGs. In fact, \citet{Cappellari2007} noted that a value of $(V/\sigma)^{\bigast}\approx0.4$ approximately separates the fast/regular and slow/non-regular rotating classes indicated by the IFS kinematics \citep{Emsellem2007,Cappellari2007}. This fact confirms and explains the early results, based on long-slit spectroscopy, reporting a connection between $(V/\sigma)^{\bigast}$ and galaxy properties \citep[e.g.,][]{Davies1983,Kormendy1996}.

All these results about the \vse\ diagram found by the \sauron\ survey \citep{deZeeuw2002}, were confirmed and strengthened, with the volume-limited sample of 260 galaxies, by the \atl\ survey \citep{Cappellari2011a}, which also provided a reliable census of the different classes of ETGs. The results were presented by \citet{Emsellem2011} and are reproduced, on top of the theoretical relations, in \autoref{fig:vs_eps}. The figure shows that the magenta line and its projections envelope, still broadly describes the location of regular rotators on the \vse\ diagram. Again a fixed $(V/\sigma)^{\bigast}$ value provides a rough separation of the two main kinematic classes, and the larger sample allows one to more accurately define the dividing line around $(V/\sigma)^{\bigast}\approx1/3$. 

The \atl\ survey uncovered 11 counter-rotating disks (\autoref{fig:jam_kdc}), which constitute 4\% of the volume-limited ETG sample \citep{Krajnovic2011}.
They can appear quite flat, as expected due to their disk-like nature. However, unlike the regular rotators, they all lie below the magenta line in the \vse\ diagram, because they do not have oblate velocity ellipsoids, but are instead dominated by tangential anisotropy ($\sigma_\phi>\sigma_R$ or $\gamma<0$). Below the magenta line one also finds some galaxies which are classified as regular rotators by \citet{Krajnovic2011}. These are likely transition objects containing an amount of counter-rotating stars which is too small to produce clear evidence for counter-rotation in the $\sigma$ field, but sufficiently large to produce a detectable decrease of their global rotation. There is in fact no reason to expect a sharp transition between the regular rotators and the counter-rotating disks. The classification into one class or the other will simply depend on the amount of externally-acquired counter-rotating gas \citep{Bois2011}.

\subsection{Quantitative kinematic classification}
\label{sec:kin_class}

\subsubsection{Early-type galaxies}
\label{sec:lam_eps_etgs}

It is useful to define a quantitative measure which approximately encodes the visual distinction between the two classes of regular and non-regular rotators illustrated in \autoref{fig:kinematic_classes}. A physically-motivated and clean separation can be obtained using dynamical models (\autoref{fig:kappa_histogram}). And this approach can be used when maximum accuracy is desired. However it may be possible to define a more economical approach. 

Another alternative is to use the dividing line $(V/\sigma)^{\bigast}\approx1/3$ on the \vse\ diagram  (\autoref{sec:vs_from_ifu}). However the \vs\ quantity has one major limitation: it makes no use of the spatial distribution in the kinematic maps. For this reason a galaxy with a KDC like NGC~5813 (\autoref{fig:kinematic_classes}c) can overlap on the \vse\ diagram with inclined regular rotators (e.g.\ NGC~3379) which have dramatically different kinematic appearance, as noted by \citet{Emsellem2007}. This was the motivation to develop a new physical parameter which retains the useful characteristics of the classic \vs\ quantity, but also includes spatial information. A natural replacement for $V$ is the magnitude of the luminosity-weighted averaged projected angular momentum $\langle\mathbf{L}\rangle=\langle \mathbf{R}\times\mathbf{V}\rangle$. To remove the need to determine vector directions and make a more easily computable quantity, this quantity was replaced by the surrogate $\langle R\,|V|\rangle$, where $R$ is the projected distance from the galaxy center. When this proxy for the angular momentum is made dimensionless and normalized with a quantity like the $V_{\rm rms}\equiv\sqrt{V^2+\sigma^2}$, which is proportional to mass, according to the scalar virial theorem \citep{Binney2008}, one obtains the parameter \citep{Emsellem2007}
\begin{equation}
\lambda_R\equiv
\frac{\langle R |V| \rangle}{\langle R \sqrt{V^2+\sigma^2} \rangle}
=\frac{\sum_{n=1}^N F_n R_n |V_n|}{\sum_{n=1}^N F_n R_n \sqrt{V_n^2+\sigma_n^2}},
\label{eq:lambda_r}
\end{equation}
where the $F_n$ are the fluxes within the $N$ spatial bins where mean stellar velocities $V_n$ and velocity dispersion $\sigma_n$ are measured. And the summation is extended out to a certain finite radius $R_{\rm max}$, within a galaxy isophote.

The distribution of the \lre\ values for ETGs for \atl\ (from \citealt{Emsellem2011}) and the SAMI Pilot survey (from \citealt{Fogarty2015}), is shown in \autoref{fig:vs_eps} (CALIFA is not included here because kinematics is not yet available). Also overlaid are the same lines shown in the \vse\ diagram, computed with the analytic expression of \autoref{eq:vs_tvt}--\autoref{eq:vs_proj}, as well as the anisotropy-shape relation of \autoref{eq:beta_vs_eps}. To calculate those lines, the {\em projected} location on the \vse\ diagram were simply converted into the corresponding ones for the \lre\ diagram using the empirical calibration below \citep{Emsellem2007,Emsellem2011}
\begin{equation}
\lambda_{R_{\rm e}} \approx \frac{k\, (V/\sigma)_{\rm e}}{\sqrt{1 + k^2 (V/\sigma)_{\rm e}^2}}
\qquad {\rm with} \qquad k=1.1
\label{eq:from_vse_to_lre}
\end{equation}

Given the close relation between \vs\ and \lr, broadly speaking, the \lre\ diagram shows the same information as the \vse\ diagram. Regular rotators are still generally described by the envelope of the magenta line and its projections, while non-regular rotators are not. The difference between the two diagrams is that, on the \lre\ diagram the regular and non-regular rotators of \citet{Krajnovic2011} are significantly better separated \citep{Emsellem2011}. 

\begin{marginnote}[120pt]
\entry{$\varepsilon_{\rm e}$}{Luminosity weighted ellipticity within the half-light isophote}
\end{marginnote}

More recently, the SAMI-pilot sample of \citet{Fogarty2015} and the CALIFA one by \citet{Falcon-Barroso2015} increased the number of slow rotators by $2.5\times$. They strongly confirm the \atl\ finding that ``genuine'' disk-less slow rotators are all rounder than $\varepsilon_{\rm e}<0.4$. Counter-rotating disks were shown in \autoref{sec:jam_results} to be structurally equivalent to regular-rotators and should not be classified as slow rotators. The new data motivate a refinement to the $\lr=0.31\sqrt{\varepsilon_{\rm e}}$ fast/slow rotator division of \citet{Emsellem2011} as follows, to reduce the risk of missing very round non regular rotators (see \autoref{fig:vs_eps})
\begin{equation}
\lr<0.08+\varepsilon_{\rm e}/4\qquad {\rm with} \qquad \varepsilon_{\rm e}<0.4.
\label{eq:fast_slow_divide}
\end{equation}
The explicit inclusion of the roundness criterion in the classification allows one to identify ETGs with disks, or fast rotators, without the need for kinematic observations.

The separation between fast and slow rotators was {\em defined} by \citet{Emsellem2007,Emsellem2011} for \lr\ measured within the half-light isophote. For this reason a galaxy cannot change its class when more extended data become available. In this classification like in most other ones, the {\em scale} is important! The same is true e.g. for Hubble's classification: a nuclear dust disk in an E galaxy may resemble a spiral galaxy from HST observations alone \citep[e.g.][]{Young2008}. But this does not make us classify the galaxy a spiral. Similarly, an E which turns out to have extended spiral arms in very deep observations \citep[e.g.][]{Duc2015}, should not be classified as a spiral.

We stress that the classification based on the \lre\ diagram only constitutes an automatic and objective proxy for the more detailed classification based on the kinematic morphology of \autoref{fig:kinematic_classes}. For reliable results it is important to verify that the two agree! This implies that, when the data are of such a low quality that no visual classification is possible, one should avoid using the \lre\ alone to kinematically classify ETGs, as this may lead to meaningless results. This is because systematic effects, like spurious fluctuations, on the velocity fields, produce systematic biases in \lr, as discussed by \citet{Emsellem2007}.

\subsubsection{Connection between dynamics and nuclear surface brightness profiles}
\label{sec:lam_eps_cores}

\begin{figure}
\centering
\includegraphics[width=0.7\columnwidth]{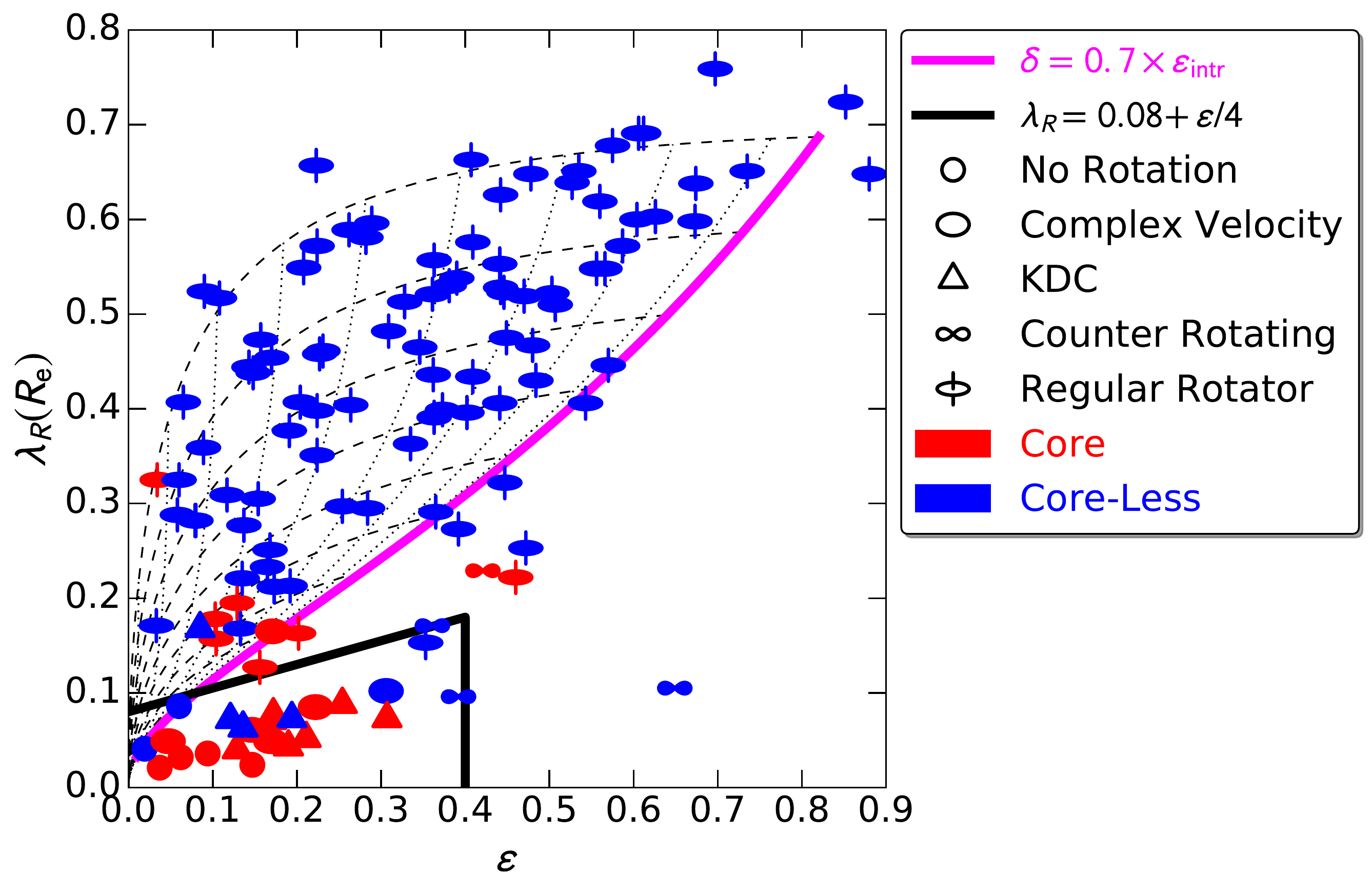}
\caption{{\bf Cusps and cores on the \lre\ diagram.} The values and symbols are the same as in \autoref{fig:vs_eps}, but only for the galaxies for which a core or core-less classification exists \citep[taken from][]{Krajnovic2013p23}. The magenta line is the same $\delta=0.7\times\varepsilon_{\rm intr}$ \citep{Cappellari2007} as in \autoref{fig:vs_eps}, while the brown line is the fast/slow rotator divide of \autoref{eq:fast_slow_divide}. There is a broad agreement between the slow rotators and the core galaxies, but not a full overlap.}
\label{fig:lam_eps_cores}
\end{figure}

As discussed in \autoref{sec:photometry}, the slow/fast rotator classes are expected to trace a similar phenomenon as the core/core-less classes respectively.
This hypothesis was tested by \citet{Emsellem2011} using a sample of 56 galaxies with measured central light deficit from either \citet{Kormendy2009} or \citet{Glass2011}. They concluded that there is indeed a close connection between being a slow rotator and having a central deficit in the surface brightness, with respect to a \citet{Sersic1968} profile extrapolated from larger radii. However the two classifications were not found to agree in all cases.

The question was revisited by \citet{Lauer2012}, using a subset of 63 galaxies in common between the \atl\ sample and the ``Nuker'' sample of \citet{lauer07prof}. He concluded that the two indicators generally agree, confirming they are both able to recognize dry merger relics. \citet{Krajnovic2013p23} doubled the size of both samples to 122 galaxies, thanks to new HST photometric determinations. They used a core versus core-less definition based on a ``Nuker'' profile fit (\autoref{eq:nuker}), but they showed that different criteria differ only at the $\sim3$\% level. The main figure from that paper is adapted here in \autoref{fig:lam_eps_cores}. In agreement with the previous studies, one can see that the general agreement is very good: Nearly all round and slow rotating ETGs do indeed have a core, and the vast majority of the regular rotator have a steep profile as expected.  However core and core-less ETGs do {\em not} overlap  perfectly with the regular versus non-regular rotator classes, and for the same reason they do not not separate precisely at the slow/fast rotator divide. 

To understand, at least in part, the small disagreements between the core/core-less and slow/fast classifications, it is instructive to consider a few individual examples. Three of the core-less slow rotators are counter-rotating disks (\autoref{fig:kinematic_classes}d). As these objects have disks and must have formed via gas accretion, the lack of a core makes physical sense and further emphasizes the fact, discussed in \autoref{sec:jam_results}, that counter-rotating disks are essentially ``misclassified'' slow rotators, which instead form a continuous sequence with the rest of the fast rotators. One of the counter-rotating disks however has a core. This galaxy is the well-studied E galaxy NGC~4473. The cored nature of its inner profile was studied in detail by \citet{Pinkney2003}, who suggested it may be due to a recent merger, which flattened a pre-existing cusp. The dynamics of NGC~4473 was modeled in detail in \citet{Cappellari2005} and \citet{Cappellari2007}. It was found to consists of two counter-rotating stellar components with approximate mass ratio 1:3. This galaxy was proposed as a prototype of the counter-rotating class. The large-scale stellar counter-rotation, and the ``lemniscate'' (which motivates the symbol used in this review to represent the kinematic class), or figure-of-eight, or ``double $\sigma$'' peaks, nature of this galaxy is beautifully illustrated by the combination of the \sauron\ and SLUGGS kinematics in \citet{Foster2013}. In summary, this galaxy does not differ from the other counter-rotating disks, and in this case the ``core'' classification does not imply the galaxy assembled by dry mergers like the core slow rotators.

Overall, as these few examples already illustrate, it is not surprising that the two core/core-less and slow/fast classifications do not agree in 100\% of the cases, although they agree most of the times. A galaxy which formed predominantly via gas accretion and originally posses a steep inner profile, may sometimes acquire a small gas-poor satellite which, in favorable conditions can destroy the inner cusp and nuclear disk \citep{Sarzi2015}, producing a core galaxy. The fact that most fast rotators are core-less indicates that this event is not a common occurrence. The small disagreement between the two classifications shows that, like in any classification, intermediate cases may exist and one should consider different characteristics of a galaxy, before making strong inferences about the formation mechanism of any individual object. The safest choice in this case is to consider as dry merger remnant only galaxies which {\em both} are classified as slow rotators and have an inner core. But unfortunately this information is not available for many galaxies. However the broad agreement indicates that, on a statistical basis either method is robust and allow one to draw conclusions about the overall galaxy population.

\subsubsection{Angular momentum across the Hubble sequence}
\label{sec:lam_eps_morph}

\begin{figure}
\centering
\begin{minipage}{.57\textwidth}
\includegraphics[width=\columnwidth]{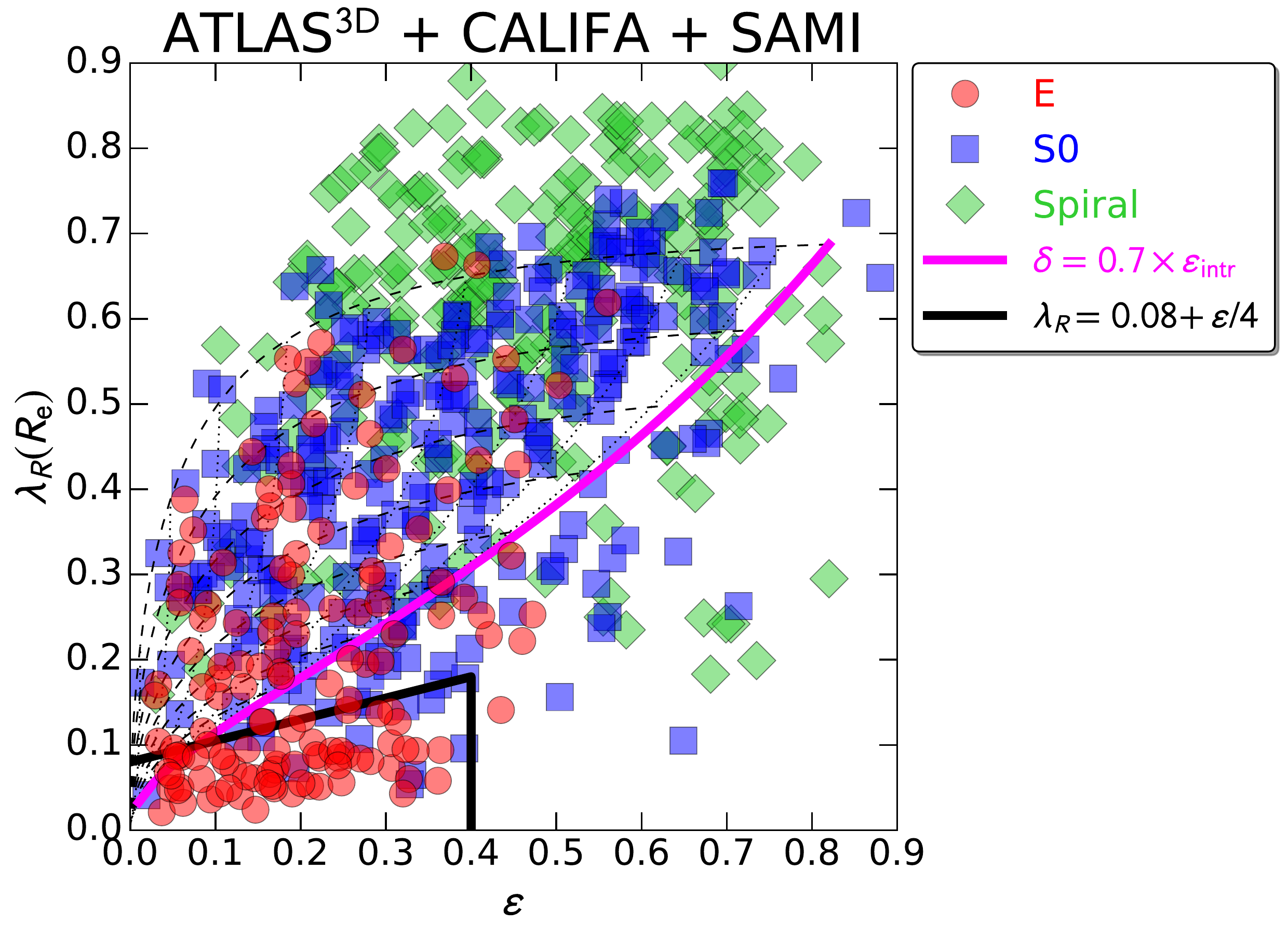}
\end{minipage}
\begin{minipage}{.42\textwidth}
\includegraphics[width=\columnwidth]{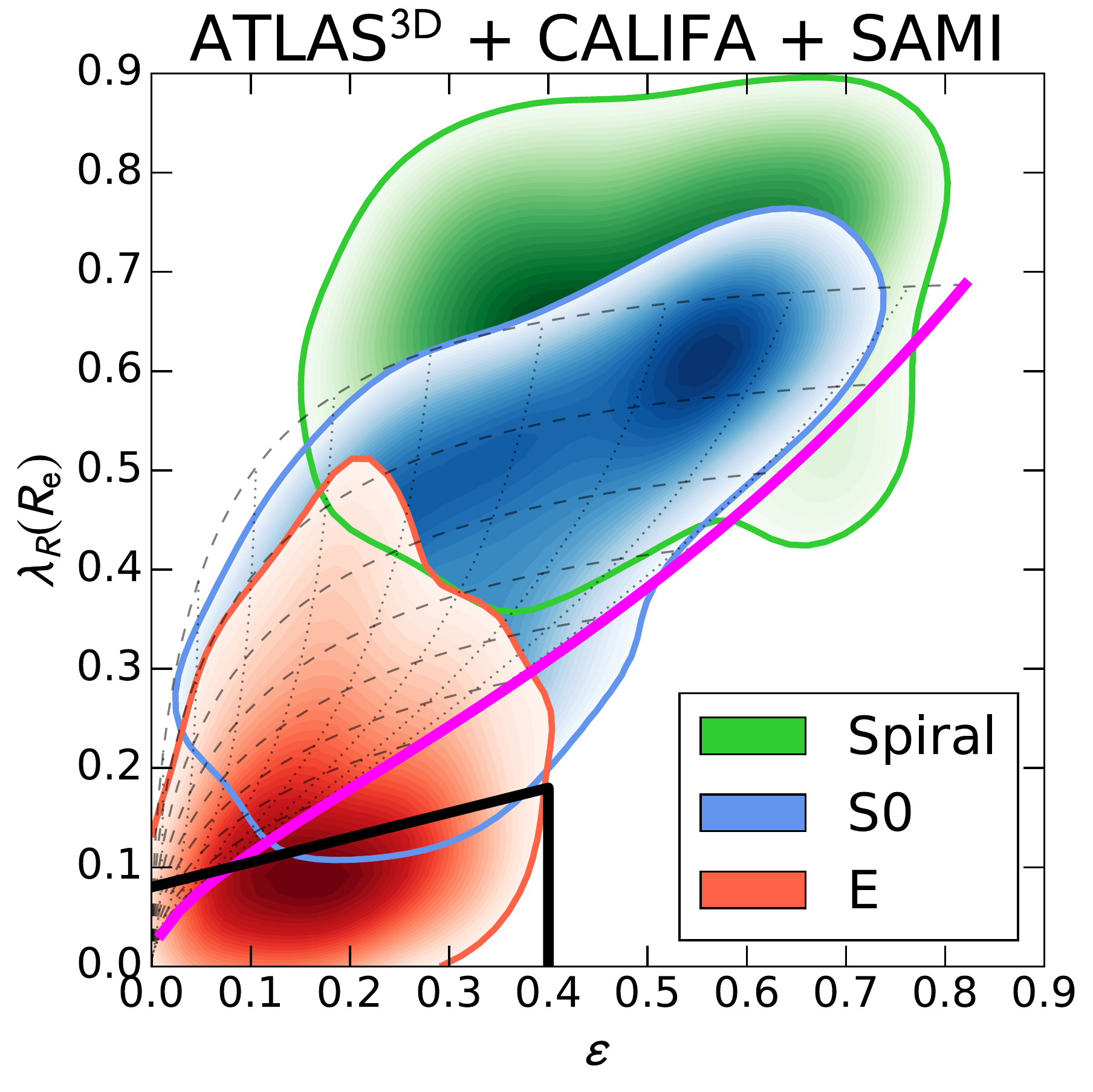}
\end{minipage}

\caption{{\bf Galaxy morphology on the \lre\ diagram.} {\em Left Panel:} Shows 666 \lre\ values, including 260 ETGs in \atl\ \citep[from][]{Emsellem2011}, 300 mostly spiral galaxies in CALIFA \citep[from][]{Falcon-Barroso2015} and 106 galaxies mostly ETGs in SAMI \citep[from][]{Fogarty2015}.
The magenta line is the same $\delta=0.7\times\varepsilon_{\rm intr}$ relation \citep{Cappellari2007} as in \autoref{fig:vs_eps}, while the brown line is the fast/slow rotator divide (\autoref{eq:fast_slow_divide}). The CALIFA sample by design lacks very round and very flat spiral galaxies and this explain their apparently different distribution in $\varepsilon$ from the ETGs. In this extended sample, which has $2.5\times$ more slow rotators than \atl, all secure slow rotators (i.e. excluding counter-rotating disks, cuspy ETGs and spirals) have $\varepsilon\la0.4$.
{\em Right Panel:} Shows a kernel density estimate of the distributions of Es, S0s and spirals in the left diagram. For each class, the thick solid line encloses 80\% of the probability. The region of slow rotators encloses almost only Es.
}
\label{fig:lam_eps_morph}
\end{figure}

In the \vse\ and \lre\ diagrams we used a single $\varepsilon$ per galaxy. However this quantity is not constant within a galaxy. In fact fast rotators are galaxies with a bulge and a disk. The flattest ones tend to be the more disk dominated ones. This means that the fast rotators near the top of the \lre\ diagram are expected to be on average the ETGs with small bulge fractions. This trend of decreasing \lr\ with increasing bulge fraction, or concentration, was shown by \citet{Krajnovic2013p17}. It has a large scatter due to projection effects, which make the observed ellipticity a poor estimator of bulge fraction, and the observed \lr\ an equally poor estimator of the intrinsic angular momentum.  In this picture one would expect spiral galaxies, which on average have smaller bulge fractions than S0s \citep[e.g.][]{Simien1986,deJong1996,Graham2001,Graham2008curv}, to overlap with the fast rotators ETGs with the largest \lr\ values. This was shown by \citet{Falcon-Barroso2015} \citep[see also][]{Querejeta2015}, including IFS kinematics of spiral galaxies from the CALIFA survey \citep{Sanchez2012}. The trend between \lr\ and galaxy concentration was also confirmed using data from the SAMI Pilot Survey by \citet{Fogarty2015}. An exception are the latest spiral galaxy types (Sd) which appear to have quite low \lr. An explanation for this fact will need to await a more detailed analysis on the kinematics of those objects, which is not yet available.

In \autoref{fig:lam_eps_morph} we illustrate the trend between galaxy morphology and \lr\ by combining data from the three largest IFS survey for which homogeneously measured parameters were published, for a total sample of 666 galaxies of all morphological types: \atl\ \citep{Cappellari2011a}, CALIFA \citep{Sanchez2012} and SAMI Pilot \citep{Fogarty2014}. We take (i) 260 values for the ETGs of the \atl\ survey from \cite{Emsellem2011}, (ii) 300 values for all morphological types, but mostly spiral a galaxies of the CALIFA survey from \citet{Falcon-Barroso2015} and Falc\'on-Barroso et al.\ in preparation, and (iii) 106 values for all morphological types, but mostly ETGs from SAMI from \citet{Fogarty2015}. One should note that, unlike for the ETGs, the distribution for the spiral galaxies, which mostly come from CALIFA, is not randomly oriented by design \citep{Walcher2014} and this explains the lack of very flat or very round spiral galaxies. Unlike the ETGs, these galaxies should not be approximately distributed like the envelope of the magenta line.

An interesting result of this plot is that the area of this diagram populated {\em only} by Es, approximately traces the region defining the slow rotator class (\autoref{eq:fast_slow_divide}). This shows that, even from images alone, the E classified as slow rotator look different from S0 galaxies and cannot be mistaken for S0s. The same is not true for the E classified as fast rotator, in the same range of ellipticity, which can be classified as either E or S0. This result
 further confirms the adopted criterion to separate the two classes of fast and slow rotators.

Overall, the plot clearly shows that the classic and still  widely used distinction between E and S0 \citep{Hubble1936,Sandage1961,deVaucouleurs1991} has little physical meaning, in fact many inclined fast rotators are classified as E. This is mainly due to the obvious difficulty of recognizing an inclined S0 from a genuinely E galaxy. The fact that E always have $\varepsilon\la0.5$ is essentially a matter of definition. For this reason flat fast rotators are always classified as S0. The misclassification of E is very significant. In the volume-limited \atl\ sample, as much as 66\% ($=2/3$) of E turns out to be a fast rotator, namely an inclined axisymmetric galaxy with a disk \citep{Emsellem2011}. Similarly, in the combined sample of \autoref{fig:lam_eps_morph}, which has twice as many E galaxies, 60\% are actually fast rotators. Most misclassifications can be corrected using other photometric indicators like isophotal shape, for edge on cases, and more in general nuclear slopes \citep{Kormendy1996}, but these are currently not applicable to large surveys or at high redshift.

\subsection{Two-dimensional Stellar population}
\label{sec:population_maps}

\begin{figure}
\centering
\includegraphics[width=\columnwidth]{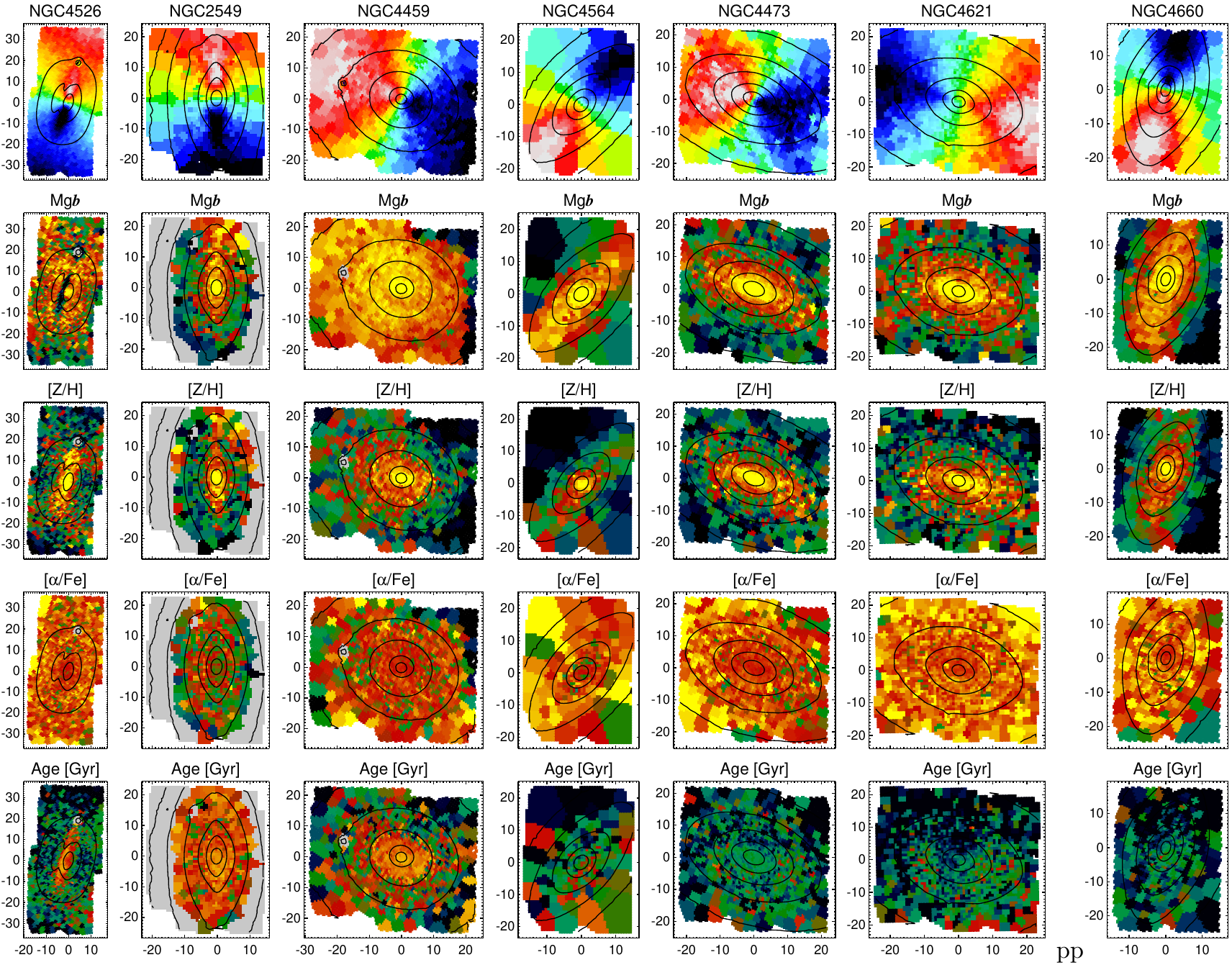}
\caption{{\bf Metallicity enhanced disks in fast rotators.} The top panels show the mean stellar velocity of seven fast rotator ETGs. Subsequent panels show the Mg$b$ line strength, the metallicity [Z/H], the elemental abundance [$\alpha$/Fe] and age. Strong nearly edge-on disks show an enhancement in the measured Mg$b$ or the inferred metallicity, which is flatter than the isophotes, while the age distribution is flat and featureless \citep[taken from][]{Kuntschner2010}. Exceptions are  NGC~4526 and NGC~4459, which contain young star forming disks, also rich of molecular gas \citep{Young2008}.
}
\label{fig:kuntschner2010_fig9}
\end{figure}

Integral field spectroscopy provides spatially-resolved stellar population information together with the stellar kinematics. Using profiles of stellar population parameters averaged over elliptical isophotes, \citet{Kuntschner2010} found a trend of increasing metallicity gradient with increasing galaxy mass for low-mass fast rotators $\mstar\la3\times10^{10}$, while the slopes decrease above that critical mass, so that the most massive systems have quite shallow logarithmic gradients. This finding agrees with long-slit observations by \citet{Spolaor2009}. The majority of the population was found to have negligible age trends. This is also true for slow rotators with KDCs, which have a population indistinguishable from the rest of the galaxy, in agreement with the SAURON observations of NGC~4365 by \citet{Davies2001} and the OASIS high-resolution observations by \citet{McDermid2006}.

It had been suggested that the local escape velocity $V_{\rm esc}$, derived from dynamical models in which mass follows light, may be a better parameter to study stellar population gradients. The key advantage is that using $V_{\rm esc}$, instead of radius, one finds local relations within a galaxy in quantitative agreement with global ones, among different galaxies. In particular \citet{Franx1990} found a good correlation between $V_{\rm esc}$ and galaxy color, while \citet{Davies1993} and \citet{Carollo1994} found good global and local correlation of $V_{\rm esc}$ with the Mg absorption line strength, for two samples of 8 and 5 ETGs respectively. \citet{Emsellem1996} first investigated the ${\rm Mg}-V_{\rm esc}$ correlation using IFS data.

The significance of the ${\rm Mg}-V_{\rm esc}$ correlation was demonstrated by \citet{Scott2009} for the 48 ETGs of the \sauron\ survey and by \citet{Scott2013p21} for the full \atl\ sample fo 260 ETGs. The dramatic improvement of these two studies was not only due to the large number of objects, but to the fact that IFS data allowed one to obtain very high quality radial gradient of the stellar population and accurate dynamical models. These studies showed that the ${\rm Mg}-V_{\rm esc}$ is truly universal in passive ETGs, and constitutes an ideal benchmark for numerical simulations trying to describe in detail the local variations of stellar population in ETGs.

The IFS data not only provided accurate radial gradients, but also fully resolved spatial distributions. This fact revealed an increase of metallicity in fast rotators which is specifically associated with the disk component, rather than being uniformly distributed within the spheroidal or bulge component \citep{Kuntschner2010,Scott2013p21}. This is indicated by the fact that all inclined fast rotators have contours of equal metallicity which are flatter than the galaxy isophotes, and generally consistent with the expected flattening of the disk component. This effect is visualized in \autoref{fig:kuntschner2010_fig9}. The figure also shows that the disk do not stand out in the age maps.

In recent times, thanks to the availability of high-quality stellar population models \citep[see][for a review]{Conroy2013} with high spectral resolution based on large empirical spectral libraries \citep{bruzual03,Vazdekis2012,Maraston2011,Conroy2012models}, the studies of the stellar population from integral-field data has moved to the full-spectrum fitting approach (using fitting codes like \textsc{ppxf} \citealt{Cappellari2004}; \textsc{starlight} \citealt{CidFernandes2005}; \textsc{steckmap} \citealt{Ocvirk2006}; \textsc{vespa} \citealt{Tojeiro2007}). Recent results from the CALIFA \citep{GonzalezDelgado2015}, MASSIVE \citep{Greene2015} and MaNGA surveys \citep{Wilkinson2015}, illustrate the power of IFS studies for stellar population. However a detailed overview of these stellar population studies goes beyond the scope of this review.


\section{LOCAL SCALING RELATIONS}

Dynamical scaling relations of ETGs relate the size, luminosity ($L$) and stellar kinematics of galaxies. Sizes are typically described by the half-light radius (\re), while kinematics is generally quantified by the stellar velocity dispersion ($\sigma$) within a given aperture, which in this review is assumed no larger than \re. Given that luminosity and size depend on distance, while kinematics do not, one of the first key applications of galaxy scaling relations was to infer galaxy distances \citep{Dressler1987,Djorgovski1987}.

Nowadays, dynamical scaling relations are a key tool to study galaxy formation. The main reasons are: (i) because they provide a statistical description for easily measurable properties of galaxies as a function of time (redshift), which can be directly compared with numerical simulations \citep[e.g.][]{Robertson2006,Boylan-Kolchin2006,Oser2012,Porter2014}; and (ii) due to the fact that scaling parameters are actually expected to evolve very differently depending on the galaxy formation mechanism \citep[e.g.][]{Naab2009,Hopkins2010}. 

In this section we review the recent development on scaling relations, focusing specifically on the advancements in our understanding of scaling relations made possible by the use of integral-field spectroscopy. \citet{Kormendy2009} and \citet{Graham2013} present two contrasting overviews focused on results derived via photometry or long slit spectroscopy.

\subsection{Classic scaling relations}

\subsubsection{Scaling relations and the Fundamental Plane}

The first dynamical scaling relation to be discovered was the one between luminosity and stellar velocity dispersion \citep{Faber1976}. The observed relation had the form $L\propto\sigma^4$ and the authors pointed out it also suggests a trend between the mass-to-light ratio ($M/L$) and galaxy luminosity. 

Soon thereafter, a correlation between galaxy surface brightness $\Sigma$ and galaxy size was also found \citep{Kormendy1977}. When one defines the surface brightness as the mean value within \re, then $\Sigma_e=L/(2\pi R_e^2)$. This means that the Kormendy relation describes a correlation between galaxy radius and luminosity. The latter form has the advantage that it does not explicitly include galaxy size on both axes of the correlation, reducing the covariance between the measured values. The $L-\re$ relation \citep[e.g.][]{Shen2003} has recently become quite popular to study galaxy evolution as a function of redshift \citep[e.g.][]{vanderWel2014}, given that it does not involve any kinematic determination and for this reason is much more ``economical'' to observe than the Faber-Jackson.

\begin{marginnote}[120pt]
\entry{Fundamental Plane (FP)}{The distribution of $(L,\sigma,\re)$ galaxy parameters}
\entry{Virial relation}{$M\propto\sigma^2\re$}
\end{marginnote}

Thanks to larger systematic surveys of ETGs it was later discovered that the Faber-Jackson and the Kormendy relations are just two special projections of a plane described by galaxies in $(\log L,\log\sigma,\log\re)$ coordinates \citep{Dressler1987,Djorgovski1987}. This plane was aptly named the Fundamental Plane (FP, see \citealt{Kormendy1989araa} for a review of the initial results). It was found to hold for all ETGs, including S0s and E galaxies, with a scatter smaller than 20\% in \re\ \citep[e.g.][]{Jorgensen1996}.

The existence of the FP was interpreted as due to galaxies satisfying virial equilibrium $M\propto\sigma^2\re$, with $M$ the galaxy mass \citep{Faber1987}. However the exponents of the FP were found to deviate significantly from the virial predictions, as confirmed by all numerous subsequent studies \citep[e.g.][]{Hudson1997,Scodeggio1998,Pahre1998,Colless2001,Bernardi2003fp}. In particular, a recent determination of the plane with $\sigma_e$ measured from IFS within \re, gives $L\propto\sigma_e^{1.25} R_e^{0.96}$ (fig.~12 in \citealt{Cappellari2013p15}). This deviation of the FP from the virial predictions is called the ``tilt'' of the FP.

\subsubsection{Why the Fundamental Plane deviates from the virial relation}

The variation of the $M/L$ of the stellar population was immediately recognized as a potential source for the tilt of the FP \citep{Faber1987}. This is because systematic changes in the galaxy population were already known, with galaxies becoming older and more metal rich with increasing mass or $\sigma$ \citep[e.g.][]{Thomas2005daniel,McDermid2015}. This variation can potentially explain a major part of the FP tilt and scatter, predicting larger $M/L$ as a function of $\sigma$ by a factor of a few, as observed (depending on the photometric band) over the full range of galaxy masses \citep{Prugniel1996fp,Forbes1998fp}.

The surface brightness profiles of ETGs also display systematic variations as a function of their luminosity, with profiles becoming more concentrated in more luminous objects (\autoref{sec:sersic_profiles}). At fixed mass, a steeper profile implies a larger $\sigma$ within the central regions \citep{Ciotti1991} where the kinematics is observed. The amount of $\sigma$ variation is again in principle sufficient to explain a major part of the FP tilt \citep{Ciotti1996,Graham1997fp,Prugniel1997fp,Bertin2002fp,Trujillo2004fp}.

A third potential cause for the FP tilt is the fraction of dark matter within the region where kinematics is observed. The dark matter fraction is expected to increase systematically with mass, for the range of interest of FP studies \citep[e.g.][]{Moster2010,Moster2013,Behroozi2010,Guo2010,Leauthaud2012}. This can cause variations in the observed total $M/L$ of an amount again sufficient to produce a significant fraction of the measured tilt \citep{Renzini1993,Borriello2003fp,Tortora2012}.

\subsection{Understanding scaling relations via integral-field spectroscopy}
\label{sec:understand_scaling_rel}

\subsubsection{The Mass Plane follows the virial relation}
\label{sec:mass_plane}

Using semi-isotropic Jeans models, and stellar kinematics of 37 galaxies from long-slit observations, it was found that the $M/L$ trend with galaxy mass remains nearly unchanged when one includes the effects of galaxy non-homology \citep{vanDerMarel91,Magorrian1998}. The model accuracy was improved by fitting models to IFS stellar kinematics. Using a sample of 25 galaxies and both the Schwarzschild and Jeans approaches, \citet{Cappellari2006} found that the $(M/L)-\sigma$ relation is extremely tight and ``can be included in the remarkable series of tight correlations between $\sigma$ e and other galaxy global observables''. The relation was found to accounts for the entire scatter and tilt of the FP. Independent confirmations of these facts came from strong lensing studies \citep{Bolton2007mp,Auger2010}. 

\begin{marginnote}[120pt]
\entry{Mass Plane (MP)}{The distribution of $(M,\sigma,\re)$ galaxy parameters}
\entry{\mdyn}{Dynamically determined total galaxy stellar mass}
\end{marginnote}

This result was dramatically strengthened by two larger studies using integral-field stellar kinematics: the modeling the 260 ETGs of the \atl\ sample \citep{Cappellari2013p15} and a similar study using 106 galaxies, of generally larger masses, from the SAMI Pilot Survey \citep{Scott2015}. Both studies employed the JAM method, which was shown using simulation to produce unbiased $M/L$ estimates \citep{Lablanche2012,Li2016}. These modeling studies showed that, when replacing luminosity with mass in the FP $(L, \sigma, \re)$, the coefficients of the resulting Mass Plane $(M_{\rm JAM}, \sigma, \re)$ matched the virial predictions $M\propto\sigma^2\re$ within the errors.  However, the studies also pointed out the significant dependence of the plane coefficients on the technique used to measure them. This sensitivity explains the apparent contrast between some  past studies of the FP tilt.

As demonstrated in detail by \citet{Cappellari2013p15}, the mass $M_{\rm JAM}$ which appear in the Mass Plane represent the most accurate empirical approximation of the galaxies total stellar mass \mdyn\
\begin{equation}
M_{\rm JAM} = L\times(M/L)_{\rm JAM}\approx 2\times M_{1/2} \approx \mdyn,
\label{eq:m_jam_definition}
\end{equation}
where $L$ is the galaxy total luminosity and $M_{1/2}$ is the total mass contained within a sphere enclosing half of the total galaxy luminosity.

\begin{figure}
\centering
\includegraphics[width=0.7\textwidth]{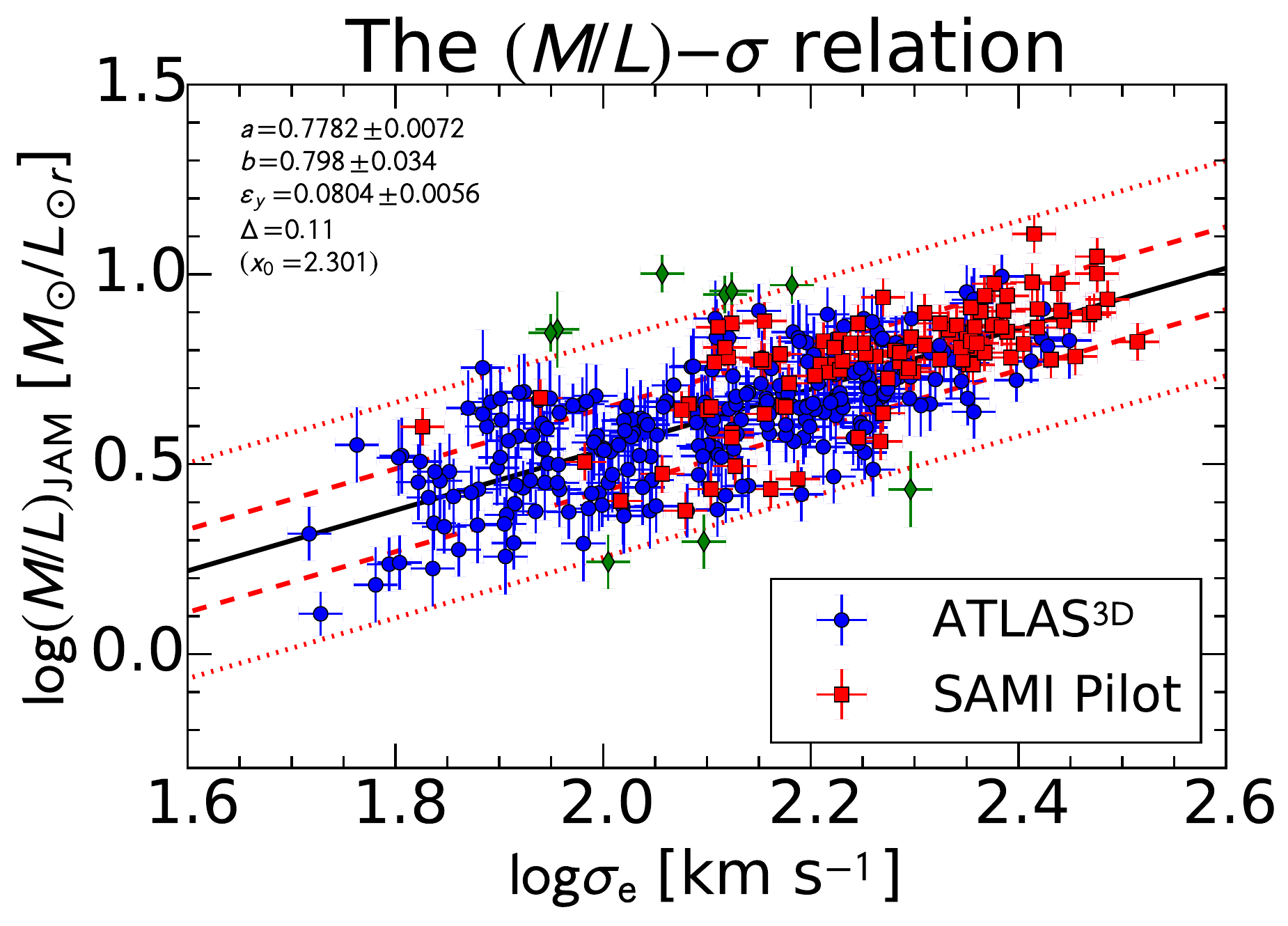}
\caption{{\bf The $(M/L)-\sigma$ relation.} The dynamical total $r$-band mass-to-light ratios and \se\ values for 366 ETGs. 260 \atl\ ETGs come from \citet{Cappellari2013p15} and 106 from the SAMI Pilot survey from \citet{Scott2015}, with an extra correction for surface brightness dimming. In both cases the values were determined from JAM modeling of the IFS stellar kinematics. The two complementary samples agree extremely well. They define a tight relation of the form $(M/L)_r\propto\sigma^{0.80}$, consistent with earlier findings, with an observed scatter of 29\%, decreasing to 20\% for  $\se>200$ \kms. The best fitting values of the linear relation $y=a+b\,(x-x_0)$, the observed rms scatter $\Delta$ in dex and the inferred intrinsic one $\varepsilon_y$ are written in the top left corner.}
\label{fig:ml_sigma}
\end{figure}

In \autoref{fig:ml_sigma} we show the combined $(M/L)-\sigma$ values taken from both the \atl\ survey \citep{Cappellari2013p15} and from the SAMI Pilot survey, in three different clusters \citep{Scott2015}. In both cases the values are computed in the SDSS $r$-band (AB mag system). Here, the published SAMI $(M/L)$ values were divided by a factor $(1+z)^3$, to account for both the $(1+z)^4$ contribution from cosmological surface brightness dimming and the $(1+z)^{-1}$ contribution due to the frequency decrease (as we are using AB mag) \citep[e.g.][]{Hogg1999}, at the typical redshift $z\approx0.06$ of the observations. This correction was not included in the original paper \citep{Scott2015}. The resulting correlation, for a combined sample of 366 galaxies, has an observed scatter of $\Delta=0.11$ dex (29\%), from which we infer an intrinsic scatter of 20\%. The scatter further decreases for the subsample of 69 galaxies with $\sigma_e>200$ \kms, for which the {\em observed} scatter becomes $\Delta=0.078$ dex (20\%). The best fitting relation was obtained with the Python version of the program \textsc{lts\_linefit}\footnote{Available from \url{http://purl.org/cappellari/software}} described in \citet{Cappellari2013p15}, which combines the Least Trimmed Squares robust technique of \citet{Rousseeuw2006} into a least-squares fitting algorithm which allows for errors in both variables and intrinsic scatter. It is given by 
\begin{equation}
\left(\frac{M}{L_r}\right)_{\rm\!\! JAM} = (6.00 \pm 0.10) \times\left(\frac{\sigma_e}{200\, \kms}\right)^{0.80\pm0.03}.
\label{eq:ml_sigma}
\end{equation}
This relation, and its scatter, are consistent within the errors with the one for the \atl\ subsample alone \citep{Cappellari2013p15}, while the slope is also consistent with the one from \citet{Cappellari2006}, albeit with much smaller errors. However both the scatter and the slope in the relation weakly depend on galaxy properties, with the relation being more shallow and having smaller scatter for galaxies in cluster and for slow rotators \citep{Cappellari2013p15}, in agreement with previous IFS studies of the FP \citep{FalconBarroso2011}.

The $M/L_r$ in $r$-band can be can be accurately converted to a different bands $B$ using the color $(B-r)_{\rm gal}$ of the galaxies and of the Sun $(B-r)_\odot$ as follows
\begin{equation}
M/L_B = (M/L_r)\times10^{0.4[(B-r)_{\rm gal} - (B-r)_\odot]}.
\label{eq:ml_conversion}
\end{equation}
We used the $B$ (Vega system) magnitudes from RC3 and the $K_s$ (Vega system) magnitudes from 2MASS, which are available for all \atl\ galaxies, to fit a tight color-$\sigma$ relation. This was used, together with the Sun colors from \citet{Blanton2007}, to provide the $(M/L)-\sigma$ relation in those two bands:
\begin{equation}
\left(\frac{M}{L_B}\right)_{\rm\!\! JAM} = (7.46 \pm 0.12) \times\left(\frac{\sigma_e}{200\, \kms}\right)^{0.87\pm0.03}
\label{eq:ml_sigma_B}
\end{equation}
\begin{equation}
\left(\frac{M}{L_{K_s}}\right)_{\rm\!\! JAM} = (1.44 \pm 0.02) \times\left(\frac{\sigma_e}{200\, \kms}\right)^{0.58\pm0.03}.
\label{eq:ml_sigma_K}
\end{equation}
The normalization of \autoref{eq:ml_sigma_B} agrees within the relative errors with the one derived by \citet{vanderMarel2007}, but with a slightly more shallow slope.

A natural question is how much of the observed trend in $M/L$ can be explained by variations in the stellar population, and to first order of its age and metallicity. The study by \citet{Gerhard2001}, using detailed dynamical models, found that the dynamically-derived \mldyn\ was related to the \mlpop\ inferred from stellar population models. This confirmed that at least part of the FP tilt is due to stellar population variations. It agrees with the fact that the scatter around the FP is also linked to variations in the stellar population \citep{Graves2009b,FalconBarroso2011,Springob2012,Magoulas2012}.
However, even improving the accuracy of the models using IFS kinematics, the relation between dynamically-derived {\em total} $M/L$ and the stellar population \mlpop\ still showed significant systematic deviations \citep{Cappellari2006}. These could only be explained by either dark matter or IMF variations among galaxies.

\subsubsection{Little dark matter and non-universal stellar Initial Mass Function}
\label{sec:imf}

\begin{figure}
\centering
\begin{minipage}[b]{.49\textwidth}
\includegraphics[width=\textwidth]{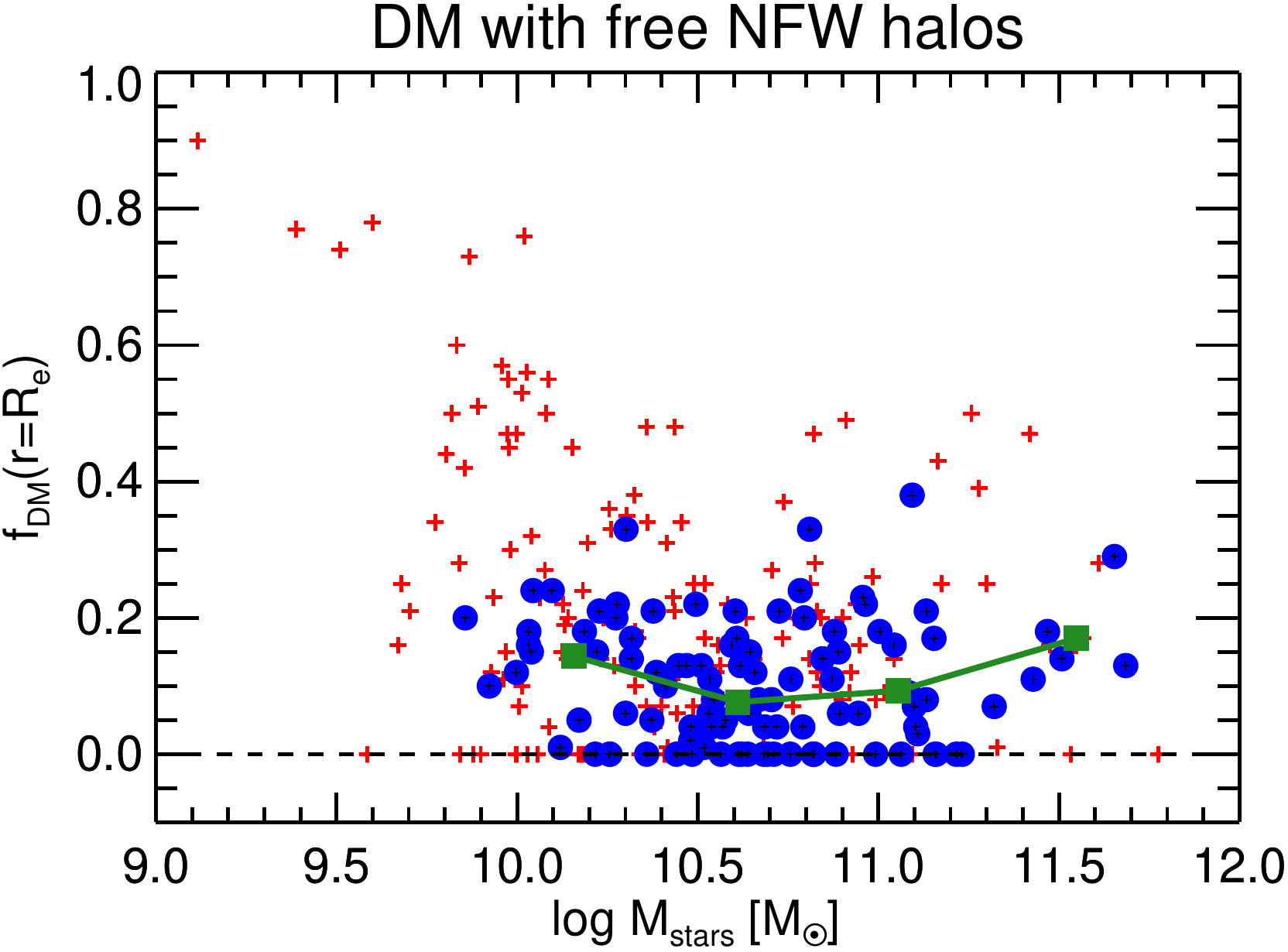} 
\end{minipage}
\begin{minipage}[b]{.49\textwidth}
\includegraphics[width=\textwidth]{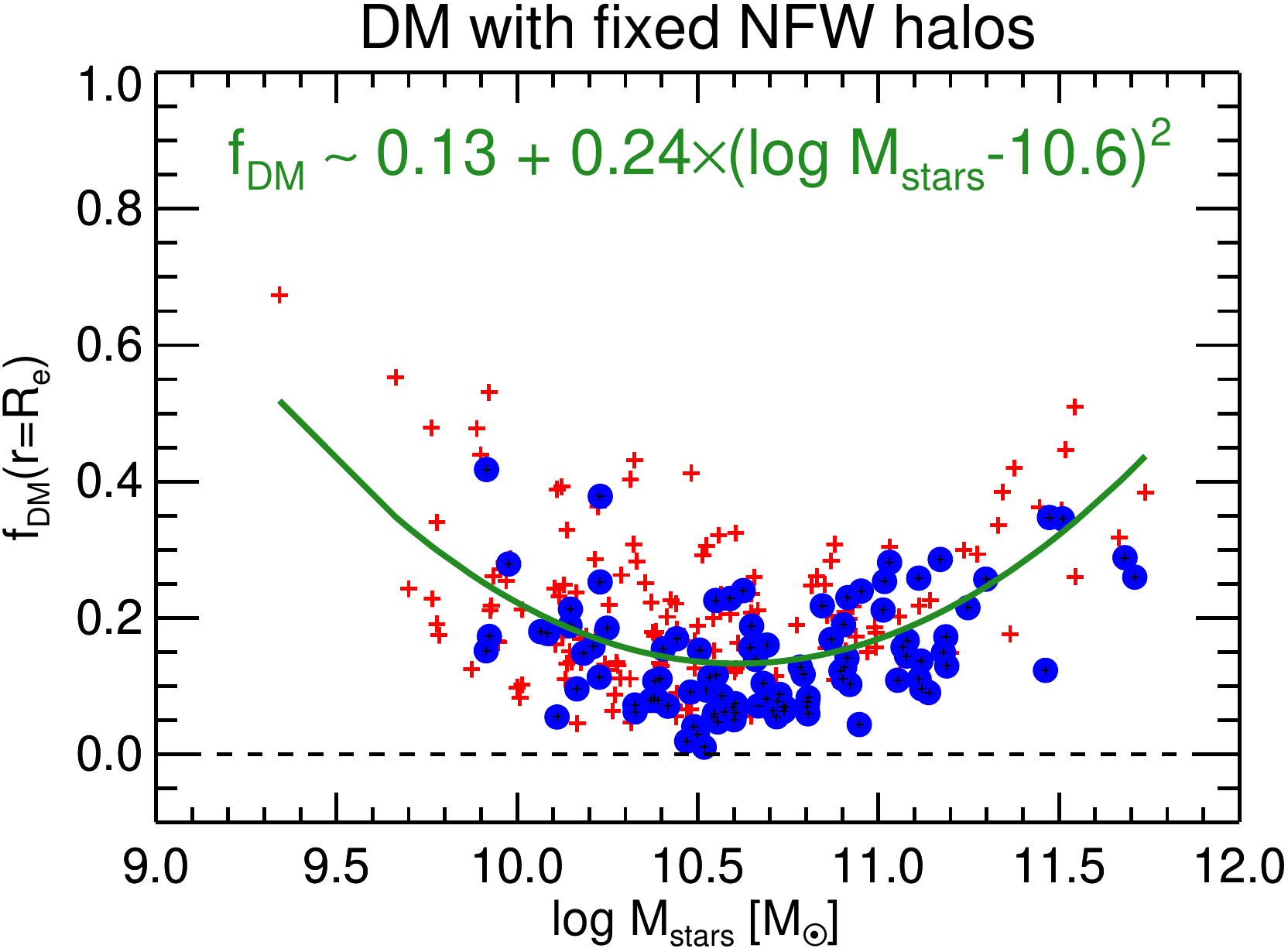} 
\end{minipage}
\caption{{\bf Dark matter from stellar dynamics.} Either leaving the halo normalization as a free parameter in the models (left panel), or constraining it to the value predicted by $\Lambda$CDM (right panel), the inferred dark matter fractions within \re\  must be small, for the models to be able to accurately fit the photometry and the IFS kinematics. Filled disks have better data, while red crosses have lower quality. (Taken from \citealt{Cappellari2013p15})
}
\label{fig:dark_matter}
\end{figure}

Using long-slit data and general models for two samples of about 20 galaxies, different studies appeared to agree that dark matter represents a minor fraction of the total, within a sphere of radius $r\sim\re$ \citep{Gerhard2001,Thomas2011}. Deviations between accurate determinations of the stellar and total masses were also found by combining the strong lensing technique with VIMOS integral-field observations \citep[e.g.][]{Barnabe2011} or using multiple long-slit observations \citep{Thomas2011}. The dark matter content within \re\ was better quantified with the modeling of the IFS kinematics for the 260 ETGs of the \atl\ sample \citep{Cappellari2013p15}. A median dark matter fraction $f_{\rm DM}(r=\re)$ as low as 13\%  was measured for the full sample (\autoref{fig:dark_matter}).

\begin{marginnote}[120pt]
\entry{IMF}{Stellar initial mass function}
\entry{\mlpop}{Stellar mass-to-light ratio measured via stellar population}
\entry{\mldyn}{Stellar mass-to-light ratio measured via dynamical models}
\entry{$f_{\rm DM}(\re)$}{Fraction of dark matter within a sphere of radius \re}
\end{marginnote}

\begin{figure}
\centering
\includegraphics[width=0.6\textwidth]{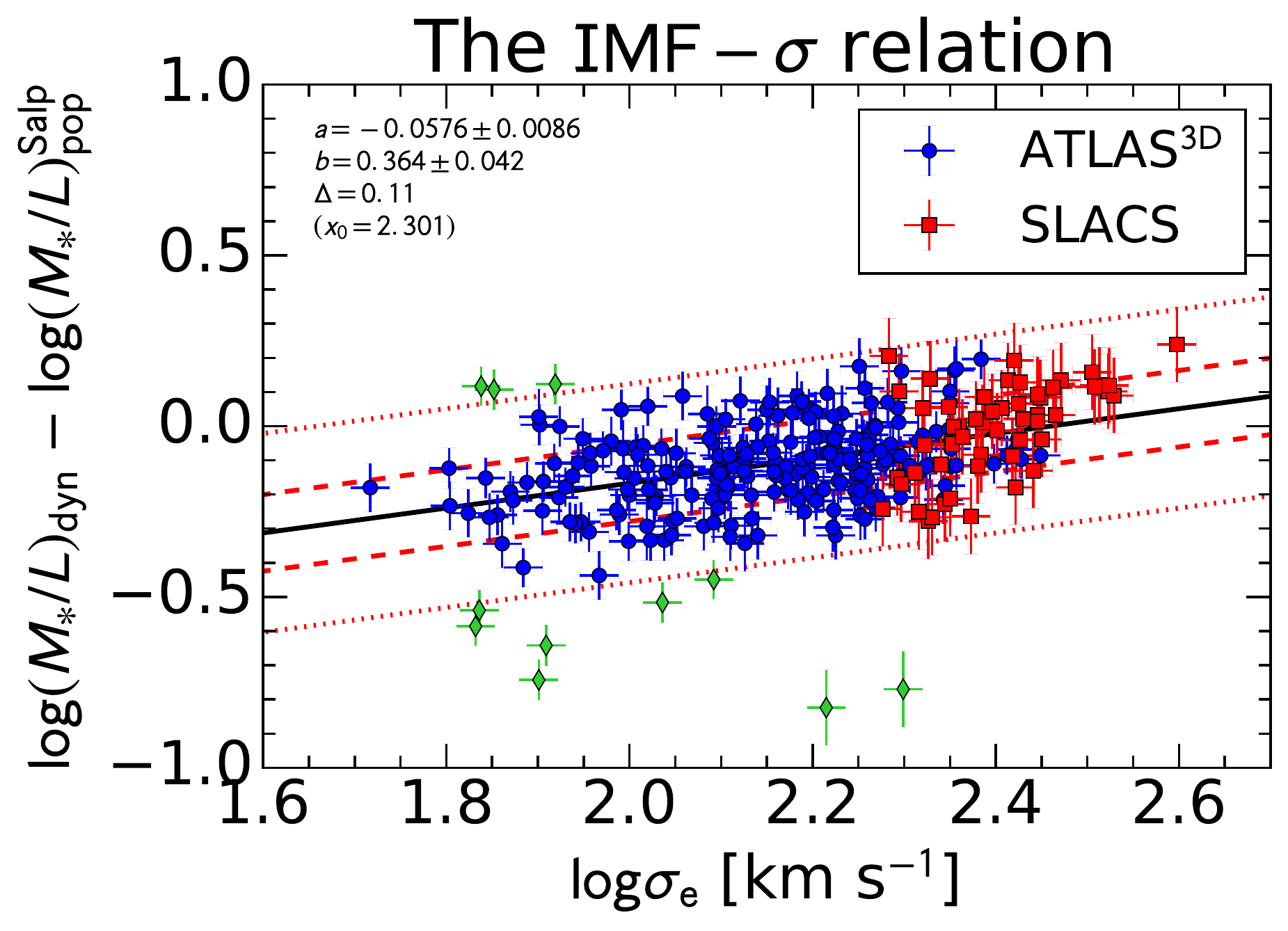}
\caption{{\bf The ${\rm IMF}-\sigma$ relation.} This is the relation between the ratio $\alpha\equiv\mldyn/\mlpop$
between the stellar $M/L$ inferred from mass dynamical or strong lensing modeling, and the same quantity inferred from full-spectrum fitting using stellar population models, assuming as a reference a \citet{Salpeter1955} IMF. The dynamical models of the IFS \atl\ data are taken from \citet{Cappellari2013p15}, while the combined lensing and dynamics models were based on the SLACS lens models of \citet{Treu2010}, but were redone using axisymmetric JAM models for maximum consistency between the two sets of determinations. The two sets of data consistently indicate a mass excess that cannot be ascribed to dark matter and can be interpreted as a variation of the IMF \citep[adapted from][]{Posacki2015}.
}
\label{fig:imf_sigma}
\end{figure}

Thanks to the large sample and two-dimensional stellar kinematics, the \atl\ study was able to show that the systematic trend in the discrepancy between \mldyn\ and \mlpop\ could not be explained by a variation in the dark matter fraction. The constraining power of the IFS data allowed the JAM dynamical models to explore, with a Bayesian approach, a range in dark matter inner slopes which included both a flat inner dark matter profile, a ``standard'' \citet*{navarro96}, and a contracted halos \citep{Gnedin2004} as special cases. Even with this freedom, the models were unable to reproduce the photometry and IFS kinematics by varying the dark matter alone \citep[see][for an illustration]{Cappellari2012}. The only remaining option for the discrepancy was then a systematic variation of the stellar initial mass function (IMF)  \citep{Cappellari2012}. This was inferred to vary in mass normalization from Milky-Way type \citep{Kroupa2001,Chabrier2003} to heavier than \citet{Salpeter1955} type, over the full mass range, with a dependence on the galaxies velocity dispersion \citep{Cappellari2012,Cappellari2013p20}. The IMF variation contrasted with previous indications of universality \citep{Bastian2010}. However, the IMF for the most massive galaxies in the sample appeared consistent with previous indications of a ``heavy'' IMF in massive ETGs from either stellar population \citep{vanDokkum2010} or strong gravitational lensing \citep{Auger2010imf}. A recent update to the ${\rm IMF}-\sigma$ relation by \citet{Cappellari2013p20}, was obtained by combining \mldyn\ determination from \atl\ with fully-consistent re-determination of \mldyn\ from SLACS \citep{Bolton2006}, combining strong lensing masses with JAM models. It is shown in \autoref{fig:imf_sigma}, from \citet{Posacki2015}. It illustrates the good agreement between the two approaches, and the need for a ``heavy'' IMF at the largest $\sigma$ values.

The trend in IMF was found to be most closely related to the bulge mass fraction, as inferred from the galaxy dynamics, rather than to the galaxy mass alone. The heaviest IMF was generally measured for the densest fast rotator galaxies and not for the most massive galaxies (\citealt{Cappellari2013p20} and \autoref{fig:mass_size_all}i). This result seems so far confirmed by individual studies of three dense ETGs using long-slit kinematics \citep{Lasker2013} and IFS data \citep{Yildirim2015}.

A large number of papers have appeared in recent years on the non-universality of the IMF in ETGs. Most of the studies used IMF sensitive spectral features in combination with stellar population models \citep[e.g.][]{vanDokkum2012,Conroy2012,Spiniello2012,Smith2012,Ferreras2013,Spiniello2015,LaBarbera2015,MartinNavarro2015}. Others used constraints provided by scaling relations \citep{Dutton2013imf_fp} or approximate dynamical models in combination with stellar population information \citep{Tortora2013imf,Conroy2013imf}. There is an encouraging consensus for the need for systematic IMF variations among different galaxies. However there are also clear tensions between different results, with stellar population models being unable to accurately predict the \mldyn\ observed by dynamical or lensing approaches \citep{Smith2014}, and possible contrast on the parameters driving the IMF variation \citep[e.g.][]{McDermid2014,MartinNavarro2015}.

Initial studies at $z\sim1$  found a Salpeter-like mass normalization of the IMF for $\mstar\ga10^{11}$ \msun\ \citep{Shetty2014,Sonnenfeld2015}. This is consistent with the IMF in the centers of massive nearby ETGs, assuming passive evolution. Similar consistency was found for the IMF at $z\sim1$ from spectral features \citep{MartinNavarro2015b}.

Integral-filed spectroscopy has the potential of revolutionizing this field once more. In particular, the ongoing MaNGA survey \citep{Bundy2015} combines a large wavelength range, which provides access to the most important IMF-sensitive spectral features \citep{Conroy2012models}, with good quality IFS data for a sample of 10,000 galaxies. The survey data can be used to try to reach a consensus, if possible, between different approach, and clarify our understanding of IMF trends in galaxies. Progress in this field can also be expected from MUSE observations, thanks to the impressive spectral data quality that the instrument can provide.

\subsubsection{Nearly isothermal mass profiles to four half-light radii}
\label{sec:total_density_profiles}

In spiral galaxies, circular velocities could be measured decades ago using ionized \citep{Rubin1980} and neutral gas \citep{Bosma1978}. Observations indicated flat circular rotation curves and consequently nearly-isothermal $\rho_{\rm tot}\propto r^{-2}$ profiles  (see review by \citealt{Courteau2014}).

Mass profiles of ETGs are much more difficult to infer, as one generally has to use expensive observations of the stellar kinematics and more complex dynamical models. For this reason, until recently, most studies did not extend to radii much larger than \re.
Long-slit observations of two different samples of $\approx20$ ETGs revealed rotation curves to be nearly flat with nearly-isothermal mass profiles, as in spiral galaxies, within the median radius $r\approx2\re$ sampled by the kinematics \citep{Gerhard2001,Thomas2011}. 
Fully consistent results were independently found using strong gravitational lensing. In particular the SLACS survey \citep{Bolton2006} found an isothermal slope, with small scatter, for the {\em total} galaxy density profile of 73 ETGs, at a typical radius of $r\approx\re/2$  \citep{Koopmans2009,Auger2010}. 

For comparison, at groups and cluster scales, mass profiles derived using X-ray modeling indicate a trend in the total density profiles varying systematically from $\rho_{\rm tot}\propto r^{-2}$ for the smallest systems, to $\rho_{\rm tot}\propto r^{-1}$ for the largest ones  \citep[e.g.][]{Humphrey2010}. A broadly consistent systematic trend was obtained by combining strong and weak lensing for massive galaxy clusters \citep{Newman2013,Newman2015}. The stacked weak-lensing approach by \citet{Gavazzi2007} indicated on average isothermal profiles out to very large radii.

A number of studies of individual ETGs exist. They used observations of extended HI disks \citep{Weijmans2008}, deep observations of the stellar kinematics at sparse locations \citep{Weijmans2009,Forestell2010,Murphy2011}, the discrete kinematics determinations of individual globular cluster \citep{Deason2012dm,Napolitano2014} and the kinematics of planetary nebulae \citep{Romanowsky2003,deLorenzi2008,deLorenzi2009,Napolitano2011,Morganti2013}. But a consistent picture did not emerge, due to inhomogeneity of the targets and observational techniques \citep[see][for a review]{Gerhard2013}.

\begin{marginnote}[120pt]
\entry{$\rho_{\rm tot}$}{Total, luminous plus dark, mass density}
\end{marginnote}

Only recently, large scale two-dimensional stellar kinematic data started to become available for statistically significant samples of ETGs \citep{Brodie2014,Raskutti2014}. \citet{Cappellari2015dm} combined the two-dimensional stellar kinematics of 14 fast rotator ETGs out to a median radius for the sample of $r\approx4\re$ from the SLUGGS survey \citep{Brodie2014}, with IFS within $\la1\re$ from \atl\ \citep{Cappellari2011a}. The data were modeled using the JAM approach. The study did not try to separate the luminous/dark matter components, as this generally requires making restrictive assumptions, but only focused on the {\em total} density, using a quite general parametrization.  The resulting {\em total} density profiles were found well described by a nearly-isothermal power law $\rho_{\rm tot}(r)\propto r^{-\gamma}$ from \re/10 to at least 4\re, the largest average deviation being 11\%. The average logarithmic slope is $\langle\gamma\rangle=2.19\pm0.03$ with {\em observed} rms scatter of just $\sigma_\gamma=0.11$. This scatter out to large radii, where dark matter dominates, is as small as previously reported by lensing studies at  $r\approx\re/2$, where the stars dominate. 

The apparent profile universality seems part of a more general trend in the total density slopes. In fact the slopes within $r\la\re$, where models using IFS were constructed for large numbers of galaxies, already indicate   (\autoref{fig:mass_size_all}c) that the total mass profiles become more shallow for galaxies with $\sigma_e\la100$ \kms, while the profile remains indeed universal with high accuracy for larger \se, with just a small decrease in the absolute slope above the critical mass $M_{\rm crit}\approx2\times10^{11}$ \msun. However, larger samples of homogeneously modeled galaxies, are needed to understand whether these central trend persist at larger radii, where dark matter dominates. Profile slopes place important constraints to galaxy formation models \citep[e.g.][]{Remus2013,Dutton2014}.

\subsection{The mass-size and mass-$\sigma$ distributions}
\label{sec:local_mass_size}

\subsubsection{Results from integral-field spectroscopy}

\begin{figure}
\centering
\includegraphics[width=0.8\textwidth]{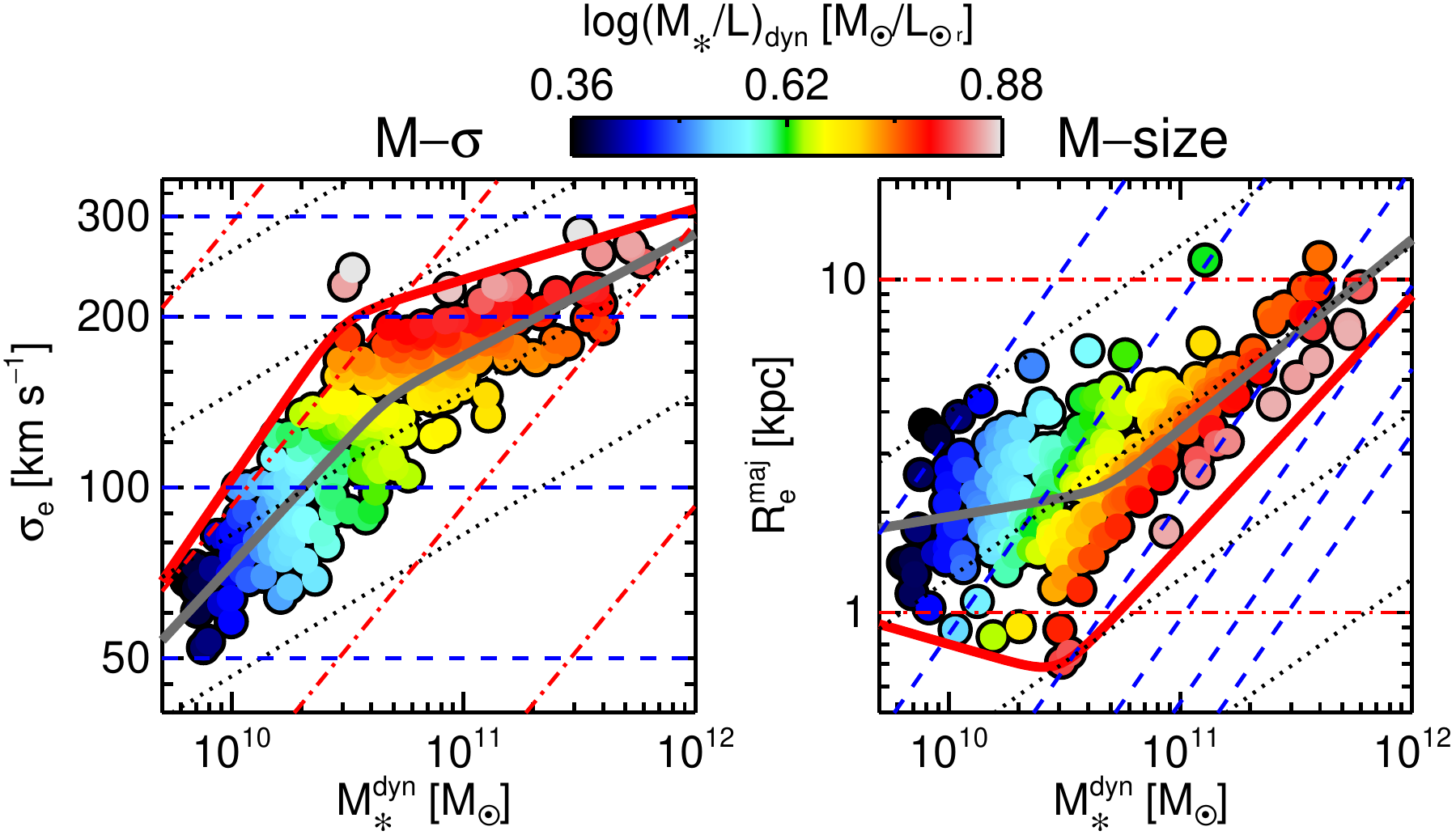}
\caption{{\bf Break in scaling relations.} The projections of the MP in the $(\mdyn,\se)$ and $(\mdyn,R_{\rm e}^{\rm maj})$ coordinates. Overlaid are lines of constant $\se=50,100,200,300,400,500$ \kms\ (dashed blue), constant $R_{\rm e}^{\rm maj}=0.1,1,10,100$ kpc (dot-dashed red) and constant $\Sigma_{\rm e}=10^8,10^9,10^{10},10^{11}$ \msun\ kpc$^{-2}$ (dotted black) transformed in different panels using the virial relation. In each panel the galaxies are colored according to the (LOESS smoothed) $\log (M/L)_{\rm JAM}$, as shown in the color bar at the top.  In both panels the thick red line shows the same ZOE relation given by \autoref{eq:zoe}. The gray line is the same $\mdyn-\se$ relation with trends $\mdyn\propto\sigma_{\rm e}^{4.6}$ for $\se\gg140$ \kms\ and $\mdyn\propto\sigma_{\rm e}^{2.3}$ for $\se\ll140$ \kms. \citep[Taken from][]{Cappellari2013p20} 
}
\label{fig:mass_plane}
\end{figure}

\begin{figure}
\centering
\includegraphics[width=\textwidth]{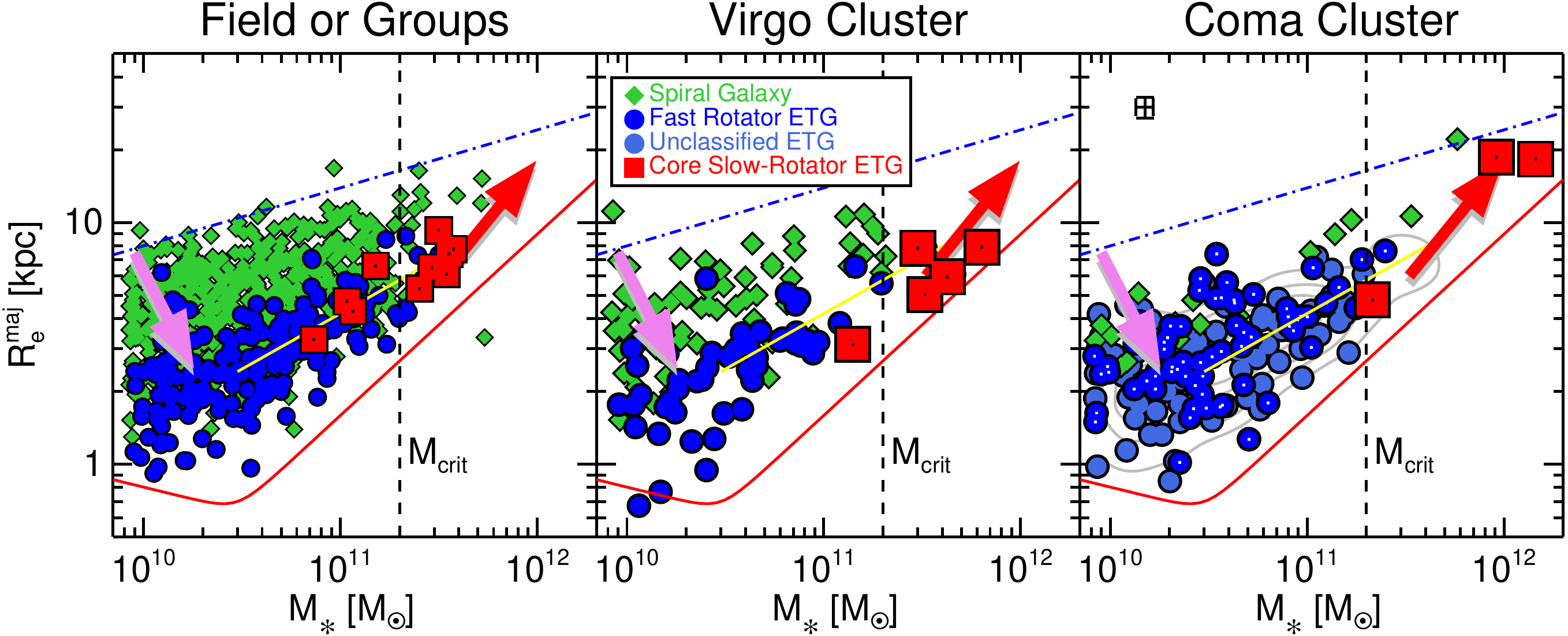}
\caption{{\bf Mass-Size versus environment.} The {\em Left Panel} shows the $(\mstar,\re)$ distribution for a sample of galaxies belonging to either the field or to small groups. The {\em Middle Panel} is for a sample in the moderately-dense Virgo cluster and the {\em Right Panel} for a sample in the Coma cluster, which is one of the densest environments in the Universe. The plots visualize the two main channels for the formation of ETGs. The magenta arrow qualitatively indicates the evolutionary channel starting from star forming spiral galaxies and producing fast rotators via gas accretion and bulge growth followed by quenching. The red arrow shows the dry merging and halo quenching channel, producing massive slow rotators with central cores in their surface brightness profiles (the left and right panels are from \citealt{Cappellari2013apjl}).
}
\label{fig:coma_mass_size}
\end{figure}

\begin{figure}
\centering
\includegraphics[width=\textwidth]{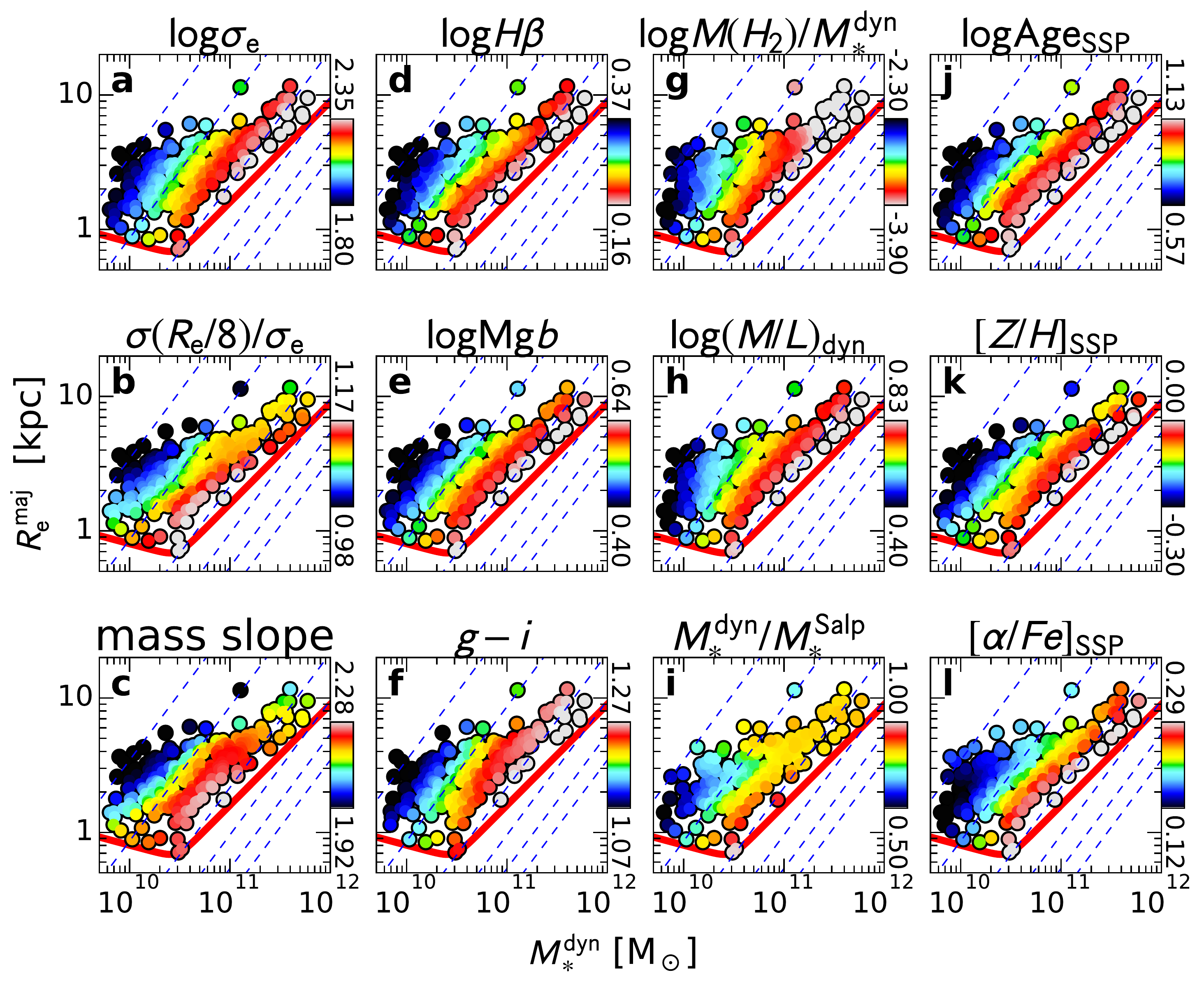}
\caption{{\bf ETGs properties on the mass-size diagram.} 
(a) Effective velocity dispersion \se\ \citep[from][]{Cappellari2013p15}; 
(b) Ratio between the central dispersion within \re/8 and \se, which is related to the steepens of the mass profile \citep[from][]{Cappellari2013p20}; 
(c) Average logarithmic slope $\langle\gamma\rangle_r=\Delta\log\rho_{\rm tot}/\Delta\log r$ in the interval 0.1--1\re, measured from models B of \citet{Cappellari2013p15};
(d) H$\beta$ line strength index within \re/2;
(e) Mg{\it b} line strength index within \re/2 from \citet{McDermid2015};
(f) $g-i$ color from SDSS \citep[from][]{Aihara2011};
(g) Ratio of the molecular hydrogen H$_2$ from \citet{Young2011} and $\mdyn\equiv\mjam$ from \citet{Cappellari2013p15};
(h) Total (luminous plus dark) dynamical $(M/L)_{\rm dyn}\equiv(M/L)_{\rm JAM}$ from \citet{Cappellari2013p15};
(i) Ratio $\alpha\equiv\mldyn/\mlpop$ between the stellar $M/L$ inferred from dynamical modeling, with the dark matter removed, and the same quantity inferred from full-spectrum fitting using stellar population models, assuming as a reference a \citet{Salpeter1955} IMF \citep[from][]{Cappellari2013p20};
(j) Means SSP age (k) metallicity [Z/H] and (l) elemental abundance [$\alpha$/Fe] within \re/2 \citep[from][]{McDermid2015}; Panels b, d, f--i were presented in \citet{Cappellari2013p20}, and panels j--l in \citet{McDermid2015}.
}
\label{fig:mass_size_all}
\end{figure}

\begin{figure}
\centering
\includegraphics[width=0.9\columnwidth, trim={2cm 4cm 0 0}, clip]{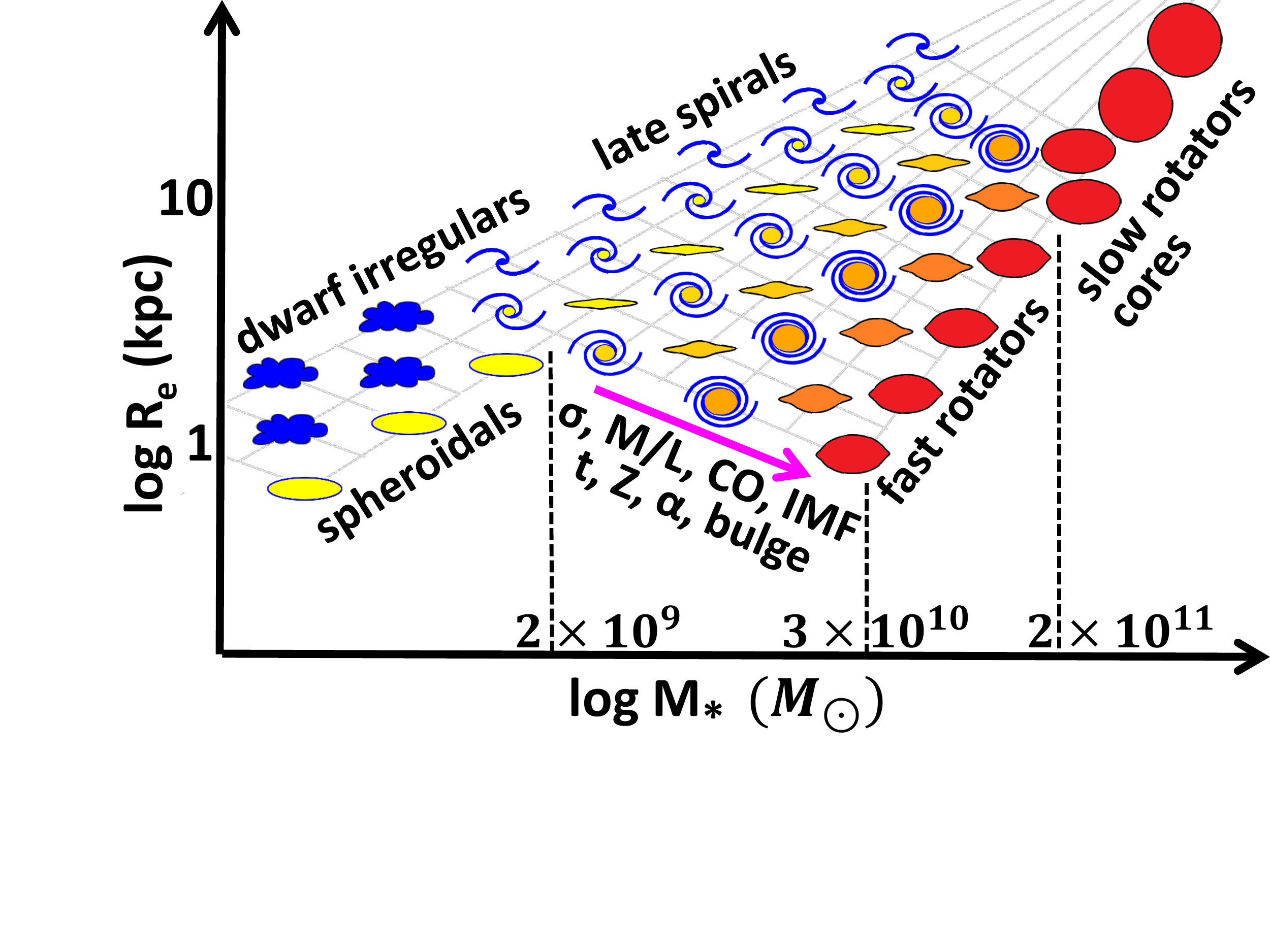}
\caption{{\bf Galaxy properties on the mass-size diagram}.  All ETGs properties related to the stellar population or gas content tend to vary along lines of nearly constant \se, which traces the bulge mass fraction, or the steepness of the mass density profile. This sequence of ETGs properties merges smoothly with the one of spiral galaxies, with little overlap between late spirals (Sc-Irr) and ETGs, a significant overlap between early spirals (Sa-Sb) and fast-rotator ETGs with low $M/L$ and no overlaps between spirals and fast-rotators with high $M/L$. Three characteristic masses are emphasized in this diagram: (i) below $M_\star\approx2\times10^9$ \msun\ there are no regular ETGs and the mass-size lower boundary is increasing; (ii) $M_\star\approx3\times10^{10}$ \msun\ is the mass at which ETGs reach their minimum size (or maximum stellar density), before a sudden change in slope $\re\propto M^{0.75}$ at larger masses; (iii) Below $M_\star\approx2\times10^{11}$ \msun\ ETGs are dominated by flat fast rotators, showing evidence for disks. Above this mass there are no spirals and the population is dominated by round or weakly triaxial slow rotators with flat (core/deficit) central surface brightness profiles. \citep[adapted from][]{Cappellari2013p20}}
\label{fig:mass_size_cartoon}
\end{figure}

In \autoref{fig:mass_plane} we show two projections of the MP. As expected, from the thinness of the plane, both projections provide essentially the same information, simply transformed into a different coordinate system. The key novelties of this plot, made possible by the use of IFS data, are (i) that the stellar $\sigma$ are not central values, but are integrated within an ellipse enclosing half of the total galaxy light, and thus more closely related to the galaxy mass appearing in the virial relation; (ii) the x-axis is not the commonly used luminosity or the stellar mass determined from stellar population models, but a dynamically-determined stellar mass $M_{\rm JAM}\approx \mdyn$ (\autoref{eq:m_jam_definition}), which includes possible variations in the stellar initial mass function (IMF). 
The accurate masses and $\sigma$ allowed one to infer the following results \citep{Cappellari2013p20}: 
\begin{marginnote}[120pt]
\entry{ZOE}{Zone of exclusion}
\entry{$M_{\rm crit}$}{Critical mass of about $2\times10^{11}$ \msun, above which passive slow rotators with cores dominate}
\end{marginnote}
\begin{enumerate}
    \item The distribution of galaxy properties on both projections is characterized by the same zone of exclusion (ZOE), which can be accurately converted from one projection to the other with the virial relation; The ZOE can be roughly approximated by a double power-law, with a break, or a minimum radius and maximum density, at a characteristic mass $M_b\approx3\times10^{10}$ \msun:
\begin{equation}\label{eq:zoe}
\rmaj=R_b \left(\frac{\mdyn}{M_b}\right)^\gamma
\left[
\frac{1}{2} + \frac{1}{2} \left(\frac{\mdyn}{M_b}\right)^\alpha
\right]^{(\beta-\gamma)/\alpha}
\end{equation}
with $R_b=0.7$ kpc, $\alpha=8$, $\beta=0.75$, $\gamma=-0.20$. The relation has an asymptotic trend $\rmaj\propto (\mdyn)^{0.75}$ above $M_b$, and a sharp transition into $\rmaj\propto (\mdyn)^{-0.20}$ below this break.    
    \item The ZOE produces a clear bend in both the {\em mean} $M-\sigma$ and $M-\re$ relations, with trends $\mdyn\propto\sigma_{\rm e}^{2.3}$  [correspondingly $\rmaj\propto(\mdyn)^{0.12}$] at small masses and $\mdyn\propto\sigma_{\rm e}^{4.7}$  [correspondingly $\rmaj\propto(\mdyn)^{0.57}$] at large masses.
    \item A second characteristic mass, at $M_{\rm crit}\approx2\times10^{11}$ \msun, separates the axisymmetric fast rotators with disks, and the spiral galaxies, at lower masses, from the rounder slow rotators with inner cores in their stellar surface brightness, the robustly-determined dry merger relics (\autoref{sec:lam_eps_cores}), at larger masses (\autoref{fig:coma_mass_size}). The near absence of spiral galaxies above $M_{\rm crit}$ produces a sudden drop in the overall specific star formation of galaxies \citep{Brinchmann2004,Salim2007}. This steep decrease in the fraction of star forming systems is well approximated by a simple model in which the quenching rate is proportional to a galaxy star formation rate \citep{Peng2010}.
    \item Below $M_{\rm crit}$, galaxy properties closely follow lines of constant velocity dispersion (dashed in \autoref{fig:mass_size_all}). These lines trace (b) equal mass concentration or (c) equal steepness of the dynamically-determined mass density profile, or bulge mass fraction. Remarkably, this trend is true for nearly every galaxy parameter related to their stellar population, like (d) H$\beta$ and (e) Mg{\em b} line strength, (f) optical color, (g) molecular gas fraction, (h) \mldyn, (i) IMF mass normalization, (j) age, (k) metallicity and (l) $\alpha$-elements enhancement. {\em This show that galaxy formation model need to reproduce this crucial dependence on the central mass density if they want to describe reality}.
    \item The sequence of ETGs properties, as a function of bulge fraction, merges smoothly with the sequence of spiral galaxies, which lie above the ETGs on the $(M,\re)$ diagram (\autoref{fig:coma_mass_size}) \citep{Cappellari2011a,Cappellari2013p20,Kormendy2012}. Late spiral galaxies (Sc) lie near the top of the diagram at any given mass and do not overlap with the ETGs, while early spirals (Sa), which have smaller \re, largely overlap with fast rotators with small $\sigma_e$. The densest and most spheroidal dominated fast rotators (including a number of disky E, \citealt{Kormendy1996}) occupy the bottom of the diagram and do not overlap with the spiral galaxies.
    \item At stellar masses below the \atl\ selection limits, the sequence of spiral galaxies and bulge-less fast rotators, smoothly continues with a sequence of dwarf spheroidals (see fig.~20 of \citealt{Kormendy2012} or fig.~9 of \citealt{Cappellari2013p20}). Interestingly, the stellar mass $M_\star\approx2\times10^9$ \msun\ where there is a sharp bend in the M-size relation of dwarf galaxies and the sequence of increasing bulge fraction starts, corresponds to the threshold for quenching of field galaxies discovered by \citet{Geha2012}.  They found that below that mass only the cluster or group environment can strip galaxies of their gas. Bulges growth cannot happen below that mass: star formation cannot be quenched by internal processes, but only by environmental ones.
\end{enumerate}
All these empirical observations are graphically summarized in \autoref{fig:mass_size_cartoon}. These results were interpreted by \citet{Cappellari2013p20} as due to the build-up of ETGs happening via two separate routes:  (a) in situ star formation: growth via gas accretion or minor gas rich mergers, which sinks towards the center and builds a bulge, which is associated with the quenching of star formation and disk fading \citep[e.g.][]{Cheung2012,Fang2013}. This moves galaxies from left to right while crossing lines of constant \se. (b) external accretion: dry merging, increasing \re\ by moving galaxies along lines of roughly constant \se\ (or steeper), while leaving the population unchanged \citep[e.g.][]{Bezanson2009,Naab2009,Hopkins2010}.

\begin{textbox}
\section{SCALING RELATIONS FOR BULGES AND DISKS?}
In the scaling relations presented in this paper (\autoref{fig:mass_plane}--\autoref{fig:mass_size_cartoon}) we treat galaxies as single entities and do not try to show separate scaling relations for bulges and disks. We prefer our global approach as it leads to more reproducible quantities. But this should not be interpreted as implying that the bulge/disk distinction is not physically meaningful. Instead, the popular photometric bulge/disk decompositions is a complementary way to approach the problems we study, which we did not have space to cover. These works reach conclusions that are remarkably consistent with the overall picture we infer from IFS observations, strengthening both conclusions \citep[see][for reviews]{Kormendy2012,Kormendy2016}. In the near future we envision an approach where photometric and IFS kinematic information are combined to bring bulge/disk decompositions to the next level. This will (i) reduce the inclination degeneracies that affect photometry alone and (ii) allow to quantify the kinematically-determined bulge/disk contributions.
\end{textbox}

\subsubsection{Connection to previous results}

The observed break in the $M-\sigma$ relation is consistent with previous reports of a change in the slope of the $L-\sigma$ relation of elliptical galaxies at low luminosities, where the relation was reported to become $L\propto\sigma^2$. An initial ``marginally significant difference'' in slope was noted by \citet{Davies1983}. But it took a couple of decades for statistically significant results \citep{Matkovic2005,deRijcke2005,Lauer2007,Forbes2008,Tortora2009}. However there is still some debate in the literature about the location and interpretation of the break. 
A break in the $\mstar-\sigma$ at $\mstar\approx2\times10^{11} \msun$ was reported by large SDSS studies, but no break was observed at lower masses \citep{Hyde2009curv,Bernardi2011mass2e11}. In the review by \citet{Graham2013}, the break in the $L-\sigma$ is also reported at $M_B\approx-20.5$, which corresponds to the same $\mstar\approx2\times10^{11} \msun$. 

Thanks to the quantitative consistency between the $(M,\sigma)$ and the $(M,\re)$ projections, and the link to other observables, \autoref{fig:mass_plane} demonstrates that the break in $L-\sigma$ reported in the earliest studies actually reflects the break in the ZOE around $\mstar\approx3\times10^{10} \msun$, and not around $\mstar\approx2\times10^{11} \msun$, where core slow rotators start dominating. In fact the break in the $\mstar-\sigma$ relation persists even when all slow rotators with core are removed. However, it is also clear from \autoref{fig:coma_mass_size} that core slow rotators approximately follow the relation defined by fast rotators with $\mstar\ga3\times10^{10} \msun$. However, the relation for core slow rotators is slightly steeper, producing a second break at $\mstar\approx2\times10^{11} \msun$ reported by SDSS. A more detailed picture above this interesting mass regime is expected from the MASSIVE IFS survey \citep{Ma2014}, but no kinematics was published yet.

A proper understanding of the break in the $\mstar-\sigma$ relations is important for studies of supermassive BHs scaling relations. The existence of a break implies that the $M_{\rm bulge}-M_{\rm BH}$ \citep{Marconi2003,Haring2004} and $\sigma-M_{\rm BH}$ \citep{Gebhardt2000bh,Ferrarese2000} cannot both be equally good predictors of BH masses \citep{Lauer2007,Graham2012,Scott2013bh}. The break provides insight into the mechanism by which BH grow. This very interesting aspect was beautifully reviewed in \citet{Kormendy2013review} and it will not be further addressed here.

The sequence defined by the most passive (old, high metallicity, heavy IMF, CO poor,...) and bulge-dominated (large \se) galaxies in the $(M,\re)$ diagram is another view of the \citet{Kormendy1977} relation between galaxy luminosity and effective surface brightness. However here the relation uses mass in place of luminosity. It is the end point of a continuous sequence, which starts from the bulge-less spirals and ends up at the ZOE. The existence of the ZOE at large stellar densities was noted before \citep{Bender1992,Burstein1997}. Importantly, the power-law $M-\re$ relation defined by the most passive ETGs (red color in \autoref{fig:mass_size_all}) stops around the break in the ZOE at the characteristic mass $\mstar\approx3\times10^{10}$ \msun. This mass was recognized by \citet{Kauffmann2003sfh} as the divide between ``two distinct families'' of galaxies: star forming and disk-like below this mass, while passive and spheroidal-like above. \autoref{fig:mass_size_all} and \autoref{fig:mass_size_cartoon} confirm and explain this result, but they also illustrate that mass, unlike \se, is actually {\em not} a good predictor of galaxy properties \citep[also see][]{Cappellari2006,Franx2008,Graves2009b}.

Unlike the bend in the $L-\sigma$ relation, the corresponding one in the more easily observable $L-\re$ relation has been known for a long time.
It was noted by \citet[their fig.~7]{Binggeli1984} and further illustrated by a number of authors \citep[e.g.][]{Kormendy1985,Graham2003,Kormendy2009,Misgeld2011}. \citet{Kormendy1985} and \citet{Kormendy2009} see dwarf spheroidal as distinct from E galaxies, but rather constituting a separate family, of gas-stripped dwarf spirals/irregulars, while \citet{Graham2003} and \citet{Graham2008curv} interpret the curvature in the $L-\re$ relation as a due to a smooth variation of the \citet{Sersic1968} index with luminosity \citep[e.g.][]{Young1994,Graham2003} in an otherwise homogeneous class of E galaxies. \citealt{Kormendy2012} and \citealt{Graham2013} illustrate two different views on this subject. The \atl\ results agree with the former interpretation, with the key difference that the class of fast rotator ETGs is now bridging the previous apparent gap between genuine E and dwarf spheroidals \citep{Cappellari2013p20}.

The important link between bulge growth and quenching has also been observed using photometric data alone, both for local galaxies \citep{Fang2013} and as a function of redshift \citep{Bell2012,Cheung2012}. The advantage of using dynamically-determined masses and density slopes is that it allows one to exclude population gradients (e.g. disk fading) as the only driver of the observed trend \citep[e.g.][]{Carollo2013}.

\subsection{Parallel sequencing of ETGs and spirals}

\begin{figure}
\centering
\includegraphics[width=\columnwidth, trim={1cm 9cm 1cm 0}, clip]{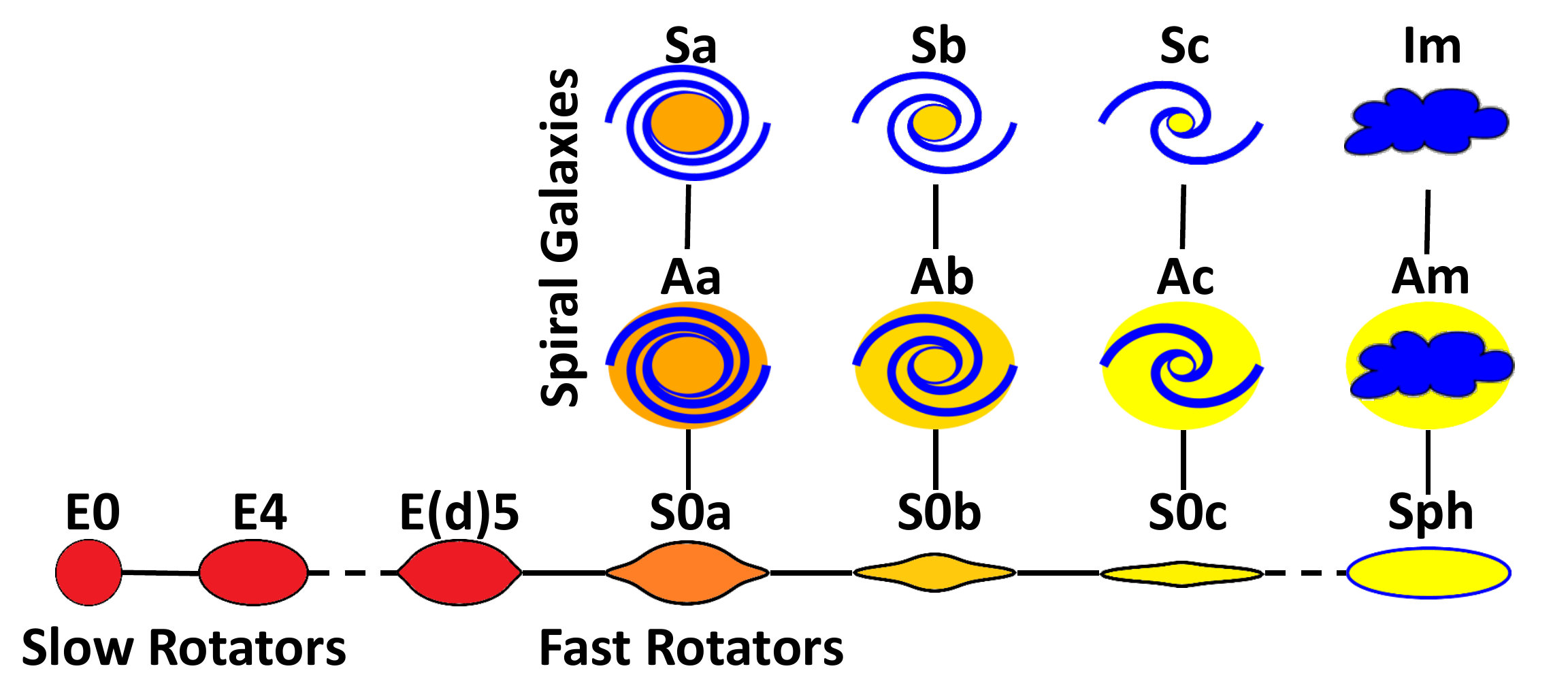}
\caption{{\bf The ``comb'' morphology diagram.} If all fast rotators could be seen edge-on, they would appear morphologically with a range of spheroid fractions ranging from thin S0s, to flat ellipticals with disky isophotes E(d) \citep{Kormendy1996}, as here illustrated. The fast rotators ETGs form a parallel sequence to spiral galaxies as already emphasized, for the subset of S0 galaxies, by \citet{vandenBergh1976}, who proposed the above distinction into S0a--S0c. Fast rotators are intrinsically flatter than $\varepsilon\ga0.4$ and span the same full range of shapes as spiral galaxies, including very thin disks. However very few Sa have spheroids as large as those of E(d) galaxies, indicating that bulges must grow in the transformation. The slow rotators are always rounder than $\varepsilon\la0.4$ and have central cores in their surface brightness. On the right-hand side of the diagram we included spheroidal Sph galaxies, following \citet{Kormendy2012}. These are bulge-less dwarf galaxies, but are significantly rounder than S0c disks. The  black solid lines connecting the galaxy images indicate an empirical continuity, while the dashed one emphasizes the dichotomy between the fast and slow rotator ETGs. (Adapted from \citealt{Cappellari2011b}, with the addition of Sph galaxies from \citealt{Kormendy2012}).
}
\label{fig:comb_diagram}
\end{figure}

The fast rotators are morphologically classified either E or S0 galaxies. All ETGs with photometric evidence for disks, including both the S0s and all galaxies classified as disky ellipticals E(d) by \citet{Bender1994}, belong to the fast rotator, or counter-rotating disks class. The reverse is not true, because disks cannot be seen in the photometry at low inclination. However the dynamical models show that all apparently round and non disky fast rotators are inclined disks, and are intrinsically still quite flat \citep{Cappellari2013p20}. This is also confirmed by statistical inversion of the observed shape distribution of the fast rotator class \citep{Weijmans2014} (\autoref{sec:intrinsic_shapes}). This indicates that {\em all} fast rotators have disks and would be classified as either S0 or disky elliptical E(d) {\em if} they could be observed edge-on. The fast rotators, like S0 galaxies \citep{Laurikainen2010,Kormendy2012}, span the full range of bulge fraction as spiral galaxies \citep{Krajnovic2013p17}, and include some extremely flat disks. They form a parallel sequence to spiral galaxies on scaling relations (\autoref{fig:mass_size_cartoon}).

The slow rotators (or equivalently the non-regular rotators) are clearly different. They are never intrinsically flatter than $\varepsilon\approx0.4$ (\autoref{sec:lam_eps_morph}). As a class, they are inconsistent with axisymmetry, as evidenced by kinematic twists (\autoref{sec:intrinsic_shapes}). They have cores or light deficits in their central surface brightness (\autoref{sec:lam_eps_cores}), and their velocity fields are qualitatively very different from simple JAM models.

All these arguments led \citet{Cappellari2011b} to propose a revision of \citet{Hubble1936} classic tuning fork diagram. The new \atl\ ``comb'' diagram combines elements from two previously proposed revisions to Hubble's diagram with the new findings from IFS observations: (i) the parallelism between spiral galaxies and S0s in the revised classification diagram by \citet{vandenBergh1976}, (ii) the link between S0s and disky ellipticals in the diagram proposed by \citet{Kormendy1996} and (iii) the distinction between slow and fast rotators, and the link of the latter with S0s, indicated by the IFS observations. 

In the revised diagram (\autoref{fig:comb_diagram}) the ETGs are moved from the handle of the tuning fork to a parallel sequence to spiral galaxies. Contrary to previous diagrams, the proposed one is not a symmetric fork or trident, but rather an asymmetric ``comb''. This is to emphasize the fact that many of the galaxy properties (e.g.\ stellar age and gas content) are shared by the two fast/slow families of ETGs, and mostly vary along the spiral sequence. The latter includes \citet{vandenBergh1976} ``anemic'' spirals, to point out that the distinction between ETGs and spirals is not well defined and includes transition objects of uncertain classification. In some cases faint spiral arms become visible in ETGs with very deep optical \citep{Duc2011,Duc2015} or HI observations \citep{Serra2012}.

The need for revising Hubble's tuning-fork diagram, and the parallelism between S0 and spiral galaxies was also pointed out, based on photometric arguments, by \citet{Kormendy2012}. The main conceptual difference is that the \atl\ diagram emphasizes the parallelism to spirals of the entire class of kinematically-classified fast rotators, not to S0 alone. Moreover, \citet{Kormendy2012} extended the S0 sequence to include bulge-less spheroidal Sph galaxies. \citet{Cappellari2013p20} reached a similar conclusion regarding Sph. We now also include Sph in \autoref{fig:comb_diagram}. The similarity of the analysis and conclusions in the two independent studies is a demonstration of the robustness and relevance of the proposed parallelism.


\section{ENVIRONMENTAL TRENDS}

Galaxy environmental trends represent a vast subject with a long history. Here we only briefly mention some of the earliest works and again focus specifically on what integral-field spectroscopy specifically adds to this topic. For more general reviews, the reader is referred to \citet{Blanton2009} and \citet{Kormendy2012} for normal galaxies and to \citet{Boselli2014}, for  emphasis on the faint end of the galaxy luminosity. 

\subsection{Kinematic morphology-density relation}
\label{sec:kinematic_morphology_density}

\begin{figure}
\centering
\begin{minipage}[b]{0.44\textwidth}
\includegraphics[width=\textwidth]{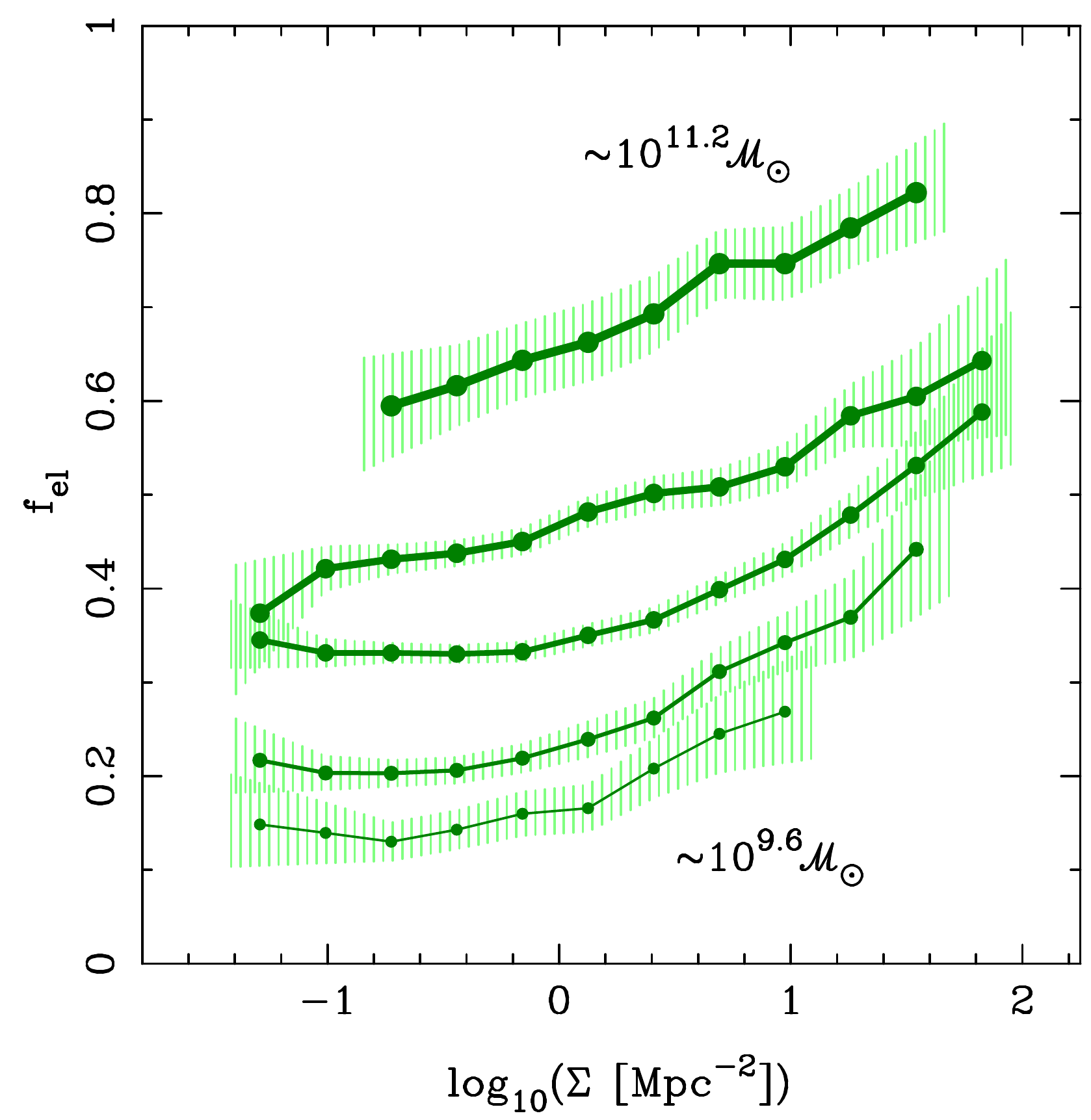}
\end{minipage}
\begin{minipage}[b]{0.55\textwidth}
\includegraphics[width=\textwidth]{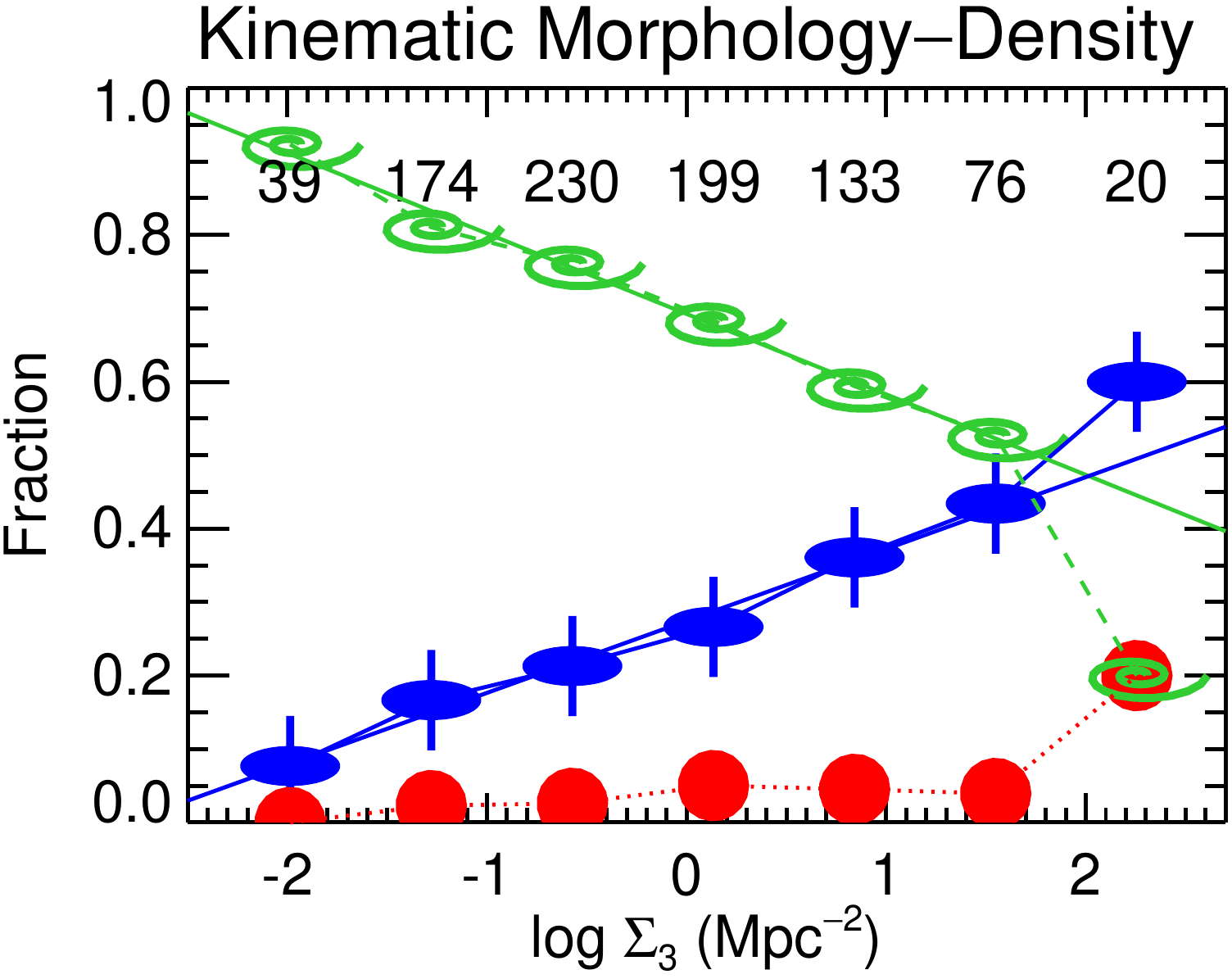}
\end{minipage}
\caption{{\bf Photometric versus kinematic morphology-density.}
{\em Left panel:}  The $T-\Sigma$ relation: the early-type fraction is plotted against local galaxy density for galaxies in selected narrow bins of stellar mass \citep[from][]{Bamford2009}. {\em Right panel:} the $kT-\Sigma$ relation: for fast rotators (blue ellipse with vertical axis), slow rotators (red filled circle) and spiral galaxies (green spiral). The numbers above the symbols indicate the total number of galaxies in each density bin. This panel was adapted from \citet{Cappellari2011b}, by including the $\varepsilon>0.4$ criterion for slow rotators (\autoref{eq:fast_slow_divide}). A key difference between the environmental trends using morphological or kinematic classifications, is that in the former case, even at the lowest densities there is always a fraction $f(E)\ga10\%$ of (misclassified) elliptical galaxies, however in the latter case, the slow rotators are virtually absent in the lowest density environments $f(SR)\la2\%$.
}
\label{fig:morpho_dens}
\end{figure}

Large optical surveys of galaxy clusters in the 70', discovered that that galaxy morphology significantly depends on environment \citep{Oemler1974,Davis1976,Melnick1977}. The classic work by \citet{Dressler1980} revealed a nearly universal morphology-density $T-\Sigma$ relation, in which the fraction of spiral galaxies systematically decreases with increasing projected galaxy number density.  This general trend was subsequently confirmed with samples of ever increasing size, reaching a peak with the Galaxy Zoo sample of $10^5$ galaxies \citep{Bamford2009,Skibba2009}. The trend was found to hold over extended density ranges \citep{Postman1984,Giovanelli1986}.

The $T-\Sigma$ relation is partly driven by the fact that galaxies become more massive in dense environments \citep{Kauffmann2004}, and galaxy mass drives galaxy properties. However the $T-\Sigma$ relation is also clearly detectable at fixed mass (\citealt{Bamford2009,Smith2012coma}; left panel of \autoref{fig:morpho_dens}) and this shows that purely environmental effects must play a role. In fact \citet{Peng2010} convincingly argued that the effect of mass and environment are separable and act independently in transforming galaxies and this naturally explain the existence of the \citet{Schechter1976} luminosity function.

Studies of the $T-\Sigma$ later extended to larger redshift, thanks to the resolving power of the Hubble Space Telescope, which allowed for morphological classification to performed initially to $z\la0.5$ \citep[e.g.][]{Dressler1997,Fasano2000,Treu2003,Wilman2009} and later out to $z\la1$ \citep[e.g.][]{Stanford1998,vanDokkum2000,Smith2005,Postman2005,Cooper2006,Capak2007,Poggianti2008}. 
These studies have found that the fraction of S0 galaxies in clusters decreases with redshift, while the fraction of spiral galaxies correspondingly increases. The trend is much reduced in the field \citep{Dressler1997,Fasano2000,Postman2005,Smith2005}. These results indicate that spiral galaxies become passive and transform into S0 due to the cluster environmental effects \citep{Boselli2006}. The morphological evolution is less significant at the largest galaxy masses, which indicates that these massive ellipticals were already passive from larger redshift and remain so during their evolution \citep{Stanford1998,Postman2005,Tasca2009}, while most of the observed evolution consists of a transformation of spirals into S0s \citep{Smith2005,Moran2007}.

\begin{marginnote}[120pt]
\entry{$T-\Sigma$}{Relation between the morphology and environmental density}
\entry{$kT-\Sigma$}{Relation between the kinematic classification and environmental density}
\end{marginnote}

\begin{figure}
\centering
\begin{minipage}[b]{0.35\textwidth}
\includegraphics[width=\textwidth]{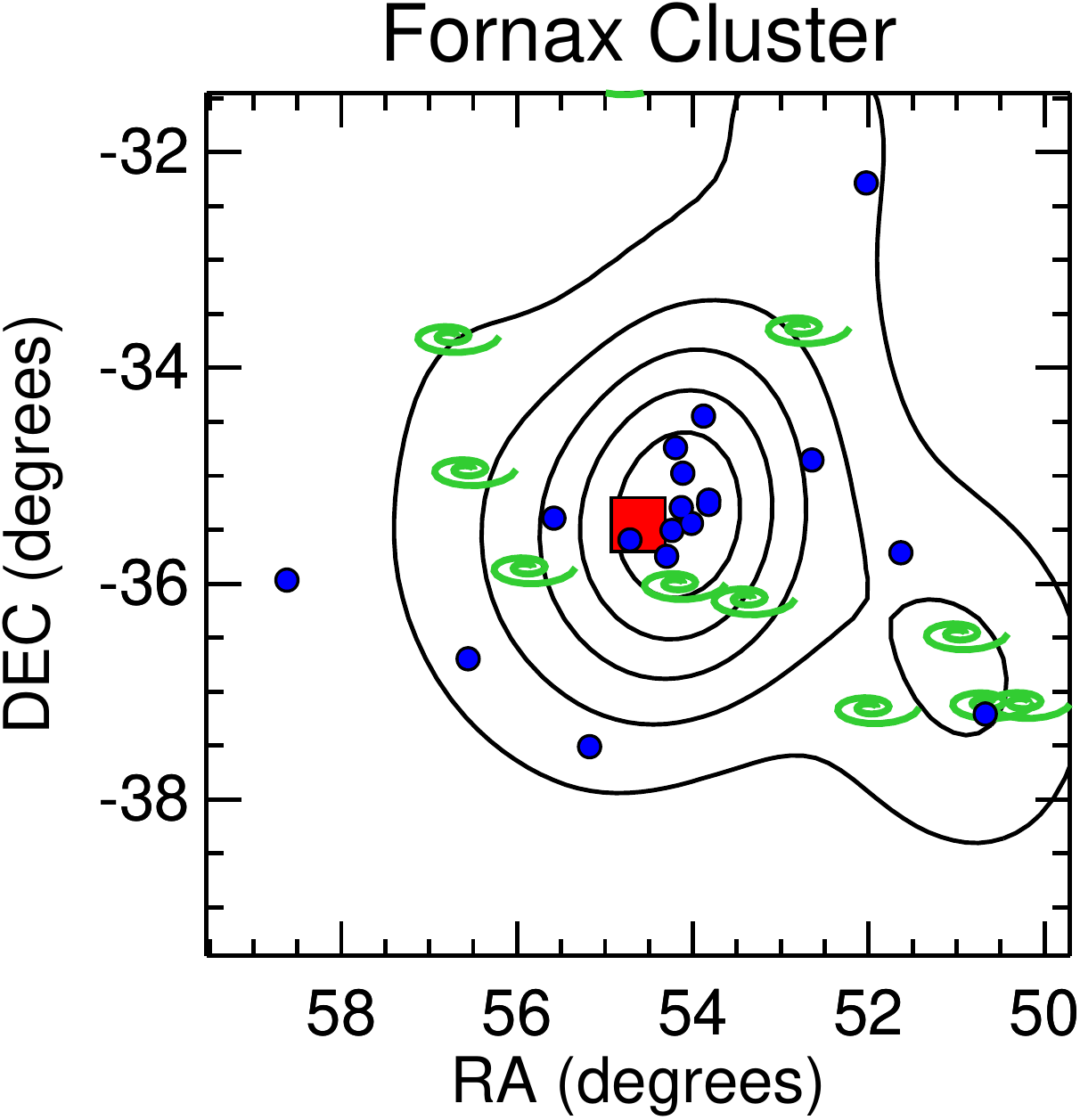}
\end{minipage}
\begin{minipage}[b]{0.31\textwidth}
\includegraphics[width=\textwidth, trim={1cm 0 0 0}, clip]{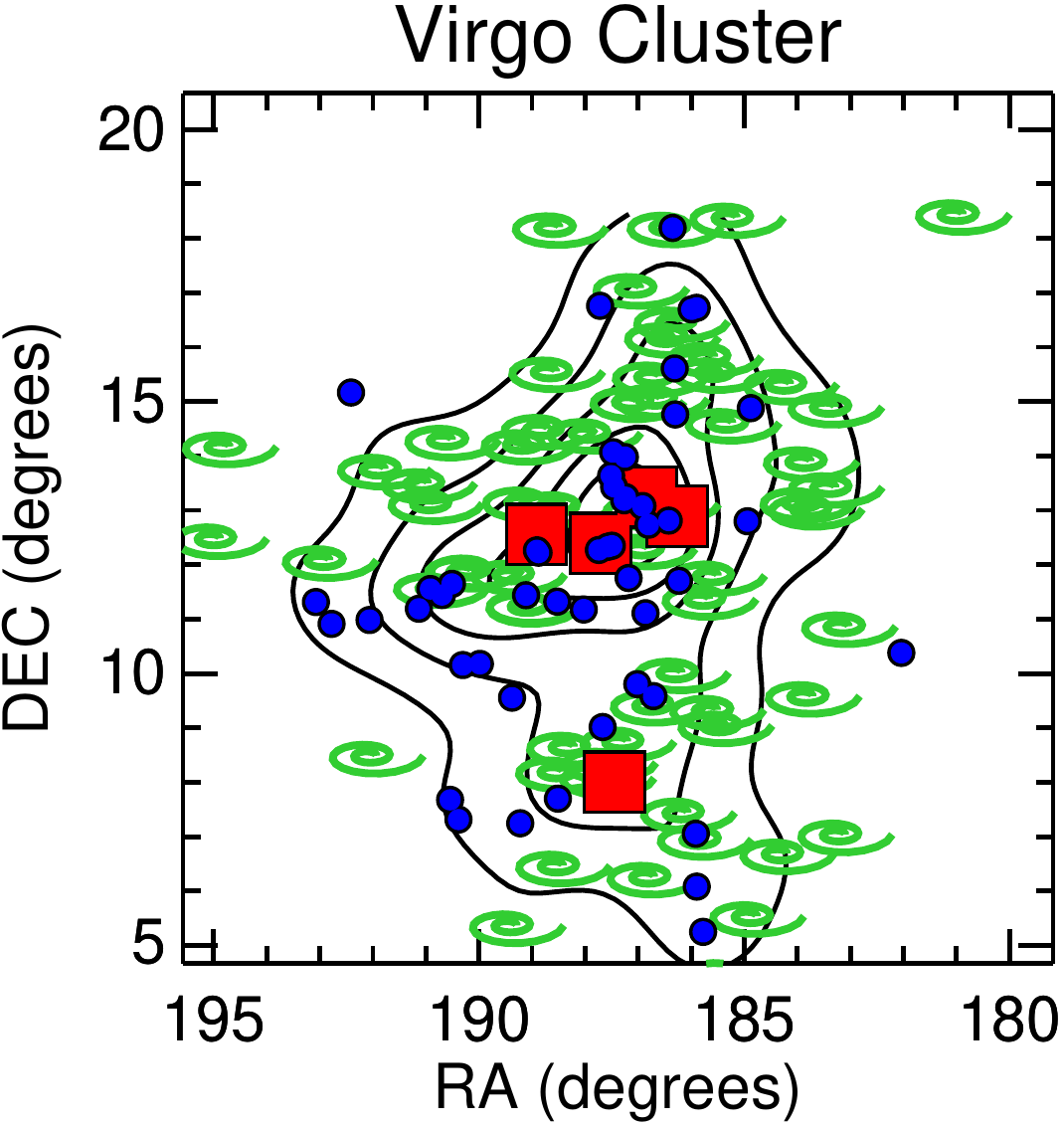}
\end{minipage}
\begin{minipage}[b]{0.325\textwidth}
\includegraphics[width=\textwidth, trim={1cm 0 0 0}, clip]{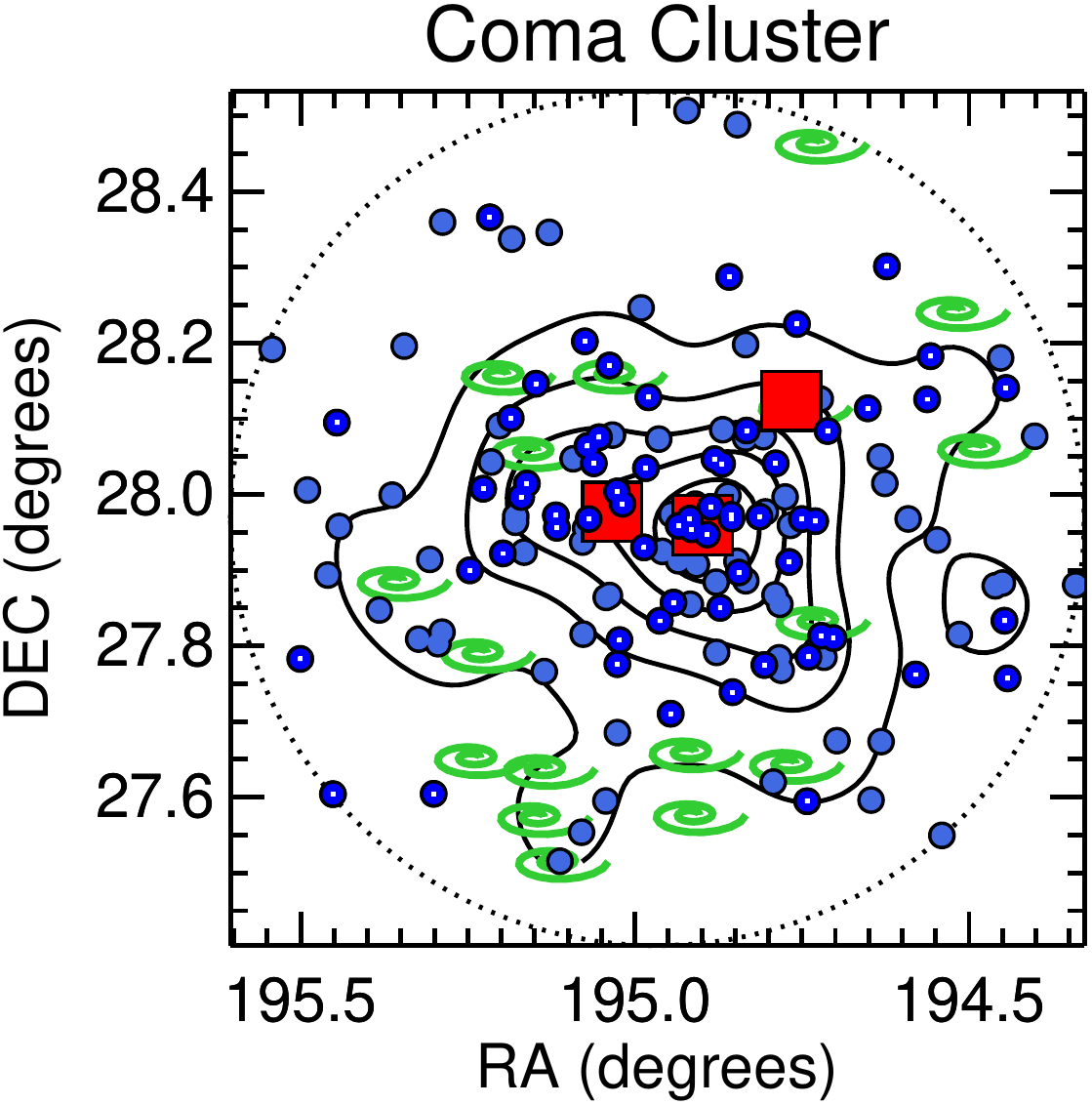}
\end{minipage}
\caption{{\bf Slow rotators in clusters.} {\em Left panel:} distribution of galaxies in the Fornax cluster of galaxies (Abell S0373). The red squares are the slow rotators with core, the blue circles are the rest of the ETGs and the green spirals are the spiral galaxies. A kernel density estimate for the galaxy distribution is overlaid with linearly spaced contours. This panel was adapted from \citet{Scott2014}, with the main difference that the slow rotator NGC~1427 is not plotted in red because it is core-less \citep{lauer07prof}. The only core slow rotator NGC~1399 sits near the peak of the cluster density.
{\em Middle panel:} Same as in the left panel, for the Virgo cluster. Four core slow rotators are found very near the center of the main cluster A and one is M49 at the center the infalling sub-cluster B at the south of the center. 
{\em Right panel:} Same as in the left panel, for the Coma cluster (Abell 1656). The cluster is dominated by two massive core slow rotators near the peak density, while another lies along a slight overdensity. The last two panels are taken from \citet{Cappellari2013apjl}.
The samples in all three clusters are selected to be accurately complete down to the same luminosity $M_{K_s}<-21.5$ as the \atl\ sample.
}
\label{fig:slow_rotators_in_clusters}
\end{figure}

Motivated by the results reviewed in \autoref{sec:kinematics}, \citet{Cappellari2011b} introduced the {\em kinematic} morphology-density relation $kT-\Sigma$, which uses the physically robust fast/slow rotator classification to replace the strongly inclination-dependent S0/E morphological classification. The key differences (\autoref{fig:morpho_dens} right) between the kinematic and morphological relations are the following: 
\begin{enumerate}
    \item The genuinely spheroidal system, the slow rotators, namely the galaxies that would be classified as E from any direction, are essentially absent in the field. Considering the lowest three density bins, out of a statistically significant sample of 443 galaxies, only 2\% are slow rotators. Slow rotators only play a significant role at the largest environmental density, which correspond to the center of cluster and groups. This contrasts with the fraction $f(E)\ga10\%$ of (misclassified) E which is inferred from morphological classification even at the lowest densities (e.g. \citealt{Dressler1980}, \autoref{fig:morpho_dens} left). 
    Given the small number statistics of slow rotators, it is revealing to see their actual distribution within clusters. \autoref{fig:slow_rotators_in_clusters} visualizes the clear tendency for slow rotators to be found either at the cluster center, or within local overdensities, due to infalling sub-cluster being assembled into the main cluster. This tendency has been seen in most clusters for which IFS data exist and where well-defined overdensities are present \citep{Cappellari2011b,Cappellari2013apjl,Houghton2013,DEugenio2013,Scott2014,Fogarty2014}. It explains why the $T-\Sigma$ relation seems to work better than the $T-R$ relation, which uses the cluster-centric radius \citep{Dressler1997}.

\item Focusing only on the ETGs subset, the ratio $f(SR)/f(FR)$ between slow and fast rotators is {\em not} a simple or monotonic function of environment. Instead, the ratio is nearly constant outside clusters, while inside clusters it becomes a strong function of the local galaxy number density \citep{Cappellari2011b,Houghton2013,DEugenio2013,Fogarty2014}. This shows that in the field, slow rotators are produced by random events, at an extremely low rate. While clusters are the natural place for the presence of slow rotators. They must form via a different process, which must be linked to the formation of the clusters themselves.

\end{enumerate}

The study of the $kT-\Sigma$ relation has just recently started. The Virgo \citep{Cappellari2011b} and Fornax \citep{Scott2014} clusters are the only two which have currently been mapped by IFS observations with good completeness. With the currently small and generally incomplete samples, a few misclassified slow rotators (e.g. counter-rotating disks, recent mergers, or uncertain kinematics), can still significantly affect the conclusions. Moreover, ideally one would like to be able to recognize the genuine dry mergers relics: slow rotators with cores. But this is currently only possible with HST, and not much farther than the Coma cluster (100 Mpc).
The situation is going to be revolutionized in the near future by IFS observations, thanks to the two large SAMI \citep{Bryant2015} and MaNGA \citep{Bundy2015} IFS surveys, which are expected to bring the study of the local $kT-\Sigma$ relation at the level of the $T-\Sigma$ one, in terms of number statistics and completeness.

\subsection{Mass-size versus environment}
\label{sec:mass_size_environment}

In recent times, the $M-\re$ relation (\autoref{sec:local_mass_size}) has become one of the most popular observational constraints to galaxy formation models. This is because the variation of galaxy sizes during the hierarchical mass assembly of galaxies, depends sensitively on the detail by which the assembly occurs \citep[e.g.][]{Nipoti2003,Naab2009,Bezanson2009,Hopkins2010}. For example, using simple virial equilibrium arguments \citep{Hausman1978}, one can show that during a gas-free merger of identical spherical galaxies on parabolic orbits, which is the most likely situation \citep{Khochfar2006c}, the radius grows proportionally to the mass and the galaxy stellar velocity dispersion (integrated over the whole galaxy) remains unchanged \citep{White1983,Barnes1992}. However, when the mass growth happens via very small gas-poor satellites, then $R$ is predicted to increase as the square of the mass fraction increase, while the $\sigma$ decreases as the square root of the mass increase \citep{Naab2009,Bezanson2009}. These predictions properly capture the trends actually observed in detailed numerical simulations \citep{Barnes1992,Hernquist1993,Nipoti2003,Boylan-Kolchin2006,Naab2009}. 

In contrast, when gas is present, it dissipates energy, falls towards the center and forms stars in a compact star burst \citep{Mihos1994burst}. Due to the shrinkage caused by the gas infall, the size of the remnant will increase more slowly than in the collisionless case, by an amount which is related to the amount of dissipation \citep[e.g.][]{Dekel2006gas,Khochfar2006}. The galaxy center will be denser and will contain a steep inner profile (cusp or extra light) as observed in many fast rotators \citep{Kormendy2009}.

To test these predictions one would ideally like to be able to follow the redshift evolution of fast rotators and core slow rotators on the $(M,\re)$ plane. This is unfortunately not yet feasible, with the current generation of instruments. However, examining the environmental dependence is a good proxy to studying the redshift evolution. This is because at high redshift the abundance of massive halos declines and fewer galaxies live in clusters.

The environmental evolution of fast and slow rotators on the $(M,\re)$ plane was investigated by \citet{Cappellari2013apjl}, using two extreme environments for which IFS data were available. The low-density sample was taken from the field/group environment of the \atl\ sample, defined as not belonging to the Virgo cluster. The high-density sample used IFS data by \citet{Houghton2013} on the Coma cluster, which is the densest environment for which resolved spectroscopic observations can be be currently taken. It has one of the largest, and carefully determined, dark halo virial masses of $M_{\rm 200}\approx1.4\times10^{15}$ \msun\ \citep{Lokas2003}, that can be expected to be found in the whole Universe \citep[e.g.][]{Springel2005nat}. Both samples were carefully selected to be nearly 100\% complete to $M_{K_s}<-21.5$ ($\mstar\ga6\times10^9$ \msun) and have fully homogeneous size and luminosity determinations from 2MASS \citep{Skrutskie2006}.

The study found that the mass-size distribution in the dense environment differs from the field/group one in two ways: (i) spiral galaxies are replaced by fast-rotator ETGs, which follow the {\em same} mass-size relation and have the {\em same} mass distribution as in the field sample; (ii) the core slow rotator ETGs are segregated in mass from the fast rotators, with their size increasing proportionally to their mass. A transition between the two processes appears around the stellar mass $M_{\rm crit}\approx2\times10^{11}$ \msun. This is illustrated in \autoref{fig:coma_mass_size}, which also includes the distribution of core slow rotators and fast rotators in the Virgo cluster, which has lower density than Coma and a virial halo mass more than $3\times$ smaller of $M_{\rm 200}\approx4\times10^{14}$ \msun\ \citep{McLaughlin1999}. The distribution of galaxies on the Virgo $(M,\re)$ diagram is intermediate between the field/groups and Coma one. The core slow rotators have slightly larger masses than the field/groups ones and 4/5 lie above  $M_{\rm crit}$. 

The environmental dependency of the distribution of spiral galaxies, fast rotators and core slow rotators, was interpreted as a direct evidence for the two channel for the build up of ETGs inferred from the distribution of galaxy properties in \autoref{sec:local_mass_size}: (i) The bulge-growth and quenching route, is illustrated by the fact that spiral galaxies are gradually replaced by fast rotator, with smaller sizes and larger \se. The latter have the same properties as those in the field, but are simply more numerous in the cluster. (ii) The merger-growth route, is illustrated by the mass increase of the core slow rotators, with $\mstar$ increasing proportionally to \re, broadly following the dry-merging model prediction.

The impressive universality of the mass-size relation for $\mstar\la M_{\rm crit}$ in these two extreme environments is excellent agreement with other studies which used larger samples but lack IFS kinematic information. Differences between the sizes of ETGs in different environments are consistently found to be $\la10\%$, which is the level of than possible systematic effects in the size determinations or sample selections \citep{Maltby2010,Huertas-Company2013,Poggianti2013,FernandezLorenzo2013,Cebrian2014}. 

The only significant difference between the low-density and high-density environments was found to be a slight asymmetry of the mass-size distribution. The peak is at the same location in the two environments, but there is a tail towards ETGs with larger sizes in clusters \citep{Cappellari2013apjl,Delaye2014}. In Coma, where galaxies are well resolved, this appear to be due to the well-known excess of passive disks in clusters \citep{vandenBergh1976,Wolf2009,Masters2010}.


\section{REDSHIFT EVOLUTION}

Only the line emission from the gas-rich star-forming galaxies can be currently spatially resolved with IFS at significant redshift \citep[e.g.][]{Wisnioski2015,Stott2016}. This is because at $z\ga1.3$ most key spectral features are red-shifted to the near infrared, where the strong contribution of the OH atmospheric lines dramatically degrades the data quality. IFS observations of the stellar continuum in ETGs are essentially non-existing at these redshifts. For this reason this section will only include an brief overview of some of the key finding which closely relate to what has been discovered with IFS observations of nearby galaxies. A comprehensive review of the redshift evolution of galaxy structure was recently provided by \citet{Conselice2014}, while the evolution of their stellar population was reviewed by \citet{Renzini2006}.

\subsection{Mass-size evolution}

\begin{figure}
\centering
\includegraphics[width=0.8\textwidth]{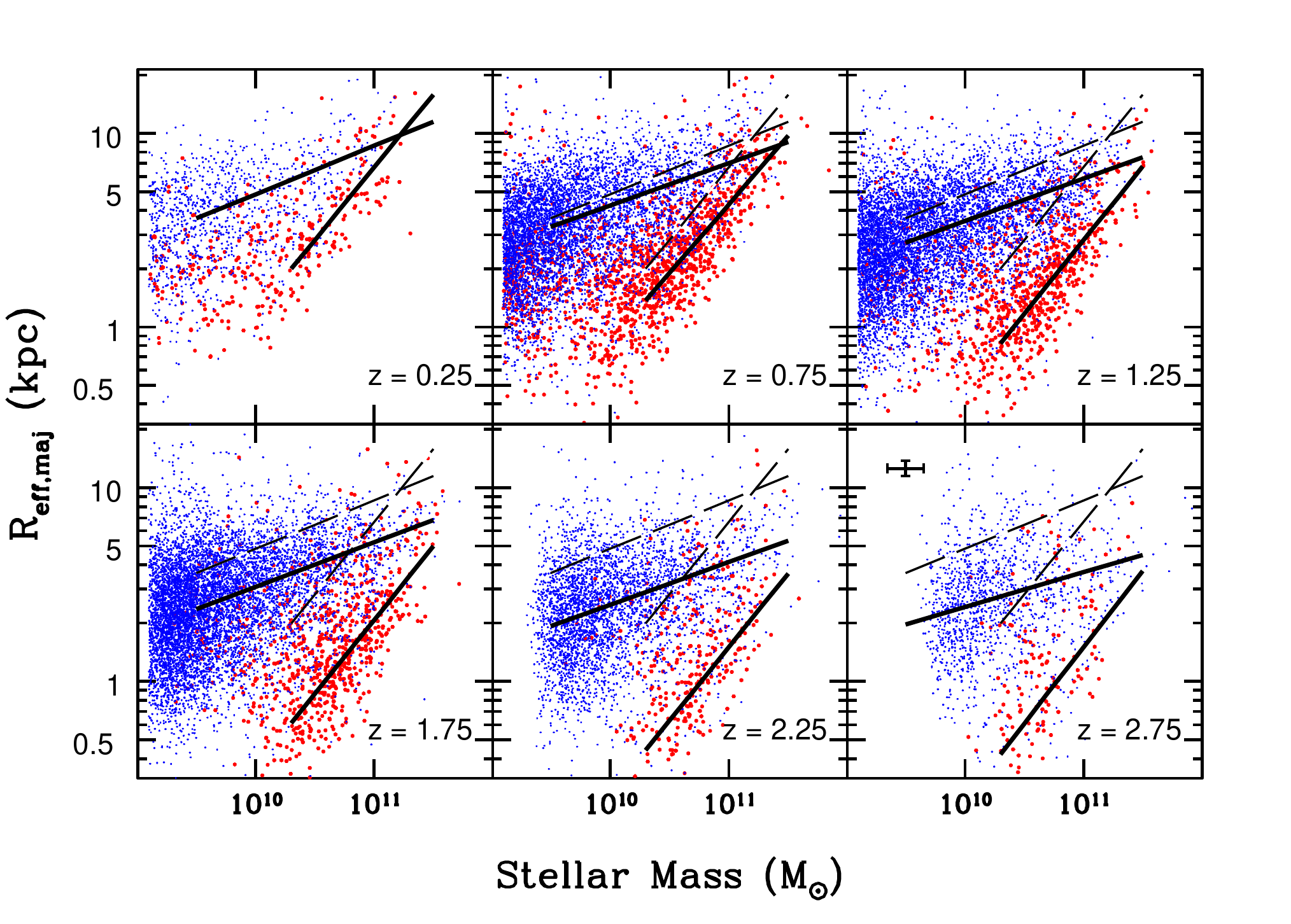}
\caption{{\bf Evolution of the mass-size distribution.} $(\mstar,\re)$ distribution of colors-selected star forming and quiescent galaxies. The lines indicate model fits to the passive and star forming galaxies. The dashed lines, which are identical in each panel, represent the model fits to the galaxies in the lowest redshift bin. The solid lines represent fits to the higher-redshift samples. Strong evolution in the intercept of the size-mass relation is seen for passive galaxies and moderate evolution for the star forming ones. \citep[Taken from][]{vanderWel2014}
}
\label{fig:mass_size_redshift}
\end{figure}

The stellar population of nearby ETGs, the scatter of the color-magnitude relation in clusters and the redshift evolution of their FP clearly indicate that ETGs formed their stars at $z\ga2$ (\citealt{Thomas2005daniel}; see review by \citealt{Renzini2006}). When samples of spectroscopically-confirmed passive galaxies beyond $z\ga1.4$ were finally discovered, they appeared to have large masses $\mstar\ga10^{11}$ \msun, comparable to some of the most massive local ETGs \citep{Cimatti2004,Cimatti2006,Glazebrook2004}. However their sizes were found to be significantly smaller that their local counterparts \citep{Daddi2005,Trujillo2006,Trujillo2006gems,Trujillo2007,Zirm2007,Toft2007,Longhetti2007,Cimatti2008,vanDokkum2008}. 

Initial concerns about the possible effect of unresolved nuclear emission from AGNs, strong stellar population gradients or inaccurate mass estimates, were cleared out via spectroscopic determinations of their velocity dispersion, which showed values broadly consistent (but with non-negligible size bias) with the virial predictions inferred from the given masses and radii \citep{Cappellari2009,Cenarro2009,Onodera2010,vandeSande2011,vandeSande2013,Toft2012,Bezanson2013,Belli2014}. 

It appears now well established that the {\em mean} size of ETGs grows significantly from $z\sim3$ to the present day, while spiral galaxies grow at a smaller rate. The precise amount of this evolution depends sensitively on how ETGs are defined, given the strong dependence of galaxy size on their properties, like color, star formation, Sersic index or morphology (\autoref{fig:mass_size_all}, \citealt{Newman2012,Cappellari2013p20,Poggianti2013}). The largest and most homogeneous study is currently the analysis of the CANDLES survey with extensive multi-band HST data \citep{Grogin2011,Koekemoer2011} by \citet{vanderWel2014}, reproduced in \autoref{fig:mass_size_redshift}. The slope of the $\mstar-\re$ relation for ETGs, for $\mstar\ga3\times10^{10}$, is found to be consistent with the slope $\mstar\propto R_{\rm e}^{0.75}$ of the ZOE (\autoref{eq:zoe}) over the whole redshift range $0\la z\la3$. Also apparent at all redshifts is break in the overall mass-size, due to the minimum size of ETGs around a few $10^{10}$ \msun, observed locally (see \autoref{fig:mass_size_cartoon}).

The evolution of the distribution of galaxies on the $(\mstar,\re)$ plane was interpreted by \citet{vanDokkum2015} using a simple statistical model which tries to capture the dominant mode of growth. They concluded that the population of ETGs progenitors likely followed two main simple evolutionary tracks in the $(\mstar,\re)$ plane: (i) a shallow $\Delta\log\re\sim0.3\Delta\log\mstar$ growth dominated by gas accretion. Along these track the originally gas rich and star forming galaxies become denser, increasing their $\sigma$, until they reach a threshold at which their stars are quenched. (ii) A steeper track dominated by (mainly dry) mergers, where their size increase proportionally to their mass. An independent analysis of the CANDLES data, reaches quite similar conclusions, and also emphasizes the need for two similar formation channels for ETGs \citep{HuertasCompany2015}.

Although the detailed mechanism is still actively debated, from high-redshift observations, a consensus is emerging that galaxy quenching is linked to the growth of the galaxies {\em central} density, quantified using either photometrically-predicted $\sigma$ \citep{Franx2008,Bell2012,vanDokkum2015} or the stellar density $\Sigma_1$ within an aperture of radius $R=1$ kpc \citep{Cheung2012,Barro2016}. Importantly, {\em central} density was found to be a much better predictor of quenching than other galaxy parameters and in particular galaxy mass. These findings are fully consistent with the link between bulge growth and quenching reviewed in \autoref{sec:local_mass_size} from IFS observations of local ETGs.

The two evolutionary channels described by \citet{vanDokkum2015} and \citet{HuertasCompany2015} from direct observations of the $(\mstar,\re)$ redshift evolution and galaxy morphology, are the same that were also proposed to explain galaxy properties on the $(\mstar,\re)$ plane using IFS observations in \autoref{sec:local_mass_size} and from the environmental dependency in \autoref{sec:mass_size_environment}. The remarkable agreement between these two sets of independent observations, based either on the fossil-record in nearby galaxies, or on redshift evolution, strengthens the conclusions of both complementary approaches.

\subsection{Profile evolution}

\begin{figure}
\centering
\includegraphics[width=0.9\columnwidth,trim={0 0 0 8.5cm},clip]{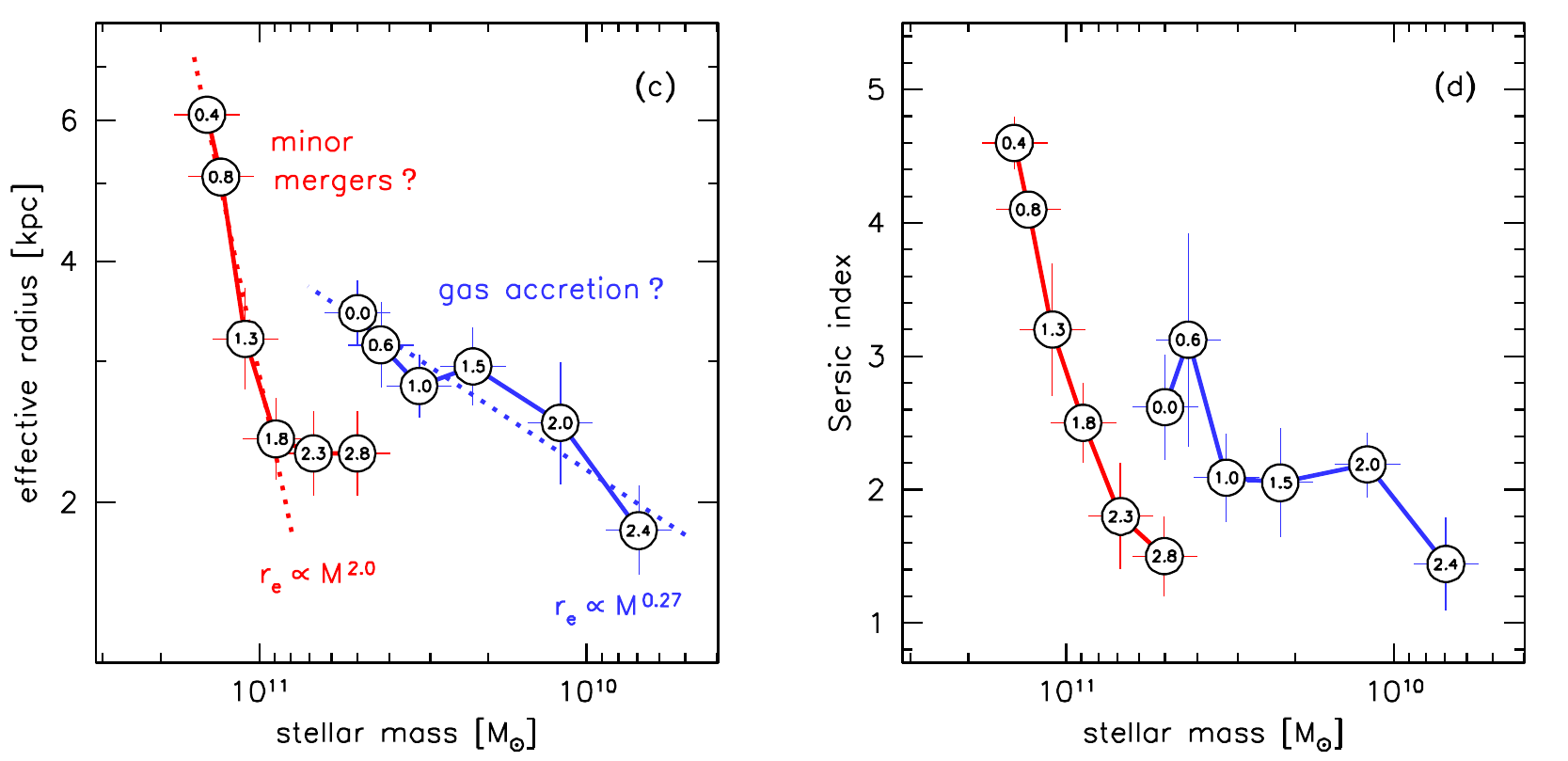}
\caption{{\bf Size and profile evolution.} \re\ and Sersic index as a function of redshift and \mstar. The blue line is for the progenitors of galaxies with present-day masses $\mstar\approx5\times10^{10} \msun$ while the red one is for present-day masses of $\mstar\approx3\times10^{11} \msun$. The less massive galaxies have have undergone much less structural evolution than the present-day giant elliptical galaxies that populate the high-mass end of the mass function. \citep[Taken from][]{vanDokkum2013}
}
\label{fig:profile_evolution}
\end{figure}

Using observations at different redshifts, one can directly trace the variation of the Sersic indices as a function of time. This analysis was performed for the progenitors of ETGs with present-day mass $\mstar\approx3\times10^{11}$ \msun\ by \citet{vanDokkum2010profiles} and followed up with a consistent approach, but for progenitors with masses $\mstar\approx5\times10^{10}$ \msun, by \citet{vanDokkum2013}. The first set of galaxies have present-day masses above the critical mass $M_{\rm crit}\approx2\times10^{11}$ \msun, dominated by the core slow rotators, while the second set have present-day masses near the break $M_b\approx3\times10^{10}$ where spiral galaxies and fast rotators overlap, and slow rotators are absent.

The results, summarized in \autoref{fig:profile_evolution}, indicate that the progenitors of slow rotators grew rapidly in size, while those of fast rotators remained almost unchanged from $z\approx2$. Similarly, the progenitors of slow rotators change little in mass from $z\sim2$, while the the progenitors of fast rotators still grow significantly \citep{Muzzin2013}.  This mass and size growth difference is accompanied by a strikingly different behavior of the Sersic indices: both classes of galaxies start with disk-like $n\sim1$, but the slow rotators progenitors rapidly increase their concentration, and reach $n\ga4$ near $z\sim0$, while the fast rotators progenitors maintain a more constant Sersic index, with $z\sim0$ values around 2--3, which is consistent with the typical value for the fast rotators in the \atl\ sample \citep{Krajnovic2013p17}. 

\section{IMPLICATIONS FOR GALAXY FORMATION}
\label{sec:galaxy_formation}

The aim of this review is to provide an overview of the empirical signatures of galaxy formation in ETGs, as they were obtained mainly by IFS observations. Here we sketch ideas on galaxy formation driven by the observations we described. A detailed review of the theoretical models of the formation of ETGs is given by \citet{Somerville2015} and Naab \& Ostriker (ARA\&A in preparation). Theoretical studies trying to specifically address the formation of the fast and slow rotator ETGs classes revealed by IFS observations were presented by \citet{Bois2011,Khochfar2011,Naab2014}.

\subsection{Galaxy evolution on the $(\mstar,\re)$ plane}

\begin{figure}
\centering
\includegraphics[width=0.6\columnwidth]{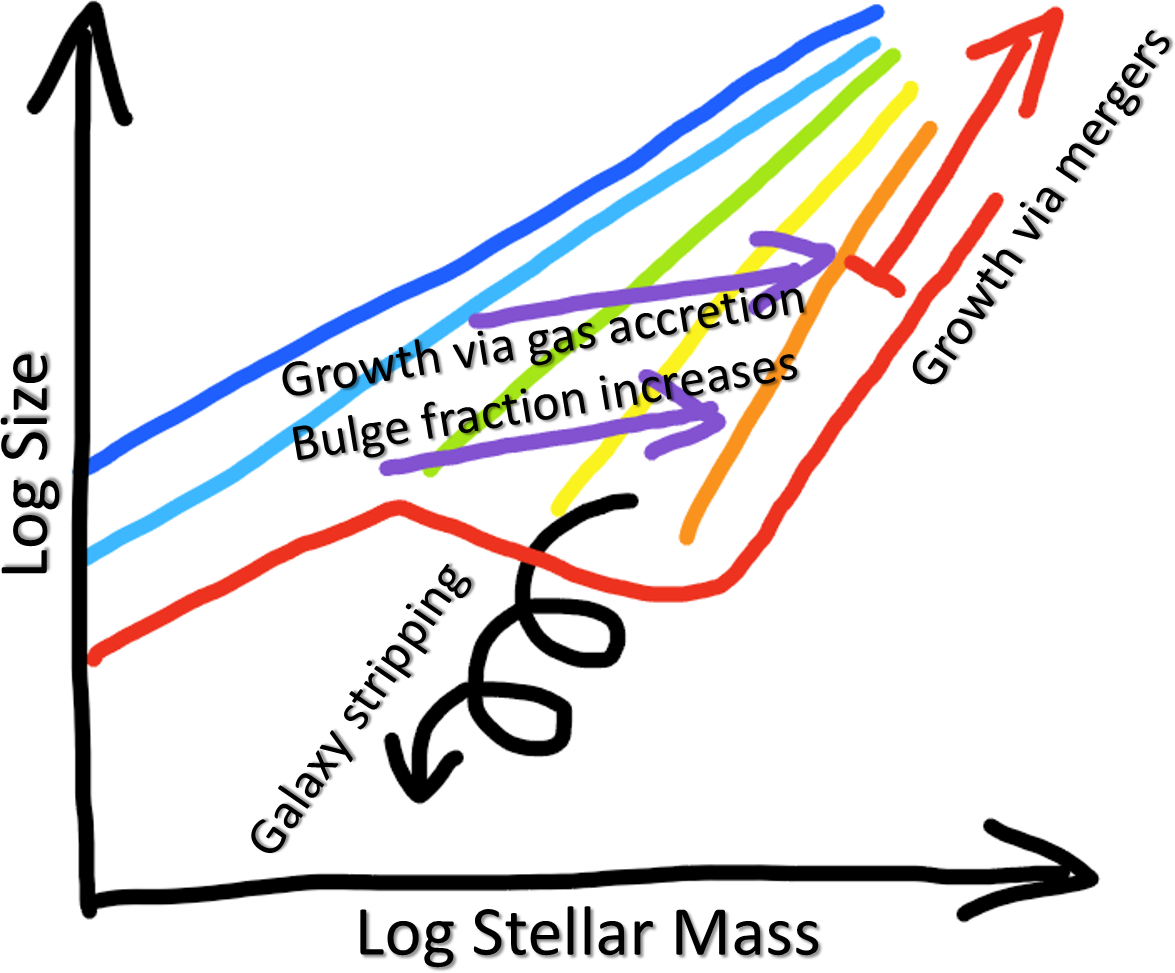}
\caption{{\bf Galaxy evolution on the mass-size plane.} Although the evolution of an individual galaxy is a complex combination of events, the observations indicate that the average evolution of ensembles of galaxies can be described by the following simple picture. The progenitors of fast rotators ETGs are star forming disks. They grow their mass, while slowly increasing their sizes, predominantly by gas accretion (purple lines). During this accretion they grow a bulge, which increases the likelihood for the galaxy to have its star formation quenched. They end up as bulge-dominated and passive fast rotators near the ZOE. Slow rotators build up most of their mass rapidly at high-z and subsequently grow mostly by gas poor (dry) merging, while varying their size nearly proportionally to their mass or more (red line). They end up in the top right of this diagram. The region below the ZOE is not empty: it includes ultra compact dwarfs (UCD), which are the likely cores of normal galaxies which had their envelopes stripped by the gravitational field of larger galaxies. (Based on fig.~15 of \citealt{Cappellari2013p20} and fig.~28 of \citealt{vanDokkum2015})}
\label{fig:mass_size_evolution_cartoon}
\end{figure}

Consistently, from either the IFS results and from the high-redshift ones, one can describe the build up of ETGs on the $(\mstar,\re)$ diagram as in \autoref{fig:mass_size_evolution_cartoon}. The {\em average} growth of the overall ETGs population can be summarized has happening trough two main channels. The first one is dominated by gas accretion. Spiral galaxies accrete gas, or experience gas rich minor mergers. The gas sinks towards the center, increases the galaxy mass and at the same time builds a bulge. The presence of a bulge increases the likelihood for the galaxy to have its star formation quenched. This makes the progenitor spiral end up as a passive fast rotator ETG, with a stellar disk and, on average, a larger bulge fractions. As a result of this first channel, all galaxy properties related to the star formation history, vary on average along lines of nearly constant \se. The fact that, at fixed mass, dense bulges have $\alpha$-enhanced stellar population (\autoref{fig:mass_size_all}) indicates that bulge formation must be a rapid process as expected during intense starbursts. The gas accretion and subsequent star formation leaves the metallicity enhancements of the disks in fast rotators as a fossil record (\autoref{fig:kuntschner2010_fig9}).
    
The second channel on the $(\mstar,\re)$ diagram is dominated by mostly-dry mergers, which move galaxies by increasing their size roughly proportionally to their mass, while leaving \se\ nearly unchanged. During dry mergers the stellar population evolves passively and varies little for old systems. For this reason no significant trend is observed in galaxy properties related to the stellar population along this second channel. However dry mergers leave an imprint in the nuclear profiles, when supermassive black holes of the merging galaxies sink towards the center via dynamical friction and eject stars in radial orbits, scouring a nuclear core or deficit \citep[e.g.][]{Faber1997,Milosavljevic2001,Kormendy2013review}.

Observations indicate that fast and slow rotator predominantly follow either one or the other channel. A key piece of information is provided by the environmental distribution (\autoref{sec:kinematic_morphology_density}). If slow rotators formed by dry merging of fast rotators: (i) they would roughly follow their distribution, like fast rotators follow the spiral distribution. Instead, their distribution in clusters is very different, with the slow rotators near the cluster/group centers and the fast rotators following spiral galaxies (\autoref{fig:slow_rotators_in_clusters}); (ii) moreover one would find core slow rotator along the whole sequence of passive ETGs, starting from $\mstar\ga3\times10^{10}$ \msun, as they are still building-up their mass, especially in the filed. Instead they suddenly appear only above $\mstar\ga10^{11}$ \msun\ (\autoref{fig:coma_mass_size}).

We mention in passing, that the region of the $(\mstar,\re)$ diagram below the ZOE, but for $\mstar\la3\times10^{10}$ \msun, is not actually devoid of stellar systems. It is populated by objects generally called dwarf ellipticals (dE) and ultra compact dwarfs (UCD)  \citep[e.g.][]{Misgeld2011,Norris2014}. Evidence suggests they may be normal ETGs which fell well into the halo of a larger galaxy and had their outer stellar envelope stripped \citep[e.g.][]{Drinkwater2003}. Integral-field observations of the kinematics and population of these galaxies have just started and appears to confirm this interpretation \citep{Rys2014,Guerou2015}.   According to this picture, dE and UCD do not follow the main route of galaxy formation. This interesting topic will not be discussed in any detail here.

\subsection{Hierarchical origins of fast and slow rotators ETGs}

\begin{figure}
\centering
\includegraphics[width=0.9\columnwidth]{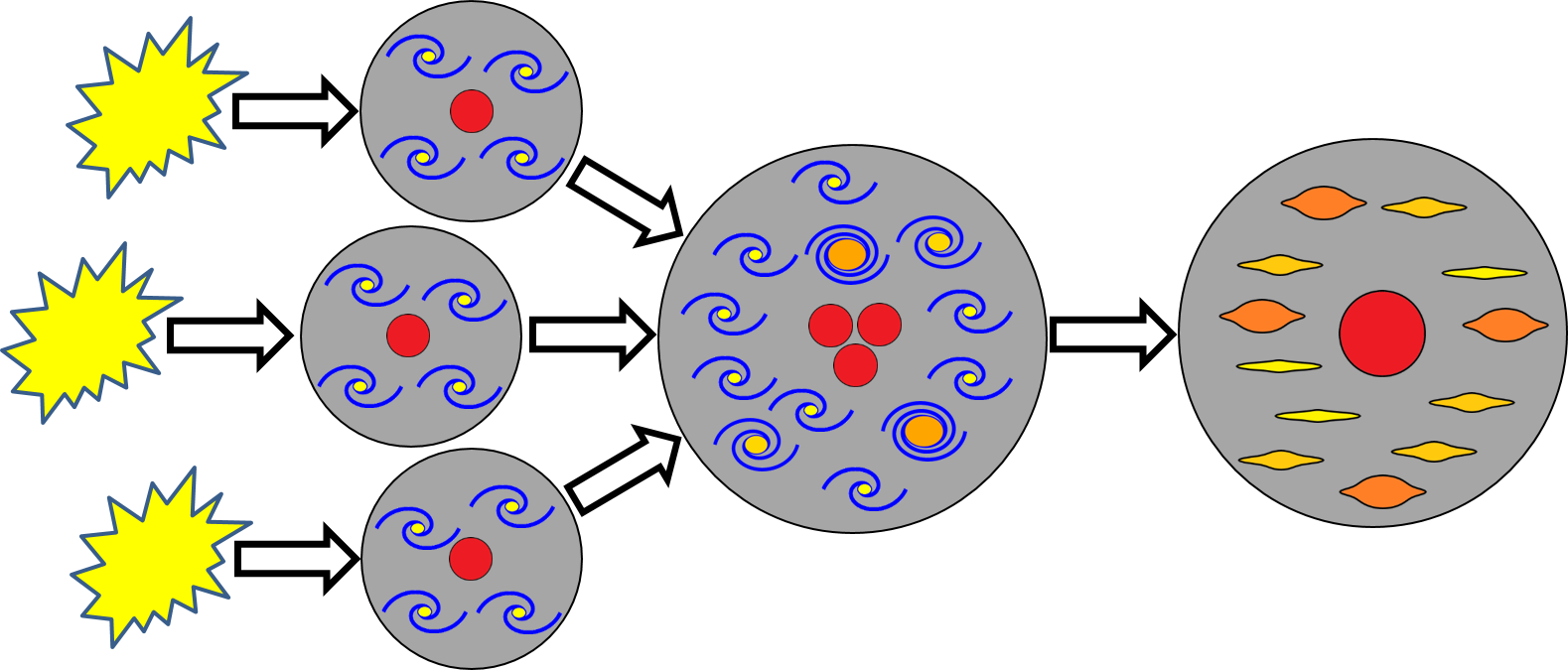}
\caption{{\bf Hierarchical origin of fast and slow rotators.} The progenitors of core slow rotators form in the high-redshift universe, at the center of the largest dark matter overdensities. They quickly grow above the critical mass required to be surrounded by a hot gaseous envelope which prevents further accretion. When groups merge to form massive clusters, slow rotators sink toward the center where they merge to form more massive slow rotators. The same cannot happen to fast rotators, which have masses too small to efficiently sink to the center and velocities too fast to merge. They are quenched by the cluster and from then on stop increasing their mass, but are only affected by tidal effects. Not included here is the internal quenching, which appears associated to the bugle growth and can act independently of environment.}
\label{fig:hierarchical_evolution_cartoon}
\end{figure}

To understand how fast and slow rotators can follow separate evolutionary channels, one needs to consider the hierarchical growth of galaxies and clusters \citep[e.g.][]{DeLucia2012,Wilman2012}. According to our current understanding of galaxy formation \citep[e.g.][]{Mo2010}, after the big bang, the primordial dark matter distribution is imprinted with small fluctuations and these inhomogeneities grow by gravitational instability to form dark matter halos. The primordial gas falls into a nearby halo and loses energy by radiating some of its energy. For this reason it sinks towards the halo center and forms a rotating disk. When the gas becomes sufficiently denser that the halo, it collapses into small clumps due to radiative cooling and starts producing stars into stellar clusters. 

The largest dark matter overdensities are able to acquire large amount of gas early-on, when they are still at the center of their own halo, dominating the overall gravitational potential. They quickly reach the critical mass above which the infalling gas is shock heated by the halo gravitational field and have their star formation suppressed \citep{Keres2005,Dekel2006}. These massive galaxies are the progenitors of the slow rotators. The hot gas is actually observed in X-ray in local massive boxy E \citep{Bender1989}, in core galaxies \citep{Kormendy2009} and in slow rotators in particular \citep{Sarzi2013p19}.  When two groups merge, during the hierarchical build-up of galaxies and clusters, the central galaxies in the two groups, which sit at rest near the center of mass of the halo, will efficiently sink to the center of the resulting larger cluster/group of galaxies via dynamical friction. These massive slow rotator will efficiently merge, due to their large mass and small relative velocities, forming a more massive slow rotator with an inner core scoured by the resulting black hole binary. In this way a typical slow rotator is able to remain as such for the rest of its evolution (\autoref{fig:hierarchical_evolution_cartoon}). 

This picture is similar to the one that motivates the separation of ``central'' (i.e.\ the most massive) and ``satellite'' (i.e.\ the rest) galaxies in dark matter halos, when building theoretical descriptions of galaxy and halo properties \citep[e.g.,][]{Zheng2005,Zehavi2005,Conroy2006}. However in the theoretical approaches every halo has a single central galaxy, by definition. While in reality a given cluster, or a given massive dark matter halo, may contain a handful of slow rotators, which were central galaxies of their respective halos when they formed, but have yet to merge into a single object. For this reason the correspondence between central galaxies and slow rotators is only approximately correct.

The regions of lower dark matter density are unable to efficiently acquire gas and grow more slowly and gradually. They form disk galaxies, which at high redshift are observed to have high gas fractions \citep{Daddi2010,Tacconi2010}, have large velocity dispersion, are turbulent compared to their local counterparts \citep{ForsterSchreiber2006,ForsterSchreiber2009,Genzel2006,Genzel2008,Law2012,Kassin2012,Stott2016} and appear unstable to clump formation \citep{Elmegreen2007,Genzel2011}. These clumps may sink toward the center and start forming the bulge \citep{Bournaud2007clumps,Dekel2009clumps}, or the gas may simply spiral toward the nucleus to grow the bulge in starbursts. During the initial stages, when the velocity dispersion of their groups is quite small, gas rich mergers can also happen. Isolated spirals are able to sustain a nearly constant star formation for a long time, possibly with the contribution of the so called ``cosmological fountain'' effect \citep{Fraternali2008,Fraternali2014}. During this phase they lie on the so-called star formation main sequence \citep{Brinchmann2004,Salim2007}.

However the bulge growth (and \se\ or $\Sigma_1$ rise) increases the likelihood for the galaxy to have its star formation quenched. Some form of feedback is observed to stop the star formation and make the galaxy passive. The link between bulge fraction (or central mass density) and quenching is currently unclear. Outflows from supernovae or from a central AGN \citep[see][for reviews]{Fabian2012,Kormendy2013review,Heckman2014} must both play a role \citep{Silk1998,Granato2004,Bower2006,Croton2006,Hopkins2006}, as may the stabilizing effect of the bulge itself \citep{Martig2009,Ceverino2010}, which may contribute to the observed decrease of the star formation efficiency in ETGs \citep{Saintonge2012,Martig2013,Davis2014}. Evidence of feedback in action within a field fast rotator was presented in \citet{Alatalo2011}. These forms of internal feedback are needed to produce fast rotators in the field. 

But environment has a more dramatic effect. In fact, as soon as one of these disk galaxies is acquired by a large halo, its gravitational pull will be unable to acquire more gas, because their small halos will be orbiting at high speed within the ambient gas, which is at rest within the main halo. Moreover, at sufficiently high gas densities, their own gas, including their corona and fountain effect,  will be stripped entirely \citep{Boselli2006}. As a result, when disk galaxies enter sufficiently massive clusters, they stop growing, due to both the lack of gas accretion and inability to merge because of the high relative velocities of satellites. The environment will now only act via tidal perturbations due to the high-speed encounters, by puffing up their disks \citep[e.g][]{Moore1996}. From then on, the galaxies will be ETGs fast rotators. Their masses will stop growing during subsequent mergers of the parent group/cluster, while slow rotators will continue to grow in mass following the first dry-merging channel. This additional growth of slow rotators during the merging of clusters/groups, tends to produce a gap between the mass distribution of the fast and slow rotators. It explains observed know gap in luminosity between the first ranked galaxy in a massive cluster, and the subsequently ranked galaxies \citep{Sandage1973,Lauer2014}, also clearly visible in \autoref{fig:coma_mass_size} (right panel).

This picture is broadly consistent will all observational evidences, but Nature is certainly not that simple. For example central galaxies in clusters are not always passive \citep{Liu2012}, however the small fraction of star forming ones is consistent with the general picture. Moreover the channel from star forming spirals to bulge-dominated passive fast rotators does not always progress monotonically in one direction. Rejuvenation events can happen in ETGs \citep{Kaviraj2007} and specifically in fast rotators \citep{Young2014}. Moreover the gas versus stars misalignment indicates external accretion \citep{Sarzi2006,Davis2011b}. Small fractions of neutral gas are detected in 40\% of ETGs in the field, especially at large radii \citep{Morganti2006,Oosterloo2010,Serra2012}. And tidal tails are not uncommon also in fast rotators \citep{Duc2011,Duc2015}.  Finally, all trends illustrated in \autoref{fig:mass_size_all} have significant scatter. But the incidence of these events is fully consistent with the general picture we described. In particular, the IFS observations allow one to exclude a scenario in which passive fast rotators can acquire major amounts of gas and become star forming spirals again. These very bulge dominated spirals are extremely rare (e.g. the Sombrero galaxy M104). Moreover, if these events were common, they would erase the clear empirical inverse dependency between bulge fraction (or \se) and star formation indicators.
Similarly one can exclude a scenario in which the slow rotators acquire a disk and become fast rotators. This is because the dynamics of bulges in fast rotators is well described by the simple axisymmetric JAM models over the whole mass range. This contrast with the strikingly different dynamics and shape of slow rotators, as revealed by the IFS data.

\section{CONCLUSIONS}

This review was written at an ideal time, when studies of ETGs using the first generation of IFS survey (\sauron\ \citealt{deZeeuw2002}, \atl\ \citealt{Cappellari2011a} and CALIFA \citealt{Sanchez2012}), those targeting one galaxies at a time, were nearly complete. The second generation of IFS surveys (SAMI \citealt{Bryant2015} and MaNGA \citealt{Bundy2015}) just started but are actively acquiring data. A limited preview of the SAMI results was included in this review. At the same time, the new MUSE IFS \citep{Bacon2010}, on the 8.2-m Very Large Telescope of ESO, started pushing the envelope of the data quality one can achieve on nearby galaxies \citep{Emsellem2014,Krajnovic2015}. 

For this reason, the goal of this review has been to define the status of our knowledge of the structure, kinematics and scaling relations of ETGs, before the arrival of these new large IFS survey. We aim to set a benchmark for assessing how much we have progressed.

Galaxy evolution can be studied via detailed observations in the nearby Universe or using evolutionary studies as a function of redshift. Too often the groups working in one or the other field are unaware of the results of the other and miss the opportunity of combining the two sets of information to advance our knowledge. This review tried to emphasize and illustrate the importance of a synergy between the two approaches.

In the near future, one will be able to combine our knowledge about ETGs from the upcoming multiplexed IFS survey, with the advances that the James Webb Space Telescope \citep{Gardner2006} will bring, thanks to its ability of obtaining deep near-infrared spectra, including IFS (but not multiplexed), targeting the rest-frame optical spectra of ETGs, free of the atmospheric absorptions. This will provide clean kinematics and stellar population for ETGs out to the peak of their assembly epoch around $z\sim2$.

In the more distant future, with the upcoming class of 40-m telescopes, like the European Extremely Large Telescope (E-ELT) and the Thirty Meter Telescope (TMT), coupled with multiplexed capabilities in the near-infrared it will become possible to perform IFS surveys like the ones we described here, out to $z\sim2$. We can only be looking forward to witness those new developments.

\begin{summary}[SUMMARY POINTS]
\begin{enumerate}
\item Using IFS kinematics, ETGs separate into two structurally homogeneous classes with/without disks: the fast/slow rotators. We showed that there is a dichotomy, not a smooth transition, between the two classes.
\item This dichotomy broadly agrees with the previously identified photometric separation into e.g.\ core/power-law ETGs. But the kinematic classification is nearly independent on inclination and does not require high spatial resolution. This makes it ideal for large IFS surveys.
\item IFS revealed a close link between ETGs and spirals. Below $M_{\rm crit}\approx2\times10^{11} \msun$, fast rotator ETGs form a parallel sequence in galaxy properties with spiral galaxies. While core slow rotators dominate above $M_{\rm crit}$. 
\item In the spirals to fast rotators sequence, suppression of star formation and molecular gas fraction are driven by the {\em central} mass density slope, or bulge mass fraction. 
\item ETGs are dominated by stellar mass within \re\ and their \mldyn\ is mainly due to systematic variations in the stellar population, including the IMF. 
\item The total mass density is well-described by $\rho_{\rm tot}\propto r^{-2.2}$, from the center out to at least $4\re$ with small scatter (in the currently explored mass range).
\item Slow rotators with cores are found near the centers of clusters/groups or subgroups within clusters. Fast rotators are distributed like spiral galaxies.
\item IFS observations, and redshift evolution studies, consistently and independently indicate a scenario where the evolution of fast/slow rotators follows two distinct (i) gas-accretion driven and (ii) dry-merger driven, evolutionary channels.
\end{enumerate}
\end{summary}

\begin{issues}[FUTURE ISSUES]
\begin{enumerate}
\item What are the trends in population and kinematics at radii well beyond $R\ga2\re$? Much information on galaxy assembly is contained in the stellar halos.
\item What is the physical mechanism making galaxies passive? Large IFS surveys, of both spirals and ETGs, can answer this long-standing question by combining spatially-resolved  gas and stellar observables with environment. 
\item Do we understand stellar population and IMF in galaxies? Can we reliably predict stellar masses from spectra, or are there fundamental unsolved issues?
\item What are the trends in {\em total} density profiles at large radii where dark matter dominates? How do they relate to spirals? And to detailed model predictions?
\item How do spatially-resolved population and kinematics evolve with redshift? Can we directly trace the assembly of mass and metals over time?
\item Can we follow the dark halo growth as a function of time using resolved IFS of the stars and gas dynamics?
\end{enumerate}
\end{issues}

\section*{DISCLOSURE STATEMENT}

The author is not aware of any affiliations, memberships, funding, or financial holdings that might be perceived as affecting the objectivity of this review. 

\section*{ACKNOWLEDGMENTS}

Many of the ideas presented in this review, originated during more than a decade of enjoyable and productive collaboration, initially with the \sauron\ team and subsequently with the \atl\ collaboration. I acknowledge with gratitude the numerous meetings and enlightening discussions with the members of the two teams. I am most grateful to my Scientific Editor John Kormendy and co-editor Sandy Faber for fruitful discussions and comments. I have been fortunate to be able to benefit from their insights and unique expertise on this subject. 
It is a particular pleasure to thank Eric Emsellem, Davor Krajnovi\'c and Tim de Zeeuw for thoughtful suggestions. I thank Jes\'us Falc\'on-Barroso for providing the CALIFA \lre\ values.
I acknowledge support from a Royal Society University Research Fellowship. I warmly thank my wife Christi Warner for making this review possible with her constant support and understanding.

\end{document}